\documentclass[a4paper,11pt]{article}
\pdfoutput=1 

\usepackage{jheppub} 
\usepackage{tikz}
\usetikzlibrary{matrix} 
\usepackage{hyperref}
\usepackage{cleveref}
\usepackage{supertabular}
\usepackage{todonotes}
\usepackage{mathrsfs}
\usepackage{graphicx}
\usepackage{array}
\usepackage{lipsum}
\usetikzlibrary{arrows.meta}

\newcommand{\newreptheorem}[2]{%
\newenvironment{rep#1}[1]{%
 \def\rep@title{#2 \ref{##1}}%
 \begin{rep@theorem}}%
 {\end{rep@theorem}}}
\makeatother

\newreptheorem{lemma}{Lemma}

\newreptheorem{conj}{Conjecture}

\usepackage{lscape} 
\usepackage{braket}

\newcommand{\be}{\begin{equation}}
\newcommand{\ee}{\end{equation}}
\newcommand{\ba}{\begin{aligned}}
\newcommand{\ea}{\end{aligned}}
\newcommand{\bea}{\begin{eqnarray}}
\newcommand{\eea}{\end{eqnarray}}

\newcommand{\FT}{{\mathcal{T}_{\mathbf{X}}^{\rm 5d}}}
\newcommand{\FTfour}{{\mathcal{T}_{\mathbf{X}}^{\rm 4d}}} 
\newcommand{\tX}{{\widetilde{\mathbf{X}}}}
\newcommand{\hX}{{\widehat{\mathbf{X}}}}

\newcommand{\EQfour}{{\text{EQ}^{(4)}}} 
\newcommand{\MQfour}{{\text{MQ}^{(4)}}} 
\newcommand{\EQfive}{{\text{EQ}^{(5)}}} 
\newcommand{\MQfive}{{\text{MQ}^{(5)}}} 

\tikzstyle{brane}=[draw]
\tikzset{D7/.style={circle, draw=black, inner sep=0pt, fill=white, minimum size=3mm}}
\tikzset{hasse/.style={circle, fill,inner sep=2pt}}
\tikzset{flavor/.style={regular polygon,fill=white,regular polygon sides=4,inner sep=2.5pt, draw}}
\tikzset{gauge/.style={circle, draw,inner sep=2.5pt}}
\tikzset{gaugeb/.style={circle, draw,fill=black,inner sep=2.5pt}}
\tikzset{gauger/.style={circle, draw,fill=cyan,inner sep=2.5pt}}
\tikzset{gaugeg/.style={circle, draw,fill=red,inner sep=2.5pt}}
\tikzset{SUd/.style={circle, draw=black, inner sep=0pt, fill=yellow, minimum size=2mm}}
\tikzset{bd/.style={circle, draw=black, inner sep=0pt, fill=black, minimum size=2mm}}
\tikzset{wd/.style={circle, draw=black, inner sep=0pt, fill=white, minimum size=2mm}}
\tikzset{Dynkin/.style={circle, draw=black, inner sep=0pt, fill=white, minimum size=2mm}}
\tikzstyle{ligne}=[draw, thick] 
\tikzset{doublearrow/.style={ draw=black!75, color=black!75, thick, double distance=3pt, }} 

\def\mb{\mathbb}
\def\mbf{\mathbf}
\def\mc{\mathcal}
\def\bp{\begin{pmatrix}}
\def\ep{\end{pmatrix}}

\def\h{\widehat}

\def\mb{\mathbb}
\def\mbf{\mathbf}
\def\mc{\mathcal}
\def\bp{\begin{pmatrix}}
\def\ep{\end{pmatrix}}

\def\beq#1\eeq{\begin{align}#1\end{align}}

\graphicspath{ {figs/} }

\title{Higgs Branch and VOA of 4d $\mc{N}=2$ SCFTs from IIB}
\preprint{}

\author[a,b]{Yi-Nan Wang}
\author[c]{Wenbin Yan}
\author[b]{Peihe Yang}

\affiliation[a]{School of Physics, Peking University,\\
Beijing, China, 100871}
\affiliation[b]{Center for High Energy Physics, Peking University, \\
Beijing, China, 100871}
\affiliation[c]{Yau Mathematical Sciences Center, Tsinghua University, \\
Beijing, China, 100084}

\abstract{We study the Higgs branch and associated vertex operator algebra (VOA) of 4d $\mathcal{N}=2$ superconformal field theories (SCFTs) from the geometric engineering of IIB superstring on canonical threefold singularities. For terminal singularities, we explain how to derive the 4d Higgs branch from their small resolution. We also investigate singularities with compact 4-cycles in their crepant resolution, and discuss different ways to compute their Higgs branch. Using a symplectic duality argument, we propose the first examples of 4d $\mathcal{N}=2$ SCFTs with the E-type Kleinian singularities as their Higgs branches, and conjecture their associated VOA to be affine E-type W-algebra. Many new VOAs with no known W-algebra descriptions are found, with conjectured associated varieties. We investigate the singularities associated with lisse VOAs and propose predictions for the BPS quivers of $D_N^N[k]$ and $E_7^{14}[k]$ from the perspective of deformed singularities. As an application, we show directly that the $D_N^N[1]$ theory is equivalent to the $(A_1,A_{N-2})$ Argyres-Douglas theory, and illustrate, in a general setting, how the Schur index is determined by the BPS quiver.}

\begin{document} 
\maketitle

\section{Introduction}
\label{sec:intro}

4d $\mc{N}=2$ superconformal field theories (SCFTs) have rich mathematical and physical structures, which are subjects of active researches in the past a few decades, see e.g. \cite{Tachikawa:2013kta,Akhond:2021xio} for a review. Due to the difficulty in extracting the full CFT data, one is often restricted to study the vacuum properties of 4d $\mc{N}=2$ SCFTs, especially the \textit{Coulomb branch} and the \textit{Higgs branch}. 

On the Coulomb branch, the effective field theory becomes a 4d $\mc{N}=2$ abelian gauge theory, whose effective action and moduli space are incorporated in the beautiful Seiberg-Witten theory~\cite{Seiberg:1994rs}. One can define the \textit{BPS quiver} on the Coulomb branch of a 4d $\mc{N}=2$ SUSY field theory that characterizes the dyonic charges of the BPS particles~\cite{Cecotti:2011rv}, which is useful for our purposes. Moreover, in the superconformal operator spectrum there is a class of short multiplets denoted by ${\cal E}_{r,(0,0)}$, whose primary is known as the Coulomb branch operator.\footnote{The notation of \cite{Dolan:2002zh} is adopted here. ${\cal E}_{r,(0,0)}$ is used to denote both the short multiplet and its primary. Here $r$ is the $U(1)_r$ charge.} The vev $\langle {\cal E}_{r,(0,0)}\rangle$ parametrizes the 4d Coulomb branch, and the scaling dimensions of these operators give important information of the 4d $\mc{N}=2$ SCFT. For instance, if there exist Coulomb branch operators with fractional scaling dimension, the 4d $\mc{N}=2$ SCFT is usually a non-Lagrangian or contains a non-Lagrangian part, e.g. the minimal Argyres-Douglas theory~\cite{Argyres:1995jj} (named as the $(A_1,A_2)$ Argyres-Douglas theory now).

On the other hand, the Higgs branch of 4d $\mc{N}=2$ SCFTs is less well-studied. In general, the Higgs branch of a SUSY field theory with 8 supercharges is a hyperk\"{a}hler (symplectic) singularity, which is not subject to quantum corrections. Recently, methods such as the \textit{Hasse diagram} were developed to characterize the foliation structure of these symplectic singularities~\cite{Bourget:2019aer}. 
When the 3d mirror dual of the 4d  $\mc{N}=2$ SCFT is a 3d $\mc{N}=4$ quiver gauge theory\footnote{The quiver of this 3d  $\mc{N}=4$ quiver gauge theory is also called the \emph{magnetic quiver} of the 4d theory in literature.}, such Hasse diagram can be derived from the \textit{quiver subtraction} method~\cite{Cabrera:2018ann} or the \textit{decay and fission} method~\cite{Bourget:2023dkj}.
Nonetheless, when the 3d mirror dual does not possess a Lagrangian description or cannot be easily derived, the explicit derivation of the Higgs branch geometry still largely remains as an open problem. An example of Higgs branch without a Lagrangian 3d mirror is the E-type Kleinian singularity $\mb{C}^2/\Gamma_{E_n}$ $(n=6,7,8)$, and we will later present examples of 4d $\mc{N}=2$ SCFTs with such Higgs branch.

Another important invariant of a 4d $\mathcal{N}=2$ SCFT is its associated vertex operator algebra (VOA)~\cite{Beem:2013sza}, which is closely related to its Higgs branch. In the SCFT/VOA correspondence, the vacuum character of the VOA coincides with the Schur index of the corresponding 4d $\mathcal{N}=2$ SCFT. Moreover, the associated variety of the VOA is identified with the Higgs branch of the $\mathcal{N}=2$ SCFT~\cite{Song:2017oew,Beem:2017ooy,Arakawa:2017aon}.

 For a general 4d $\mathcal{N}=2$ SCFT, the corresponding VOA is conjectured to be quasi-lisse, which implies that the VOA is strongly finitely generated and that its associated variety admits only finitely many symplectic leaves~\cite{Beem:2017ooy,Arakawa:2016hkg}. The Schur index of certain class S theories can be computed using a topological quantum field theory (TQFT) approach~ \cite{Gadde:2011uv,Lemos:2014lua,Buican:2015ina,Song:2015wta,Gadde:2011ik,Buican:2015hsa,Buican:2017uka,Mekareeya:2012tn,Lemos:2012ph}. For more general 4d $\mathcal{N}=2$ theories, the Schur index can also be evaluated via an infrared formula on the Coulomb branch~\cite{Cordova:2015nma} through the BPS quiver associated with the 4d $\mathcal{N}=2$ theory.
	

The classification of all 4d $\mc{N}=2$ SCFTs is still a difficult and open problem, while certain subclasses have been explored. In this paper, we will restrict ourselves to the geometric engineering approach, where the 4d $\mc{N}=2$ SCFT $\FTfour$ is constructed from IIB superstring theory on an isolated canonical threefold singularity $\mbf{X}$~\cite{Shapere:1999xr}. The classification of such singularities $\mbf{X}$ and the corresponding $\FTfour$ is carried out for the cases of isolated hypersurface singularities~\cite{Xie:2015rpa,Closset:2021lwy}, complete intersection singularities~\cite{Chen:2016bzh,Wang:2016yha} and rigid singularities~\cite{Chen:2017wkw}. The SCFT data such as the Coulomb branch spectrum, rank $\h r$, flavor rank $f$, central charges $a$ and $c$ can be computed from the mini-versal deformation of $\mbf{X}$, denoted as $\hX$. The BPS quiver of $\FTfour$ can be read off from the intersection numbers of the $2\h{r}+f$ 3-cycles in the deformed threefold $\h{\mbf{X}}$, and results are available for certain classes of isolated hypersurface singularities~\cite{Klemm:1996bj,Aspinwall:2004jr,Alim:2011ae,Alim:2011kw}.

Furthermore, in the recent program of 5d/4d correspondence~\cite{Closset:2020scj,Closset:2020afy,Closset:2021lwy,Mu:2023uws}, relations between the Coulomb branch of $\FTfour$ and the Higgs branch of the 5d SCFT $\FT$ defined as M-theory on $\mbf{X}$ were proposed, since they should be both connected to the deformed threefold $\widehat{X}$. In short, we first consider a 3d $\mc{N}=4$ theory that is the $S^1$ reduction of $\FTfour$, named the electric quiver(ine)\footnote{In \cite{Closset:2020scj} an electric or magnetic quiver is called a ``quiverine'' if it is non-Lagrangian.} $\EQfour$. $\EQfour$ has the same Higgs branch as $\FTfour$ due to the RG-flow invariance of the Higgs branch. The 3d mirror of $\EQfour$ is denoted as the magnetic quiver(ine) $\MQfour$ for the 4d theory~\cite{Intriligator:1996ex}. Similarly, we also introduce the $\EQfive$ as the $T^2$ reduction of $\FT$ and its 3d mirror $\MQfive$. Then it was proposed in \cite{Closset:2020scj} that the magnetic quiver $\MQfive$ for the 5d theory should be obtained by the gauging of $U(1)^f$ from the 4d \textit{electric quiver} $\EQfour$, thus relating the Coulomb branch of $\FTfour$ with the Higgs branch of $\FT$. 


A natural and largely unaddressed question is then how to compute the Higgs branch of $\FTfour$ from the geometry of $\mbf{X}$. In contrast to the Coulomb branch of $\FTfour$ encoded in the deformation $\hX$, from IIB perspective the Higgs branch of $\FTfour$ should be associated with the crepant resolution $\tX$. For the cleanness of our discussions, we only allow compact 2-cycles and 4-cycles in the resolved space $\tX$\footnote{3-cycles in the resolved space $\tX$ give rise to free vector multiplets~\cite{Closset:2020scj}, implying that $\tX$ does not precisely describe the Higgs branch of $\FTfour$.}. Denoting the Coulomb branch rank of $\FT$ by $r$, the total number of independent 2-cycles is given by $\dim(H_2(\tX,\mb{Z}))=r+f$, and the total number of independent 4-cycles is given by $\dim(H_4(\tX,\mb{Z}))=r$. There is then a relation between $r$, $f$ and the Higgs branch quaternionic dimension $\h{d}_H$ of $\FTfour$:
\be
\h{d}_H=\dim(H_2(\tX,\mb{Z}))=r+f\,.
\ee
We would then like to derive the 4d Higgs branch information either from the magnetic quiver $\MQfour$ or alternative methods when such magnetic quiver description is unavailable. 

The other goal of this paper is to initiate the study of the VOA $V_\mbf{X}$ associated to $\FTfour$, for isolated canonical threefold singularities $\mbf{X}$. As mentioned before, the canonical approach to derive the VOA $V_\mbf{X}$ is via the 4d $\mc{N}=2$ SCFT/VOA correspondence. $a$ and $c$ anomaly coefficients of $\FTfour$ give the central charge and asymptotic growth of vacuum characters of $V_\mbf{X}$, while the Higgs branch of $\FTfour$ gives the associated variety of $V_\mbf{X}$. If  the intersection pairing $n_{ij}=\Sigma_i\cdot\Sigma_j$ of compact 3-cycles in the deformed Calabi-Yau threefold $\hX$ is further known, one can compute the Schur index of $\FTfour$, hence the vacuum character of $V_\mbf{X}$. For particular $\mbf{X}$s, the corresponding $V_\mbf{X}$s can be identified with known VOAs. 


We now summarize the main results and structure of this paper.

\paragraph{Magnetic quiver for terminal singularities.}

We discuss the cases of terminal singularities $\mbf{X}$ with a small crepant resolution $\tX$, i.e. $\tX$ has no compact 4-cycle. Thus 5d SCFT $\FT$ constructed as M-theory on $\mbf{X}$ is a rank $r=0$ theory. In Section~\ref{subsec:MQ-resolution} we present the general construction of the 4d magnetic quiver $\MQfour$ of $\FTfour$ from the T-dual description of IIA superstring on $S^1\times \tX$. Namely, the $U(1)^f$ gauge nodes in $\MQfour$ are one-to-one corresponds to the non-compact divisors $D_\alpha$ $\alpha=1,\dots,f$ in $\tX$. The charged hypermultiplets in $\MQfour$ are then exactly correspond to D2-branes wrapping 2-cycles $C$ that contribute to the Gopakumar-Vafa (GV) invariants of $\tX$, which was studied extensively for the resolution of canonical singularities recently~\cite{Collinucci:2021wty,Collinucci:2021ofd,DeMarco:2021try,Collinucci:2022rii,DeMarco:2022dgh}. In $\MQfour$, the charge of such hypermultiplet corresponding to $C$ under the $\alpha$-th $U(1)$ is then given by the intersection number $D_\alpha\cdot C$. With this relation we confirmed the relation between $\MQfour$ and GV-invariants for type $(A_k,A_l)$, $(A_1,D_N)$ and even the $(A_{2n-1},E_7)$ Argyres-Douglas theory\footnote{We abbreviate the type $(G,G')$ Argyres-Douglas theory as $(G,G')$ theory later.}.

\paragraph{Higgs branch from IIB superstring for terminal singularities.}

For a resolved $\tX$ of a terminal singularity $\mbf{X}$, the extended K\"{a}hler cone is a real $f$-dimensional space, while the 4d Higgs branch has hyperK\"{a}hler (symplectic) dimension $\h{d}_H=f$. We explain how does the hyperk\"{a}hler dimension of the 4d Higgs branch arise from the $B_2$, $C_2$, $C_4$ fields in IIB superstring theory, for the generalized conifolds $\mbf{X}_{(A_1,A_{2N-1})}:\ xy+z^2+w^{2N}=0$ corresponding to type $(A_1,A_{2N-1})$ Argyres-Douglas theory in Section~\ref{subsec:HBfromresolution}. The Higgs branch of the corresponding 4d SCFT is the Kleinian singularity $\mb{C}^2/\mb{Z}_N$ in this case, due to the presence of $N$ exceptional 2-cycles in $\tX$ that shrink to zero volume in the singular limit. In Section~\ref{sec:A1A2N-1} we reproduce this result from the explicit resolution of $\mbf{X}_{(A_1,A_{2N-1})}$, and confirm its consistency with the GV invariants. Such argument is applied to other $(G,G')$ theories to explain the appearance of Kleinian singularity $\mb{C}^2/\mb{Z}_N$ in the Hasse diagram of the 4d Higgs branch.

\paragraph{$\FTfour$ for $\mbf{X}$ with $r>0$, $f=0$ and low-dimensional Higgs branch.} 

In presence of compact 4-cycles in the resolved space $\tX$, it is not straightforward to derive the 4d Higgs branch or magnetic quiver from IIB perspective. We attempt to unveil the 4d Higgs branch for cases with small $\h{d}_H=r+f$. In Section~\ref{subsec:r=1} we first define the singularities sing$(E_n)$ $(n=6,7,8)$ in Table~\ref{tab:singularities}, with $\h{d}_H=1$, $r=1$, $f=0$:
\be
\ba
\label{E6E7E8-sing}
&\mathrm{sing}(E_8):\ x_3^7+x_4^5 x_3+x_2^3+x_1^2=0\,,\cr
&\mathrm{sing}(E_7):\ x_2^5+x_3^3 x_2+x_3 x_4^3+x_1^2=0\,,\cr
&\mathrm{sing}(E_6):\ x_2^4+x_3^2 x_2+x_1^3+x_3 x_4^2=0\,.
\ea
\ee
$\mbf{X}=$sing$(E_n)$ is closely related to the local $dP_n$ del-Pezzo singularity (denoted as $\mbf{X}'$) as they have the same resolution sequence. Nonetheless, in the resolved $\tX$ the exceptional 4-cycle develops a type $E_n$ du Val surface singularity.

We argue that the 5d SCFTs $\FT$ and $\mc{T}_{\mbf{X}'}^{\rm 5d}$ have the same Higgs branch, which is the minimal nilpotent orbit of $E_n$, which means that the $\MQfive$ for $\FT$ is the unitary quiver with the shape of the affine $E_n$ Dynkin diagram. Then as $f=0$ for $\mbf{X}=$sing$(E_n)$, we have $\EQfour=\MQfive$, thus $\EQfour$ is also a unitary quiver with the shape of the affine $E_n$ Dynkin diagram. Using the symplectic duality~\cite{Braden:2014iea, Kamnitzer:2022nzd} known as the \textit{inversion} in physics literature~\cite{Grimminger:2020dmg}, we deduce that the Higgs branch of $\FTfour$ should be an $E_n$-type Kleinian singularity $\mb{C}^2/\Gamma_{E_n}$. These are the first examples of 4d $\mc{N}=2$ SCFT with $\mb{C}^2/\Gamma_{E_n}$ as its Higgs branch.

In Section~\ref{subsec:r=2} we further discuss a number of examples $\mbf{X}$ with $\h{d}_H=2$, $r=2$, $f=0$, listed in Table~\ref{t:rank-2}. We propose their Higgs branch Hasse diagrams by the same inversion argument.
\paragraph{W-algebras correspond to singularities}
In Sections~\ref{subsec:r=1} and~\ref{subsec:r=2}, we identify certain singularities whose associated VOA are W-algebras. For example, the associated VOAs of singularities sing$(E_i)$ in (\ref{E6E7E8-sing}) are expected to be the affine W-algebra
\begin{equation}
	\mathcal{W}_{k_{2d}}(E_i,\mathcal{O}_{\text{subregular}}),~~k_{2d}=-h^\vee+\frac{h^\vee}{h^\vee+1},~~i=6,7,8\,.
\end{equation}
Using results from 5d SCFT and the inversion procedure, we  propose that the $S^1$ reduction of the corresponding 4d SCFT flows to a three-dimensional $\mathcal{N}=4$ electric quiver of $E_6$, $E_7$, or $E_8$ type whose Higgs branch matches the associated variety of the corresponding affine W-algebra, which are the $\mb{C}^2/\Gamma_{E_n}$ Kleinian singularities.

For the $r=2$ cases, we also identify the VOA associated with the singularity
\begin{equation}
	\mbf{X}:\ x_1^2+x_2^5+x_3^{11}+x_3x_4^3=0
\end{equation}
to be the affine W-algebra
\begin{equation}
	W_{-30+\frac{30}{31}}(\mathfrak{e}_8,E_8(a_2))\,,
\end{equation}
whose associated variety is consistent with the Higgs branch Hasse diagram from the inversion:
\begin{equation}
	\begin{tikzpicture}[x=.5cm,y=.5cm]
		\node[bd] at (4,0) {};
		\node[bd] at (4,2) {};
		\node[bd] at (4,4) {};
		\draw[ligne, black](4,0)--(4,2) node[midway,left] {$E_7$};
		\draw[ligne, black](4,2)--(4,4) node[midway,left] {$E_8$};
	\end{tikzpicture}
\end{equation}

\paragraph{Higgs branch for $\mbf{X}$ with smooth resolution and smooth divisors.} 

In Section~\ref{subsec:smoothresolution} we discuss a few examples of $\mbf{X}$ with smooth crepant resolution and smooth divisors, namely the ones corresponding to $(D_{2m},D_{2n})$ Argyres-Douglas theories. It is interesting that the magnetic quiver $\MQfour$ of these theories were found in \cite{Carta:2021dyx}, despite of the difficulty in a direct IIB derivation. From these magnetic quivers we derive and discuss the Higgs branch Hasse diagram, from the decay and fission approach~\cite{Bourget:2023dkj}. For example, the $(D_4,D_4)$ theory (which coincides with the local $E_6$ singularity) has the magnetic quiver
\begin{equation}
	\begin{tikzpicture}[scale=1.2, baseline=-0.5ex]
    \node at (-2.0, 0) {$\MQfour = $}; 
		\draw (0,0) -- (1,0);
		\draw (0,0) -- (0.3,0.9);
		\draw (0,0) -- (0.3,-0.9);
		\draw (0,0) -- (-0.8,0.6);
		\draw (0,0) -- (-0.8,-0.6);
		
		\node[draw,circle,fill=black,inner sep=1.5pt] at (0,0) {};
		\node[draw,circle,fill=black,inner sep=1.5pt] at (1,0) {};
		\node[draw,circle,fill=black,inner sep=1.5pt] at (0.3,0.9) {};
		\node[draw,circle,fill=black,inner sep=1.5pt] at (-0.8,0.6) {};
		\node[draw,circle,fill=black,inner sep=1.5pt] at (0.3,-0.9) {};
		\node[draw,circle,fill=black,inner sep=1.5pt] at (-0.8,-0.6) {};
		
		\node at (-0.1,-0.3) {$~C_2$};
		\node at (0.5,0.9) {$~D_1$};
		\node at (0.5,-0.9) {$~D_1$};
		\node at (-1,0.6) {$D_1~$};
		\node at (-1,-0.6) {$D_1~$};
		\node at (1.2,0) {$~D_1$};
	\end{tikzpicture}
	\ / \ \mathbb{Z}_2
\end{equation}
and the Higgs branch Hasse diagram
\begin{equation}
		\begin{tikzpicture}[x=1cm,y=1cm]
			\node (1) [hasse] at (0,1) {};
\node (2) [hasse] at (0,0) {};
\node (3) [hasse] at (0,-1) {};
\node (4) [hasse] at (0,-2) {};

\draw (1) edge [] node[label=left:$d_4$] {} (2);
\draw (2) edge [] node[label=left:$A_1$] {} (3);
\draw (3) edge [] node[label=left:$A_3$] {} (4);
		\end{tikzpicture} 
	\end{equation}
consistent with the Higgsing pattern of the quiver gauge theory 
\be
\begin{array}{rcl}
& [1] &\\
& \vert & \\
& SU(2) & \\
& \vert & \\
\FTfour= {[1]}-SU(2)- & SU(3) & -SU(2)-{[1]}
\end{array}\,.
\ee

\paragraph{BPS quiver and the Schur index from singularities.}

As reviewed in Section~\ref{subsec:IntersectionBPSquiver}, BPS quivers associated to $\FTfour$ can be computed for the following two classes of $\mbf{X}$:

(1) The direct sum singularities of the form $f(z_1,z_2)+g(z_3,z_4)=0$, including the $(G,G')$ Argyres-Douglas theories  \cite{gabrielov1973intersection}. 

(2) The following singularities under a conjecture of Orlik and Randell~\cite{orlik1977monodromy}:
\begin{equation}
	\begin{aligned}F(z_0,\ldots,z_n)=z_0^{a_0}+z_0z_1^{a_1}+\ldots+z_{n-1}z_n^{a_n},\quad n\geq1\,.\end{aligned}
\end{equation}

With the BPS quiver, in section~\ref{subsec:Schurindex} we discuss the computation of the Schur index \ref{Schur-conj}, using the quantum torus algebra generated by BPS particles in a wall-crossing chamber conjectured in \cite{Cordova:2015nma}.
In Section~\ref{subsec:lisseVOA} we apply the method to compute the Schur index for $\mbf{X}$ that corresponds to a lisse VOA $V_{\mbf{X}}$, i.e. $r=f=0$. These include the first computation of the $D_N^N[k]$ theory associated to the singularity $\mbf{X}:\ x_1^2+x_2^{N-1}+x_2 x_3^2+z^k x_3=0$ and the $E_7^{14}[k]$ theory associated to the singularity $\mbf{X}:\ x_1^2+x_2^3+x_2 x_3^3+z^k x_3=0$.

Finally, we summarize the singularities and their associated VOA data discussed in this paper in Table~\ref{tab:singularity-voa} (The method presented in this paper is applicable to a variety of cases. Here, we list only the examples considered in this work).

\begin{table}[h!]
	\centering
	\rotatebox{90}{\begin{tabular}{|c|c|c|c|}
		\hline
		Singularity 
		& Central charge $c_{2d}$ 
		& MQ/Associated Variety
		& W-algebra realization \\
		\hline
		$x_1^2+x_2^2+x_3^2+x_4^{2N}$ 
		& $-\frac{2 \left(3 N^2-N-1\right)}{N+1}$ 
		& $\mathbb{C}^2/\mathbb{Z}_N\cong\mathcal{N}_{\mathfrak{sl}_N}\cap S_{[N-1,1]}$
		& $W_{-\frac{N^2}{N+1}}(\mathfrak{sl}_N,[N-1,1])$ \\
		\hline
		$x_1^2+x_2^2+x_3^{k+1}+x_4^{k+1}$ 
		& $-\frac{1}{2} \left(k^3+3 k^2-2 k\right)$ 
		& Complete graph \cite{Giacomelli:2020ryy,Xie:2021ewm}
		&  \\
		\hline
			$x_1^2+x_2^2+x_3(x_3^{N-2}+x_4^2)$ (odd N) 
		& $-3(N-1)$ 
		& $\mathbb{C}^2/\mathbb{Z}_2=\mathcal{N}_{\mathfrak{sl}_2}$
		&  $L_{-2+\frac{2}{N}}(\mathfrak{sl}_{2})$\\
		\hline
		$x_1^2+x_2^{4n}+x_3(x_3^{5}+x_4^2)~(\gcd(3,n)=1)$  
		& Table \ref{acforAD}
		& \eqref{MQAD}
		&  \\
		\hline
		$x_1^2+x_2^2+x_3(x_3^{N-2}+x_4^2)$ (even N) 
		& $-3N+4$ 
		& $\overline{\mathcal{O}}_{[k-1,1]}\cap S_{[k-2,1^2]},$
		&  $W_{-k+\frac{k}{k-1}}(\mathfrak{sl}_k,[k-2,1^2])$\\
		\hline
			$x_1^2+x_2^{2N}+x_3(x_3^2+x_4^3)$ 
		& Table~\ref{acAE}
		& \eqref{MQAE}
		&  \\
		\hline
			$x_3^7+x_4^5 x_3+x_2^3+x_1^2$
		& $-\frac{13978}{31}$
		& $\mathbb{C}^2/\Gamma_{E_8}$
		&  $\mathcal{W}^{-30+\frac{30}{31}}(E_8,\mathcal{O}_{\text{subregular}})$\\
		\hline
			$x_2^5+x_3^3 x_2+x_3 x_4^3+x_1^2$ 
		& $-\frac{3706}{19}$
		& $\mathbb{C}^2/\Gamma_{E_7}$
		&  $\mathcal{W}^{-18+\frac{18}{19}}(E_7,\mathcal{O}_{\text{subregular}})$\\
		\hline
			$x_2^4+x_3^2 x_2+x_1^3+x_3 x_4^2$ 
		& $-\frac{1342}{13}$
		& $\mathbb{C}^2/\Gamma_{E_6}$
		&  $\mathcal{W}^{-12+\frac{12}{13}}(E_6,\mathcal{O}_{\text{subregular}})$\\
		\hline
		$x_1^2+x_2^5+x_3^{11}+x_3x_4^3$
		& $-\frac{33212}{31}$
		& $\mathcal{N}_{E_8}\cap S_{E_8(a_2)}$
		&  $W_{-30+\frac{30}{31}}(\mathfrak{e}_8,E_8(a_2))$\\
		\hline
			$x_1^5+x_2^3+x_2 x_3^3+x_3 x_4^2$
		& $-\frac{4532}{13}$
		& \eqref{hasse1}
		&  \\
		\hline
			$x_1^4+x_2x_4^3+x_2^2x_3+x_3^2x_4$
		& $-244$
		& \eqref{MQr22}
		&  \\
		\hline
			$x_{1}^{2}+x_{2}x_{4}^{4}+x_{2}^{3}x_{3}+x_{3}^{5}x_{4}$
		& $-548$
		& \eqref{MQr23}
		&  \\
		\hline
				$x_{1}^{2}+x_{2}^{3}+x_{2}x_{3}^{5}+x_{3}x_{4}^{7}$
		& $-\frac{21068}{19}$
		& \eqref{MQr24}
		&  \\
		\hline
				$x_1^{3}+x_2^{3}+x_1 x_3^2+x_2x_4^2$
		& $-58$
		& \eqref{Hasse from Lagrangian}
		&  \\
		\hline
					$x_1^{5}+x_2^{5}+x_1 x_3^2+x_2x_4^2$
		& $-208$
		& \eqref{MQDD2}
		&  \\
		\hline
	\end{tabular}}
	\caption{Summary of singularities and their associated VOA data.}
	\label{tab:singularity-voa}
\end{table}

\section{Setups}
\label{sec:setup}

\subsection{4d $\mathcal{N}=2$ SCFTs from IIB on threefold singularity}\label{subsec:4dfromIIB}

A four-dimensional $\mathcal{N}=2$ theory can be obtained by compactifying Type IIB string theory on a Calabi--Yau threefold $\mathbf{X}$. 
In this paper, we always consider non-gravitational quantum field theories. To decouple  gravity, the volume of the Calabi--Yau threefold should be taken to infinity, i.e. the Calabi-Yau threefold is always non-compact. 
In addition, to obtain an interacting 4d $\mc{N}=2$ superconformal field theory, one should consider Type IIB string theory compactified on a Calabi–Yau threefold with an isolated singularity at the origin. In this setup, all degrees of freedom of the SCFT are localized at the singularity. Moreover, these singularities should be \emph{canonical} to avoid infinite distance in the moduli space~\cite{Alvarez-Garcia:2023gdd,Alvarez-Garcia:2023qqj}.

Among Calabi--Yau threefolds with singularities, we focus on a subclass of canonical \emph{isolated hypersurface singularities} (IHS) in $\mathbb{C}^4$~\cite{yau2005classification,Xie:2015rpa}:
\begin{equation}
	\mathbf{X}:=\{F(x_1,x_2,x_3,x_4)=0\}\subset\mathbb{C}^4\,,
\end{equation}
where the equations $F = dF = 0$ admit a unique solution at $x_i = 0$. 
To engineer a four-dimensional $\mathcal{N}=2$ SCFT, the following additional conditions are required \cite{Shapere:1999xr}:
\begin{itemize}
	\item The polynomial $F$ must be \emph{quasi-homogeneous}, i.e. there exists a $\mathbb{C}^*$ action with positive weights $q_i>0$ such that
	\begin{equation}
		\mathbb{C}^*:F(\xi^{q_i}x_i)=\xi F(x_i)\:,\:q_i>0\:,\:i=1,2,3,4\:.
	\end{equation}
	This $\mathbb{C}^*$ action can be associated to the $U(1)_r$ R-symmetry in the four-dimensional $\mathcal{N}=2$ superconformal field theory.
	\item In addition to the quasi-homogeneity condition, the canonical condition requires
	\begin{equation}
		\sum_i q_i>1\,.
	\end{equation}
\end{itemize}

We now summarize the relations between the Coulomb and Higgs branches of the 4d $\mc{N}=2$ SCFT and the Calabi--Yau threefold geometry~\cite{Closset:2020scj}. 


\paragraph{Coulomb branch}
The Coulomb branch data are encoded in the mini-versal deformation of the IHS $\mbf{X}$, whose deformation parameters correspond to independent Coulomb branch operators~\cite{Shapere:1999xr}. We will review this correspondence below.

The mini-versal deformation $\widehat{\mathbf{X}}$ of IHS $\mbf{X}$ takes the form
\begin{equation}
	\widehat{\mathbf{X}}\cong\left\{\widehat{F}(x)\equiv F(x)+\sum_{l=1}^\mu t_lx^{\mathfrak{m}_l}=0\right\}\,,
\end{equation}
where the monomials $x^{\mathfrak{m}_l}$ form a basis of the Milnor ring
\begin{equation}
	\mathcal{M}(F)=\mathbb{C}[x_1,x_2,x_3,x_4]/(dF)\,,
\end{equation}
and $\mu$ is the Milnor number of the singularity, expressed in terms of the weights $q_i$ as
\begin{equation}
	\mu=\prod_{i=1}^4(q_i^{-1}-1)\,.
\end{equation}
The Seiberg--Witten differential of the 4d theory is identified with the holomorphic canonical three-form on the deformed geometry,
\begin{equation}
	\begin{aligned}\Omega_3=\frac{dx_1\wedge dx_2\wedge dx_3\wedge dx_4}{d\widehat{F}}\,.\end{aligned}
\end{equation}
BPS states on the Coulomb branch arise from D3-branes wrapping three-cycles in the deformed geometry, 
whose masses are given by the period integrals of $\Omega_3$. The intersection pairing of these three-cycles leads to the BPS quiver of the 4d SCFT, which will be discussed in more details in Section~\ref{sec:VOA-BPS}.

The parameters $t_l$ parametrize the complex structure deformations and correspond to operators on the Coulomb branch. 
Requiring the holomorphic three-form $\Omega_3$ to have scaling dimension one fixes the scaling dimension of $t_l$ to be
\begin{equation}
	\Delta[t_l]=\frac{Q[t_l]}{\sum_{i=1}^4q_i-1}\,,
\end{equation}
where $Q[t_l]$ denotes the $\mathbb{C}^*$ weight (charge) of $t_l$. 
According to their scaling dimensions, the deformation parameters $t_l$ can be classified as follows:
\begin{itemize}
	\item Deformations with $\Delta > 1$ correspond to vacuum expectation values of Coulomb branch operators $\phi_{l}$, which belong to the short multiplets ${\cal E}_{r,(0,0)}$ of the superconformal algebra,
\begin{equation}
	t_l=\langle \phi_l \rangle\,.
\end{equation}
	Their total number is denoted by $\widehat{r}$. 
	\item Deformations with $\Delta < 1$ correspond to chiral coupling constants, 
	arising from F-term deformations of the form $\delta S=t_l\int d^4x\int d^4\theta(\mathcal{E}_{\mathrm{r},(0,0)})_{\mu-l}$. 
	These operators appear in pairs with the $\Delta>1$ operators, satisfying
	\begin{equation}
		\Delta[t_l]+\Delta[t_{\mu-l+1}]=2\,,\quad l=1\,,\cdots\widehat{r}\,.
	\end{equation}
	\item Deformations with $\Delta = 1$ are mass deformations, and the total number of these parameters is equal to the rank $f$ of the flavor symmetry Lie algebra.
\end{itemize}
This pairing structure between the $\Delta<1$ and $\Delta>1$ operators implies that the Milnor number can be written as
\begin{equation}
	\mu=2\widehat{r}+f\,.
\end{equation}
Hence, the Coulomb branch of the four-dimensional $\mathcal{N}=2$ theory can be identified with the complex structure deformations of the IHS. 


\paragraph{Central charges}
In a generic four-dimensional $\mathcal{N}=2$ superconformal field theory, the conformal central charges $a$ and $c$ are determined by the formula \cite{Shapere:2008zf}
\begin{equation}\label{acfromIHS}
	a=\frac{R(A)}{4}+\frac{R(B)}{6}+\frac{5\widehat{r}}{24}\,,\quad c=\frac{R(B)}{3}+\frac{\widehat{r}}{6}\,.
\end{equation}
For an IHS in $\mathbb{C}^4$, the quantities $R(A)$ and $R(B)$ can be computed as \cite{Xie:2015rpa}
\begin{equation}
	R(A)=\sum_{\Delta_l>1}(\Delta_l-1)\,,\quad R(B)=\frac{\mu}{4(\sum_{i=1}^4q_i-1)}=\frac{1}{4}\mu\Delta_{\max}\,,
\end{equation}
where $\Delta_{\text{max}}$ denotes the maximal scaling dimension among the Coulomb branch operators $(\mathcal{E}_{\mathrm{r},(0,0)})_l$.

\paragraph{Resolution}
We now briefly describe the resolution of the Calabi--Yau threefold. A crepant resolution $\pi:\widetilde{\mathbf{X}}\to\mathbf{X}$ contains an exceptional locus 
\begin{equation}
	\pi^{-1}(0)=\bigcup_{a=1}^rS_a\:,
\end{equation}
consisting of $r$ exceptional divisors that intersect along curves\footnote{Note that there may be Gorenstein terminal singularities that cannot be crepantly resolved. In this paper we by default focus on the cases of $\mathbf{X}$ with a crepant resolution, giving rise to a smooth non-compact Calabi-Yau threefold $\widetilde{\mathbf{X}}$.}, which keeps the Calabi-Yau condition. 

For a generic crepant resolution, one finds the following relations between the topological data of $\tX$ and the rank $r$/flavor rank $f$ of the 4d theory~\cite{Closset:2020scj}:
\begin{equation}\label{resolution}
	\begin{aligned}\dim H_1(\widetilde{\mathbf{X}},\mathbb{R})&=0\,,&\dim H_3(\widetilde{\mathbf{X}},\mathbb{R})&=b_3\,,\\\dim H_2(\widetilde{\mathbf{X}},\mathbb{R})&=r+f\,,&\dim H_4(\widetilde{\mathbf{X}},\mathbb{R})&=r\,.\end{aligned}
\end{equation}
Note that some of the Kähler parameters correspond to effective curves that are dual to non-compact divisors.

Although it was expected that the Higgs branch of 4d $\mc{N}=2$ SCFT should be determined by the resolution, little detail has been worked out in the literature. 

In later sections, we will present a general framework for describing the Higgs branch in terms of the Kähler moduli space of the Calabi--Yau manifold.

\subsection{4d/5d correspondence}\label{subsec:5d4d}

In this subsection, we explain the correspondence between the 4d $\mathcal{N}=2$ SCFT arising from Type IIB compactification and the 5d SCFT obtained from M-theory compactification, as proposed in \cite{Closset:2020scj, Closset:2020afy, Closset:2021lwy}.

Let us start from an isolated canonical threefold singularity $\mathbf{X}$, in particular an IHS in this paper.  
As discussed in the previous subsection, Type IIB compactification on $\mathbf{X}$ gives rise to a 4d $\mathcal{N}=2$ SCFT, which we denote by $\mathcal{T}_{\mathbf{X}}^{4\mathrm{d}}$.  
On the other hand, M-theory compactified on the same singularity yields a 5d SCFT $\mathcal{T}_{\mathbf{X}}^{5\mathrm{d}}$:
\begin{equation}
	\begin{aligned}
		&\mathcal{T}_{\mathbf{X}}^{5\mathrm{d}}\quad\longleftrightarrow\quad\text{M-theory on }\mathbb{R}^{1,4}\times\mathbf{X}\\\nonumber
		& \mathcal{T}_{\mathbf{X}}^{4\mathrm{d}}\quad\longleftrightarrow\quad\text{Type IIB on }\mathbb{R}^{1,3}\times\mathbf{X}\,.
	\end{aligned}
\end{equation}

The basic properties of $\mathcal{T}_{\mathbf{X}}^{4\mathrm{d}}$ were reviewed previously.  
For the 5d theory $\mathcal{T}_{\mathbf{X}}^{5\mathrm{d}}$, its low-energy Coulomb branch information is encoded in the data of the crepant resolution of the Calabi--Yau threefold, in particular the triple intersection numbers of compact divisors~\cite{Intriligator:1997pq}.  
Denoting the compact divisors in $\widetilde{\mathbf{X}}$ by $S_1,\dots,S_r$, reducing the M-theory three-form $C_3$ along the compact divisor  $S_i$ gives rise to the 5d $U(1)$ gauge fields $A_i$:
\begin{equation}
\label{C3-A-gaugefield}
	C_3 = \sum_{i=1}^r\omega^{(1,1)}_i \wedge A_i\,,
\end{equation}
where $\omega^{(1,1)}_i$ are the $(1,1)$-forms dual to $S_i$.
The rank of the 5d SCFT equals $r$.
The gauge coupling of $A_i$ is determined by the volume of the anticanonical divisor of the corresponding divisor $S_i$:
\be
\frac{1}{g_i^2}\propto \mathrm{Vol}(-K_{S_i})\,.
\ee
 Effective curves that are dual to non-compact divisors correspond to non-dynamical vector multiplets, and thus determine the flavor symmetry rank of the 5d SCFT. Note that the $f$ computed from the CB data of $\FTfour$ may be smaller than the actual flavor rank of the $\FT$, for instance in the examples of rank-$r$ Seiberg $E_n$ theories with a non-zero $b_3$~\cite{Closset:2020scj,Closset:2020afy}. 
When all exceptional divisors and curves shrink to zero volume, the resulting theory describes the 5d UV fixed point SCFT $\FT$.

The UV Higgs branch of a 5d SCFT is a hyper-Kähler cone.  
Let its quaternionic dimension be
\begin{equation}
\dim_{\mathbb{H}}\mathcal{M}_H\!\left[\mathcal{T}_{\mathbf{X}}^{5\mathrm{d}}\right]= d_H.
\end{equation}
Naively, complex structure deformations of the Calabi--Yau are associated with three-cycles $S_l^3$, $l=1,\dots,\mu$, and the corresponding hypermultiplet moduli are given by
\begin{equation}
	t_l=\int_{S_l^3}\Omega_3\:,
\end{equation}
where $\Omega_3$ is the holomorphic three-form.  
However, not all of these hypermultiplets are dynamical \cite{Gukov:1999ya}, and therefore the actual Higgs branch dimension $d_H$ is smaller than $\mu$.  
It was shown in \cite{Closset:2020scj} that
\begin{equation}
	d_H=\widehat{r}+f\,.
\end{equation}
Here $\widehat{r}$ denotes the dimension of the mixed Hodge structure (MHS) group $H^{1,2}(\widehat{\mathbf{X}})$ on the vanishing cohomology of $\mathbf{X}$, and for IHS one finds \cite{arnold2012singularities, Xie:2015rpa}
\begin{equation}
	\begin{aligned}\mu=2\widehat{r}+f,\quad&f=\dim H^{2,2}(\widehat{\mathbf{X}}),\quad&\widehat{r}=\dim H^{1,2}(\widehat{\mathbf{X}})=\dim H^{2,1}(\widehat{\mathbf{X}}).\end{aligned}
\end{equation}
For IHS singularities, it is expected that $f$ coincides with the flavor rank of the 4d SCFT \cite{Closset:2020scj}, and $\widehat{r}$ matches the Coulomb branch rank of the 4d $\mathcal{N}=2$ SCFT, hence we adopted the same label.

Now we discuss the correspondence between the HB and CB of  $\mathcal{T}_{\mathbf{X}}^{5\mathrm{d}}$ and $\mathcal{T}_{\mathbf{X}}^{4\mathrm{d}}$, which should be seen from the dimensional reduction to 3d. 
More precisely, we define the 3d $\mc{N}=4$ electric quivers obtained by compactifying the 5d and 4d SCFTs on $T^2$ and $S^1$ (or \emph{quiverine} if the resulting 3d $\mc{N}=4$ theory is non-Lagrangian):
\begin{equation}
	\begin{aligned}
		\mathrm{EQ}^{(5)}[\mathbf{X}] &\;\overset{3\mathrm{d}}{\simeq}\;
		D_{T^2}\mathcal{T}_{\mathbf{X}}^{5\mathrm{d}},\\[4pt]
		\mathrm{EQ}^{(4)}[\mathbf{X}] &\;\overset{3\mathrm{d}}{\simeq}\;
		D_{S^1}\mathcal{T}_{\mathbf{X}}^{4\mathrm{d}}.
	\end{aligned}
\end{equation}
Since the Higgs branch is preserved under dimensional reduction, the Higgs branch of the 5d SCFT can equivalently be described as the Coulomb branch of the corresponding \emph{magnetic quiver} in 3d $\mathcal{N}=4$ theory \cite{Ferlito:2017xdq, Cabrera:2018jxt}.

\begin{equation}
\begin{aligned}
	\mathcal{M}_H\bigl[\mathcal{T}_{\mathbf{X}}^{5\mathrm{d}}\bigr]
	&\;\cong\;
	\mathrm{HB}\bigl[\mathrm{EQ}^{(5)}\bigr]
	\;\cong\;
	\mathrm{CB}\bigl[\mathrm{MQ}^{(5)}\bigr],\\[4pt]
	\mathcal{M}_H\bigl[\mathcal{T}_{\mathbf{X}}^{4\mathrm{d}}\bigr]
	&\;\cong\;
	\mathrm{HB}\bigl[\mathrm{EQ}^{(4)}\bigr]
	\;\cong\;
	\mathrm{CB}\bigl[\mathrm{MQ}^{(4)}\bigr].
\end{aligned}
\end{equation}
The central statement of the 4d/5d correspondence~\cite{Closset:2020scj} is that the magnetic quivers can be obtained via the S-type gauging \cite{Kapustin:1999ha, Witten:2003ya} of the $U(1)^f$ flavor symmetry from the 4d and 5d electric quiver:
\begin{equation}
	\mathrm{MQ}^{(5)}[\mathbf{X}]\cong\mathrm{EQ}^{(4)}[\mathbf{X}]/\mathrm{U}(1)^f\qquad\mathrm{~MQ}^{(4)}[\mathbf{X}]\cong\mathrm{EQ}^{(5)}[\mathbf{X}]/\mathrm{U}(1)^f.
\end{equation}
From this construction, one can infer the dimensions of the Higgs and Coulomb branches of the magnetic quiver $\mathrm{MQ}^{(4)}$:
\begin{equation}
	\begin{aligned}\dim\mathrm{HB}[\mathrm{MQ}^{(4)}]=d_H-f=\widehat{r},\quad&\dim\mathrm{CB}[\mathrm{MQ}^{(4)}]=r+f=\widehat{d}_H\end{aligned}
\end{equation}
where $\widehat{d}_H$ denotes the Higgs branch dimension of the 4d SCFT.  
Similarly, for the magnetic quiver $\mathrm{MQ}^{(5)}$ we find
\begin{equation}
	\dim\mathrm{HB}[\mathrm{MQ}^{(5)}]=\widehat{d}_H-f=r,\quad\dim\mathrm{CB}[\mathrm{MQ}^{(5)}]=d_H-f+f=d_H,
\end{equation}
which is consistent with the dimensions of the Coulomb and Higgs branches of the original 5d SCFT.

\subsection{4d Higgs branch from resolution - schematics}\label{subsec:HBfromresolution}

In Section~\ref{subsec:4dfromIIB}, the crepant resolution $\tX$ of $\mathbf{X}$ is expected to capture the Higgs branch structure of $\mathcal{T}_{\mathbf{X}}^{4\mathrm{d}}$. 
Nevertheless, the (extended) K\"ahler cone of $\tX$ is a real space~\cite{Witten:1996qb,Morrison:1996xf}, and one must introduce additional real coordinates in order to reproduce the hyperk\"ahler structure of the Higgs branch of $\mathcal{T}_{\mathbf{X}}^{4\mathrm{d}}$.

From the topological data of $\tX$ in~\eqref{resolution} and the expansion of M-theory $C_3$ field in (\ref{C3-A-gaugefield}), we have 
\begin{equation}
	r=b_4(\tX)
\end{equation}
compact divisors that are Poincaré dual to $(1,1)$-forms $\omega_i^{(1,1)}$, which in turn are Hodge dual to $(2,2)$-forms $\widetilde{\omega}_i^{(2,2)}$. Expanding $C_3$ on these forms as in (\ref{C3-A-gaugefield}) yields $r$ dynamical $U(1)$ vector multiplets, whereas expanding on the $(1,1)$-forms dual to the $f$ non-compact divisors yields non-dynamical vector multiplets; consequently, $f$ can be interpreted as the flavor rank.

For the non-compact Calabi--Yau threefold $\tX$, the number of compact 2-cycles differs from $r$, as given by~\eqref{resolution},
\begin{equation}\label{b2:rf}
	b_{2}(\tX)=r+f\,.
\end{equation}
Here, $f$ denotes the number of additional 2-cycles corresponding to $(1,1)$-forms $\omega_\alpha^{(1,1)}$ that do not have Hodge dual $(2,2)$-forms in the non-compact $\tX$. They also correspond to $f$ linearly independent non-compact divisors $D_\alpha$ $(\alpha=1,\dots,f)$ in $\tX$\footnote{For non-compact divisors $D_\alpha$, they give rise to $U(1)$ gauge fields with zero-gauge coupling on the Coulomb branch, thus these are interpreted as non-dynamical background gauge field for flavor symmetries.}.

The real dimension~\eqref{b2:rf} of the extended K\"ahler cone of $\tX$ matches the quaternionic dimension of the Higgs branch of $\mathcal{T}_{\mathbf{X}}^{4\mathrm{d}}$, 
\begin{equation}
	\dim\mathrm{HB}[\mathcal{T}_{\mathbf{X}}^{4d}]=\widehat{d}_H=r+f
\end{equation}
in agreement with the $5\mathrm{d}/4\mathrm{d}$ correspondence reviewed in Section~\ref{subsec:5d4d}.

Using the bases $\omega_i^{(1,1)}$ and $\omega_\alpha^{(1,1)}$, the Type IIB NS-NS and R-R sector gauge fields can be expanded as
\begin{equation}
	\begin{aligned}B_2&=\sum_i b_i\wedge \omega^{(1,1)}_i+\sum_\alpha b_\alpha\wedge \omega^{(1,1)}_\alpha\cr
		C_2&=\sum_i c_i\wedge \omega^{(1,1)}_i+\sum_\alpha c_\alpha\wedge \omega^{(1,1)}_\alpha\cr
		C_4&=\sum_i d_{2,i}\wedge \omega^{(1,1)}_i+\sum_i d_i\wedge \widetilde{\omega}^{(2,2)}_i+\sum_\alpha d_{2,\alpha}\wedge \omega^{(1,1)}_\alpha\,.\end{aligned}
\end{equation}
Moreover, the K\"ahler (1,1)-form expands as
\begin{equation}
	J=\sum_iJ_i\omega_i^{(1,1)}+\sum_\alpha J_\alpha\omega_\alpha^{(1,1)}\,.
\end{equation}

Now let us group the fields into quaternionic coordinates of the Higgs branch. First for the fields with indices $i = 1, \dots, r$, since $C_4$ is a self-dual four-form, $d_{2,i}$ and $d_i$ are not independent. Thus, the corresponding four-dimensional scalar degrees of freedom are $b_i$, $c_i$, $d_i$, and $J_i$. 

For the fields with indices $\alpha = 1, \dots, f$, the four-dimensional scalar degrees of freedom are $b_\alpha$, $c_\alpha$, $d_\alpha$ (the 4d electromagnetic dual scalar of $d_{2,\alpha}$, which we denote by $d_\alpha=\star_{\mathrm{4d}} d_{2,\alpha}$), and $J_\alpha$. 

The detailed map between such real scalars from IIB and the Higgs branch coordinates is not straight-forward. 
We now illustrate this construction using the simplest example--the conifold singularity.

\paragraph{Example: the conifold.}
The conifold singularity is defined by 
\begin{equation}
	\mathbf{X}:\ xy + zw = 0\,.
\end{equation}
The associated theory $\mathcal{T}_{\mathbf{X}}^{4\mathrm{d}}$ is simply a free hypermultiplet. 

After performing a small resolution by blowing up either $(x,z)$, $(x,w)$, $(y,z)$, or $(y,w)$ in the $\mb{C}^4$ ambient space, we obtain a smooth exceptional 2-cycle $C \cong \mathbb{P}^1$ with normal bundle $N_{C|X} = \mathcal{O}(-1) \oplus \mathcal{O}(-1)$. For example, the blow-up of $(x,z)$ can be denoted as $(x,z;\delta)$ in the notations of \cite{Lawrie:2012gg,Apruzzi:2019opn}. After the blow-up, the old coordinates $x$ and $z$ are replaced as $x\rightarrow x\delta$, $z\rightarrow z\delta$, where with abusing of notations $[x:z]$ are now projective coordinates of $\mb{P}^1$. After the replacement and proper transformation, the resolved equation is 
\be
\label{conifold-resolved}
\tX:\ xy\delta+zw\delta=0\rightarrow xy+zw=0\,,
\ee
having the same form as before. The exceptional $C=\mb{P}^1$ is given by the equation $\delta=y=w=0$. After $y=w=0$ is plugged into the resolved equation (\ref{conifold-resolved}). 

In this case, there is no compact 4-cycle, and we have
\begin{equation}
	r = \widehat{r} = 0\ , \quad f = 1\ , \quad \widehat{d}_H = 1\,.
\end{equation}
The expected Higgs branch of $\mathcal{T}_{\mathbf{X}}^{4\mathrm{d}}$ is $\mathbb{C}^2 \cong \mathbb{R}^4$, corresponding to the one of a free hypermultiplet.

\paragraph{Example: generalized conifold.}
The conifold singularity can be generalized to 
\begin{equation}
	xy + z^2 + w^{2N} = 0\,.
\end{equation}
IIB superstring theory on such singularity exactly gives rise to type $(A_1,A_{2N-1})$ generalized Argyres-Douglas theory.

The small resolution involves blowing up either $(x, z \pm i w^N)$ or $(y, z \pm i w^N)$, yielding an exceptional curve $C \cong \mathbb{P}^1$ with normal bundle $N_{C|X} = \mathcal{O} \oplus \mathcal{O}(-2)$.

This singularity has no compact divisor, and the relevant  data are
\begin{equation}
	r = 0, \quad \widehat{r} = N - 1, \quad f = 1, \quad \widehat{d}_H = 1.
\end{equation}
The magnetic quiver $\mathrm{MQ}^{(4)}$ of $\mathcal{T}_{\mathbf{X}}^{4\mathrm{d}}$ is already studied in \cite{Xie:2012hs}:
\begin{equation}
	 \begin{tikzpicture}[x=.5cm,y=.5cm]
		\draw[ligne, black](0,0)--(2,0) ;
		\node[] at (-2,0) {$\mathrm{MQ}^{(4)}= $};
		\node[bd] at (0,0) [label=below:{{\scriptsize$1$}}] {};
		\node[bd] at (2,0) [label=below:{{\scriptsize$1$}}] {};
		\node[] at (1,-0.4) {\scriptsize$N$};
	\end{tikzpicture} 
\end{equation}
from which we can read off the expected Higgs branch of $\mathcal{T}_{\mathbf{X}}^{4\mathrm{d}}$, given by the Kleinian singularity $\mathbb{C}^2 / \mathbb{Z}_N$.

After identifying the coordinates for the Higgs branch moduli space, we can determine the hyperk\"ahler metric. In general, the Higgs branch metric receives instanton corrections and is difficult to compute explicitly. 
For the generalized conifold singularity, the corresponding metric has been analyzed in~\cite{Ooguri:1996me,Saueressig:2007dr}.

In this case there is only a single compact curve $C$ in the resolved space, and the real coordinates of the hypermultiplet moduli space are given by 
\be
J_1=\int_C J\ ,\ b_1=\int_C B_2\ ,\ c_1=\int_C C_2\ ,\ d_1=\star_{\rm 4d}\int_C C_4\,.
\ee

In general, the hypermultiplet moduli space receives quantum corrections from D$p$-brane instantons with $p = -1, 1, 3, 5$, as well as from NS5-brane instantons.  
The leading NS5-brane contribution behaves as~\cite{Alexandrov:2007ec}
\begin{equation}
	\sim e^{{-2\pi|k|V/g_{s}^{2}}},
\end{equation}
where $V$ is the Calabi--Yau volume. Hence, no corrections arise in the non-compact limit.  
The conifold limit in IIB theory corresponds to
\begin{equation}
	J_1\to0,\quad b_1\to0,\quad\tau_2\to\infty,\quad\mathrm{with}\quad\tau_2|b_1+iJ_1|\mathrm{=finite.}
\end{equation}
In this limit, the D3 and D5-brane corrections vanish, and the D$(-1)$-instanton correction behaves as
\begin{equation}
	\sim \exp(-|m|\tau_2)
\end{equation}
which also vanishes near the conifold limit. The remaining contribution comes from $D1$-instantons.  
As shown in~\cite{Ooguri:1996me,Saueressig:2007dr}, the metric is expressed in terms of a potential $V$, which takes the following form when $N$ two-cycles collapse at the conifold singularity:
\begin{equation}
	\begin{aligned}V=N\left(\frac{1}{4\pi}\ln\left(\frac{1}{z\bar{z}}\right)+\frac{1}{2\pi}\sum_{m\neq0}K_0(2\pi\tau_2|mz|) \mathrm{e}^{2\pi\mathrm{i}m(c_1-\tau_1b_1)}\right)\end{aligned}
\end{equation}
where $z=b_1+iJ_1$. Here $K_0$ is the Bessel function.  
The hypermultiplet metric takes the form
\begin{equation}
	\mathrm{d}s^2=\tau_2^{-2}[V^{-1}(\mathrm{d}d_1-\vec{A}\cdot\mathrm{d}\vec{y})^2+V|\mathrm{d}\vec{y}|^2]\,,
\end{equation}
with $\vec{y} = (-(c_1 - \tau_1 b_1), z \tau_2, \bar{z} \tau_2)$. When the $N$ vanishing two-cycles shrink to zero size, the resulting geometry reduces to the $\mathbb{C}^2/\mathbb{Z}_N$ singularity \cite{Ooguri:1996me,Greene:1996dh}, which is consistent with the result obtained from the magnetic quiver. We confirm this from the explicit crepant resolution in Section~\ref{sec:AkAl}.

Since instanton corrections are generally difficult to compute, for more complicated singularities we will instead try to derive the magnetic quiver directly from the singularity, which can be generally done for $r=0$ cases. 

For further convenience, we briefly review some basic concepts of the Gopakumar--Vafa (GV) invariants \cite{Gopakumar:1998jq,Gopakumar:1998ii}. 
GV invariants are related to BPS particles arising in string compactifications. 
From the M-theory perspective, M2-branes can wrap compact two-cycles in a Calabi--Yau threefold, giving rise to BPS particles in the resulting low-energy field theory. 
The volumes of these two-cycles are interpreted as the masses of the corresponding BPS particles.

Upon reducing M-theory to Type IIA string theory, an M2-brane reduces to an infinite tower of D2-branes labeled by their momentum along the M-theory circle $S^1.$ The effective field theory obtained from Type IIA compactification is captured by the topological A-model with target space given by the Calabi--Yau threefold. 
In particular, the four-dimensional effective action contains couplings of the form
\begin{equation}
	\int \mathrm{d}^4xF_g(\mathbf{t})R_+^2F_+^{2g-2}\,,
\end{equation}
where $\mathbf{t} = B + iJ$ is the complexified Kähler form, $R_+$ denotes the self-dual part of the Riemann tensor, and $F_+$ is the self-dual part of the graviphoton field strength. 
Here $F_g(t)$ is the genus-$g$ amplitude of the topological string. All of these terms can be described by the topological string partition function $F(\mathbf{t},g_s)$.

The partition function of topological string theory receives both perturbative and non-perturbative contributions. 
The non-perturbative contribution from D2-branes wrapping two-cycles takes the general form
\begin{equation}
	\begin{aligned}F_{\text{non-per}}(\mathbf{t},g_s)&=\sum_{k=1}^\infty\sum_{g=0}^\infty\sum_{\mathbf{d}}\frac{n_\mathbf{d}^g}{k}(2\sin(kg_s/2))^{2g-2}e^{-k\mathbf{d}\cdot\mathbf{t}}\,,\end{aligned}
\end{equation}
where $n_{\mathbf{d}}^g$ denotes the GV invariant associated with genus-$g$ curves of degree $\mathbf{d}$, 
$g_s$ is the string coupling, and $k$ labels the D2-brane momentum along the circle $S^1$. 
These invariants are conjectured to be integers and count BPS states arising from D2-branes wrapping two-cycles in the Calabi--Yau threefold.

In \ref{subsec:MQ-resolution}, we focus on cases with $r=0$ and employ string
duality to show that the magnetic quiver can be obtained from Type IIA
compactification on $\tX \times S^1$.
The resulting theory is precisely the magnetic quiver, whose
hypermultiplet arises from D2-branes wrapping genus-zero curves.
Hence the magnetic quiver is captured by the genus-zero GV invariants
$n^{0}_{\mathbf{d}}$.
\subsection{Magnetic quiver, 3d mirror symmetry and inversion}
\label{sec:inversion}

For many Lagrangian SUSY field theories with 8 supercharges, there does not exist a Lagrangian description of its magnetic quiver(ine). We encounter many examples of this type later, and we would like to discuss how to compute its Higgs branch Hasse diagram~\cite{Bourget:2019aer}.

A particularly important class of examples is the 3d $\mc{N}=4$ unitary quiver, modding out a diagonal $U(1)$ subgroup. If such quiver acts as the $\EQfour$ for a 4d SCFT $\FTfour$, its Higgs branch would be the same as that of the $\FTfour$.

\begin{figure}
\centering
\includegraphics[height=5cm]{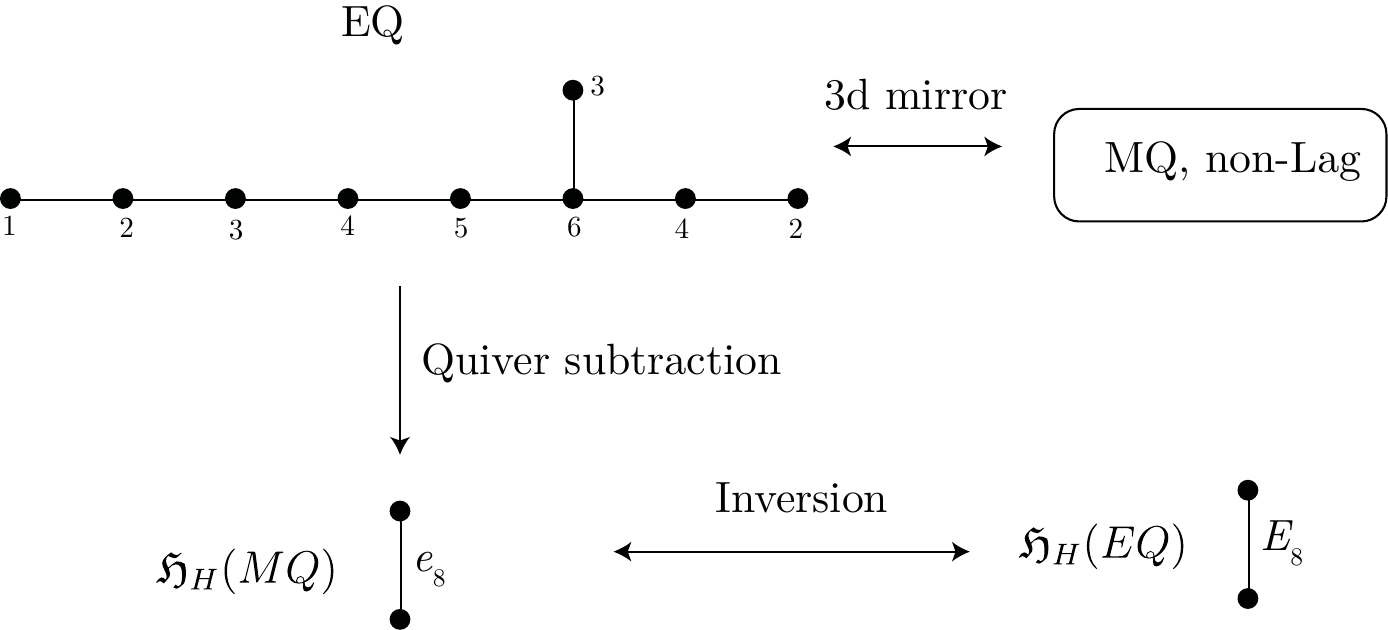}
\caption{The inversion on the Coulomb branch Hasse diagram of an affine $E_8$ type unitary quiver EQ, leading to the Higgs branch Hasse diagram of EQ. One can conclude that the Higgs branch of EQ is a $\mb{C}^2/\Gamma_{E_8}$ Kleinian singularity.}\label{f:Inversion}
\end{figure}

A simple class of examples is the unitary quiver of the affine $E_n$ shape. In \cite{Grimminger:2020dmg} an \emph{inversion} phenomenon was observed. We assume that the  affine $E_n$ shaped quiver $\mathrm{EQ}$ is the 3d mirror of a non-Lagrangian 3d theory $\mathrm{MQ}$, such that CB(EQ)$=$HB(MQ) and HB(EQ)$=$CB(MQ). From the  Hasse diagram $\mathfrak{H}_C(\mathrm{EQ})=\mathfrak{H}_H(\mathrm{MQ})$ obtained via quiver subtraction~\cite{Cabrera:2018ann}, the Coulomb (Higgs) branch of $\mathrm{EQ}$ ($\mathrm{MQ}$) should be the closure of minimal nilpotent orbit of $E_n$ with $\dim(\mathrm{CB(EQ)})=\dim(\mathrm{HB(MQ)})=h^\vee_{E_n}-1$. On the other hand one can compute that $\dim(\mathrm{HB(EQ)})=\dim(\mathrm{CB(MQ)})=1$. The Hasse diagram $\mathfrak{H}_H(\mathrm{EQ})\equiv \mathfrak{H}_C(\mathrm{MQ})$ can be derived via the inversion - one turns the Hasse diagram $\mathfrak{H}_C(\mathrm{EQ})=\mathfrak{H}_H(\mathrm{MQ})$ upside down, and replaces each symplectic leaf via the map $A_n\leftrightarrow a_n$, $D_n\leftrightarrow d_n$, $E_n\leftrightarrow e_n$, see figure~\ref{f:Inversion}. In fact, closures of minimal nilpotent orbits and Kleinian surface singularities are expected to form a symplectic dual pair \cite{Braden:2014iea, Kamnitzer:2022nzd}. After the inversion, $\mathfrak{H}_H(\mathrm{EQ})\equiv \mathfrak{H}_C(\mathrm{MQ})$ only has a single leaf of an $E_n$ Kleinian singularity, and thus we read off that $\mathrm{HB(EQ)}\cong \mb{C}^2/\Gamma_{E_n}$.

For more complicated 3d $\mc{N}=4$ unitary quiver $\mathrm{EQ}$ with good nodes only, we conjecture that the inversion procedure always works\footnote{We would like to thank Zhenghao Zhong for pointing out this point.}. The key point is that each step of the quiver subtraction of $\mathrm{EQ}$ exactly corresponds to a partial Higgsing of the unitary quiver gauge theory $\mathrm{EQ}$, and the inversion can be applied stepwise. 

As an example let us study the following unitary quiver $\mathrm{EQ}$:
\begin{equation}
	\begin{tikzpicture}[x=.5cm,y=.5cm]
		\draw[ligne, black](0,0)--(2,0);
		\draw[ligne, black](2,0)--(4,0);
		\draw[ligne, black](4,0)--(8,0);
		\draw[ligne, black](8,0)--(10,0);
		\draw[ligne, black](10,0)--(12,0);
			\draw[ligne, black](12,0)--(14,0);
		\draw[ligne, black](8,0)--(8,2);
		\node[bd] at (0,0) [label=below:{{\scriptsize$2$}}] {};
		\node[bd] at (2,0) [label=below:{{\scriptsize$4$}}] {};
		\node[bd] at (4,0) [label=below:{{\scriptsize$6$}}] {};
		\node[bd] at (6,0) [label=below:{{\scriptsize$8$}}] {};
		\node[bd] at (8,0) [label=below:{{\scriptsize$10$}}] {};
		\node[bd] at (10,0) [label=below:{{\scriptsize$7$}}] {};
		\node[bd] at (12,0) [label=below:{{\scriptsize$4$}}] {};
	\node[bd] at (14,0) [label=below:{{\scriptsize$1$}}] {};
		\node[bd] at (8,2) [label=left:{{\scriptsize$5$}}] {};
	\end{tikzpicture}
\end{equation}
 Subtraction of an affine $E_7$ sub-quiver from the above one exactly corresponds to Higgsing the $U(1)$ subgroup of the $U(4)$ node on the left and removing two charged hypermultiplets under the $U(1)$ on the right. After the quiver subtraction, the leftmost $U(2)$ node should still have 4 fundamental hypermultiplets, and thus we should add a new $U(1)$ node to the left, corresponding to the rebalancing process. This partial Higgsing leads to an $E_7$ leaf in the Higgs branch Hasse diagram $\mathfrak{H}_H(\mathrm{EQ})$. 

\begin{figure}
\centering
\includegraphics[height=6cm]{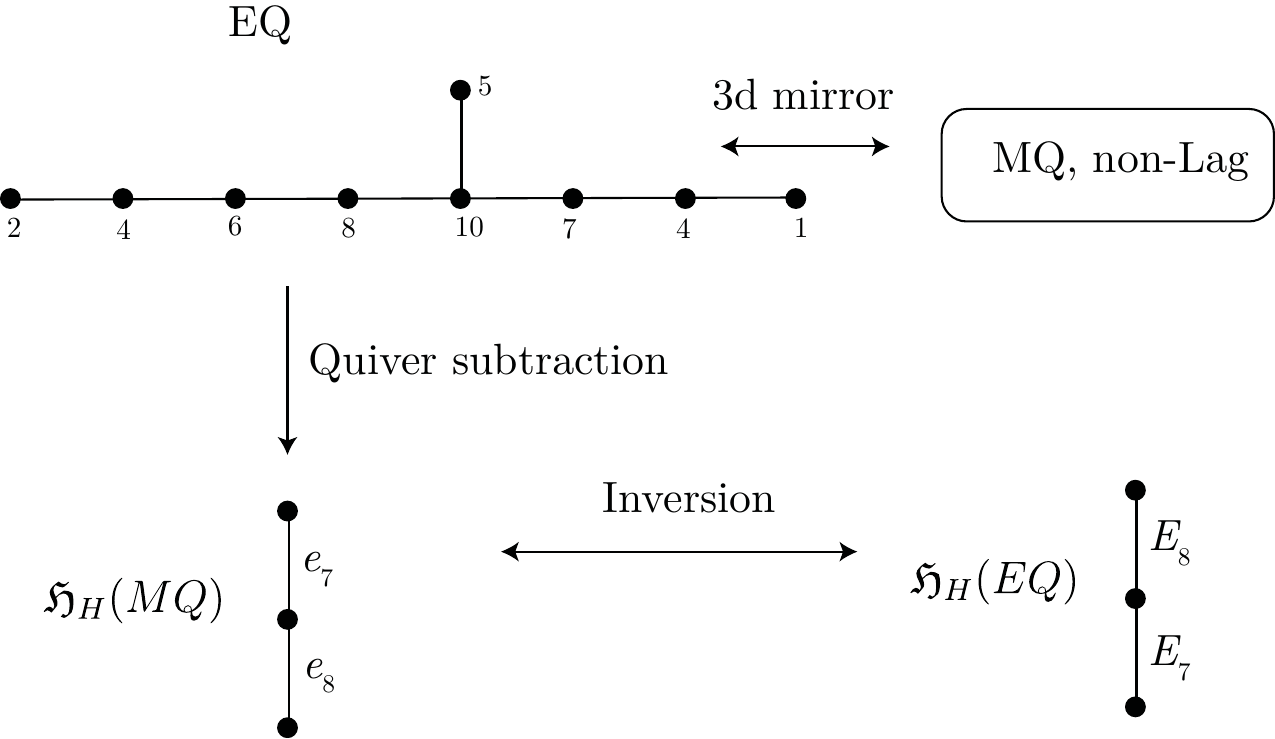}
\caption{The inversion on the Coulomb branch Hasse diagram of a more complicated unitary quiver EQ, leading to the Higgs branch Hasse diagram of EQ. }\label{f:Inversion-2}
\end{figure}

After this partial Higgsing, we arrive at a type affine $E_8$ unitary quiver
\begin{equation}
	\begin{tikzpicture}[x=.5cm,y=.5cm]
    \draw[ligne, black](-2,0)--(0,0);
		\draw[ligne, black](0,0)--(2,0);
		\draw[ligne, black](2,0)--(4,0);
		\draw[ligne, black](4,0)--(8,0);
		\draw[ligne, black](8,0)--(10,0);
		\draw[ligne, black](10,0)--(12,0);
		\draw[ligne, black](8,0)--(8,2);
        \node[bd] at (-2,0) [label=below:{{\scriptsize$1$}}] {};
		\node[bd] at (0,0) [label=below:{{\scriptsize$2$}}] {};
		\node[bd] at (2,0) [label=below:{{\scriptsize$3$}}] {};
		\node[bd] at (4,0) [label=below:{{\scriptsize$4$}}] {};
		\node[bd] at (6,0) [label=below:{{\scriptsize$5$}}] {};
		\node[bd] at (8,0) [label=below:{{\scriptsize$6$}}] {};
		\node[bd] at (10,0) [label=below:{{\scriptsize$4$}}] {};
		\node[bd] at (12,0) [label=below:{{\scriptsize$2$}}] {};
		\node[bd] at (8,2) [label=left:{{\scriptsize$3$}}] {};
	\end{tikzpicture}
\end{equation}
The only possible Higgsing of this quiver is to completely Higgs all the gauge groups, which leads to another $E_8$ leaf in $\mathfrak{H}_H(\mathrm{EQ})$. One should be careful about the order of leaves $\mathfrak{H}_H(\mathrm{EQ})$, because the bottom node corresponds to the unhiggsed quiver and the top node corresponds to the completely Higgsed quiver, we should be putting the $E_7$ leaf on the bottom and the $E_8$ leaf on the top. We summarize this case in figure~\ref{f:Inversion-2}.

\subsection{Vertex operator algebras associated with CY3 singularities}
\label{sec:VOAfromsingularity}

In this subsection, we briefly summarize the procedure for extracting
VOA data from a given singularity.

For any 4d $\mathcal{N}=2$ SCFT, one can identify a protected subsector called the Schur sector, which corresponds to a vertex operator algebra (VOA)~\cite{Beem:2013sza}.
This SCFT/VOA correspondence implies the following dictionary:
\begin{itemize}

\item The central charge of the VOA and the level of each AKM subalgebra relate to the corresponding 4d quantities via
\begin{equation}
	c_{2d} = -12c_{4d}\,, \qquad k_{2d} = -\frac{1}{2}k_F\,,
\end{equation}
where the normalization of $k_F$ follows the conventions of~\cite{Beem:2013sza,Beem:2014rza}.

\item The associated VOA always contains an affine vertex subalgebra $V_{k_{2d}}(\mathfrak{g}_F)$, where $\mathfrak{g}_F$ is the flavor symmetry Lie algebra of the 4d $\mc{N}=2$ SCFT.

\item The (normalized) vacuum character of the VOA coincides with the Schur index of the 4d SCFT.
\item The Higgs branch of the 4d theory is identified with the associated variety of the VOA~\cite{Beem:2017ooy,Song:2017oew}.
\end{itemize}

Usually it is not easy to identify the corresponding VOA of a given 4d theory with a known VOA. However, for regular class-$S$ theories \cite{Gaiotto:2009we, Gaiotto:2009hg}, the corresponding VOA can be constructed abstractly \cite{Arakawa:2018egx, Yanagida:2020kim}. Moreover, the corresponding VOA of generalized Argyres-Douglas (AD) theories are also well-studied and are usually affine vertex algebras, affine W-algebras or their extensions. There are also systematic constructions of VOAs labeled by complex reflection groups which are usually related to 4d $\mc{N}=4$ SYM~\cite{Bonetti:2018fqz, Arakawa:2023cki}.

In this paper, we focus on the connection between IHSs and the VOAs associated with
four-dimensional SCFTs. For the 4d $\mc{N}=2$ SCFT  $\mathcal{T}_{\mathbf{X}}^{4d}$, we
expect an associated VOA $V_{\mathbf{X}}$.
The central charge $c_{2d}$ of the VOA $V_{\mathbf{X}}$ can be
determined from the deformation data of the singularity
\eqref{acfromIHS} via
\begin{equation}
	c_{2d}=-12c.
\end{equation}
The difference of the four-dimensional central charges,
$ a - c $, is related to the asymptotic growth of the character
of the associated VOA $V_{\mathbf{X}}$. More precisely, the asymptotic behavior of the vacuum character takes the form \cite{kac1989classification,Xie:2019vzr,Li:2022njl}
\begin{equation}
	X_{V_{\mathbf{X}}}(\tau)\sim\mathcal{A}(V_{\mathbf{X}})e^{\frac{\pi i\mathcal{G}(V_{\mathbf{X}})}{12\tau}}\,.
\end{equation}
The precise relation between the asymptotic growth
$\mathcal{G}(V_{\mathbf{X}})$ and the difference of the central charges is given by
\begin{equation}
	a-c=-\frac{1}{48}\mathcal{G}(V_{\mathbf{X}})\,.
\end{equation}

The associated variety of a VOA, which is identified with the Higgs
branch of $\FTfour$, can be described in terms of
GV invariants in the absence of compact divisors
after resolution.
As briefly reviewed in Subsection~\ref{subsec:HBfromresolution} and
discussed in detail in Section~\ref{sec:r=0resolution}, this framework
allows us to obtain many class-$S$ VOAs directly from
singularities.

For singularities containing compact divisors, discussed in
Section~\ref{sec:r>0resolution}, we mainly focus on smoothable cases with
$b_3=f= 0$.
In this situation, we propose a method to determine the magnetic
quiver using string dualities.
More precisely, we first identify the corresponding five-dimensional
SCFT arising from M-theory compactified on a singularity
$\mbf{X}^\prime$ that shares the same resolution sequence with $\mbf{X}$.
The four-dimensional magnetic quiver $\MQfour$ can then be obtained
by the inversion of magnetic quiver of the five-dimensional SCFT
$\MQfive$.
In this way, we discover a new class of quasi-lisse VOAs arising from
singularities, distinct from class-$S$ VOAs and affine
$W$-algebras.
From the perspective of geometric engineering, most singularities do
not admit a class-$S$ realization and correspond to
new quasi-lisse VOAs, making the extraction of VOA data from
singularities a challenging and interesting problem.

The IR behavior of the Coulomb branch of a four-dimensional $\mathcal{N}=2$ theory is encoded in the Seiberg--Witten curve \cite{Seiberg:1994rs}. 
Massive BPS particles are characterized by the phases of their central charges $Z(\gamma)$. 
BPS particles carry electric and magnetic charges, as well as flavor charges; consequently, the charge vector $\gamma$ takes values in a lattice of rank $N=2\widehat{r}+f$. 
These particles satisfy the BPS bound 
\begin{equation}
	M_\gamma\geq|Z(\gamma)|;\quad Z(\gamma_1+\gamma_2)=Z(\gamma_1)+Z(\gamma_2)
\end{equation} 
where $M_\gamma$ denotes the mass of the particle. 
In 4d $\mathcal{N}=2$ theories, the BPS quiver can be generated by a finite set of generators,
\begin{equation}
	\gamma=\sum_{i=1}^Nn_i\gamma_i\,.
\end{equation}
One may further decompose the complex plane of the central charge $Z(\gamma)$ into two half-planes: one corresponding to BPS particles and the other to anti-BPS particles. 
After an appropriate choice of basis, the BPS quiver consists of $2\widehat{r}+f$ hypermultiplets, with one node associated to each charge vector $\gamma_i$, and oriented arrows connecting between pairs of nodes. For example the BPS quiver of $(A_1, A_4)$ theory is 
\begin{equation}
	\begin{tikzpicture}[x=1cm, y=1cm]
		
		\node[circle, draw] (1) at (0,0) {};
		\node[circle, draw] (2) at (3,0) {};
		\node[circle, draw] (3) at (-3,0) {};
		
		\node[below=0.3cm] at (1) {$\gamma_1$};
		\node[below=0.3cm] at (2) {$\gamma_2$};
		\node[below=0.3cm] at (3) {$\gamma_3$};
		
		\draw[->] (3) -- (1);
		\draw[->] (2) -- (1);
		
	\end{tikzpicture}
\end{equation} 

In geometric engineering, the mini-versal deformation of a singularity $\widehat{\mathbf{X}}$ has $\mu = 2\widehat{r}+f$ three-cycles $\Sigma_i$, with $i=1,\ldots,\mu$. 
The hypermultiplets arise from D3-branes wrapping these three-cycles $\Sigma_i$. 
The number of arrows from $\gamma_i$ to $\gamma_j$ in the BPS quiver is determined by the intersection pairing of the corresponding three-cycles in $\widehat{\mathbf{X}}$ \cite{Klemm:1996bj,Aspinwall:2004jr,Alim:2011ae,Alim:2011kw}
\begin{equation}
	n_{ij}=\Sigma_i\cdot\Sigma_j=-n_{ji}\,.
\end{equation}
At the singular point in moduli space, the BPS particles become massless and the corresponding three-cycles shrink to zero size; such cycles are referred to as vanishing cycles. 
Consequently, determining the BPS quiver is equivalent to computing the intersection matrix of the vanishing cycles. 
In Section~\ref{subsec:IntersectionBPSquiver}, we review several known results for the intersection matrices associated with specific singularities.

The choice of basis for the vanishing cycles is not unique and is defined only up to reflections and braid group actions,
\begin{equation}
	\begin{aligned}
		& r_i:(\Sigma_1,\ldots,\Sigma_i,\ldots,\Sigma_\mu)\to(\Sigma_1,\ldots,-\Sigma_i,\ldots,\Sigma_\mu)\\
		& \mathcal{B}: (\Sigma_1,\ldots,\Sigma_i,\Sigma_{i+1}\ldots,\Sigma_\mu)\to(\Sigma_1,\ldots,\Sigma_{i+1}+(\Sigma_{i+1}\cdot\Sigma_i)\Sigma_i,\Sigma_i,\ldots,\Sigma_\mu)\,.
	\end{aligned}
\end{equation}
Once the BPS quiver is determined, the vacuum character can be systematically extracted using the infrared formula on the Coulomb branch \cite{Cordova:2015nma}. 
In particular, the Schur index of a 4d $\mc{N}=2$ SCFT can be computed from the BPS quiver via the IR formula
\begin{equation}
	\mathcal{I}(q)=(q)_\infty^{2\widehat{r}}\mathrm{~Tr}\left[\mathcal{O}(q)\right],
\end{equation}
where $\mathcal{O}(q)$ is the Kontsevich--Soibelman operator. 
A detailed definition of this operator and applications of the above formula can be found in Sections~\ref{subsec:Schurindex} and~\ref{subsec:lisseVOA}.

In additional to the Schur index that corresponding to the vacuum module of 4d VOA, one can also extract the modular data of this VOA.  A recent approach uses the BPS quiver to construct a 3d wall-crossing invariant, which captures the ellipsoid partition
function of a $U(1)_r$ -twisted circle compactification of the holomorphic-topological twist of 4d $\mathcal{N}=2$ theories \cite{Gaiotto:2024ioj}. It has been shown \cite{Dedushenko:2018bpp} that when the Coulomb branch spectrum of the 4d theory consists only of operators with fractional scaling dimensions, the resulting theory possesses an empty Coulomb branch and therefore corresponds to a 3d rank-zero 
theory. These three-dimensional rank-zero SCFTs typically admit a UV description
with $\mathcal{N}=2$ supersymmetry and exhibit supersymmetry enhancement
to an infrared 3d $\mathcal{N}=4$ SCFT.
A rank-zero SCFT admits both $A$- and $B$-topological twists, each of which
gives rise to a generally non-unitary topological field theory (TFT).
When the three-dimensional rank-zero theory further lacks any flavor
symmetry and is associated with a lisse VOA \cite{Arakawa:2010dtu,Xie:2019vzr},
the resulting TFT is semisimple.
Such TFTs admit holomorphic boundary conditions, and the two-dimensional
VOA supported on the boundary is precisely the VOA associated to the
corresponding four-dimensional $\mathcal{N}=2$ SCFT \cite{Ferrari:2023fez,Kim:2025rog,Kim:2024dxu,Go:2025ixu,Nishinaka:2025ytu,ArabiArdehali:2024ysy,Gaiotto:2024ioj,ArabiArdehali:2024vli}.

On the other hand, when the three-dimensional theory possesses a nontrivial
Higgs branch, as is typically the case for VOAs associated to general
four-dimensional $\mathcal{N}=2$ SCFTs, the corresponding VOA is conjectured
to be quasi-lisse \cite{Beem:2017ooy,Arakawa:2016hkg}.
In this situation, the resulting topological field theory is non-semisimple,
and the supporting VOA may admit logarithmic modules.
Nevertheless, this framework is still expected to be applicable and to allow
one to extract basic information about the representation theory of the VOA \cite{Gaiotto:2024ioj,Go:2025ixu,Kim:2025rog,Nishinaka:2025ytu}.

In summary, in this paper we establish a correspondence between IHS and VOA.
Within this framework, we extract the central charges, asymptotic
growth, associated varieties, and vacuum characters for a class of
VOAs.
A systematic analysis starting from the BPS quiver to derive the
modular properties of lisse VOAs, as well as its extension to theories
with non-trivial Higgs branches, is left for future work.

\section{$r=0$ cases with a small resolution}
\label{sec:r=0resolution}

In this section, we focus on singularities that do not contain a compact divisor after resolution---that is, singularities which only admit a small resolution $\tX$, with 5d CB rank $r=0$. Although we have derived the hyperk\"ahler metric for the case of the generalized conifold in Section~\ref{subsec:HBfromresolution}, it is generally difficult to compute instanton corrections and obtain explicit hyperk\"ahler metrics in more general cases. Instead, in this section, we derive the magnetic quiver for the corresponding four-dimensional $\mathcal{N} = 2$ SCFT $\FTfour$.

\subsection{Magnetic quiver from resolution geometry}
\label{subsec:MQ-resolution}

For such theories, the magnetic quiver arises naturally from Type IIA compactification. More precisely, to provide a string-theoretic interpretation of this structure, we interpret the 3d mirror with T-duality between IIA and IIB superstring, as discussed, for example, in \cite{Seiberg:1996ns,Hori:1997zj}. Specifically, the three-dimensional $\mathcal{N} = 4$ theory $\EQfour$ obtained from Type IIB string theory on $\hX \times S^1_R$ is mirror to the three-dimensional $\mathcal{N} = 4$ theory  obtained from Type IIA string theory on $\tX \times S^1_{\alpha'/R}$. The latter is exactly the 3d magnetic quiver theory $\MQfour$ corresponding to the 4d $\mathcal{N} = 2$ theory $\FTfour$.

Now consider the four-dimensional effective field theory arising from Type IIA on $\tX$. Expanding the three-form potential $C_3$ yields $f$ $U(1)$ gauge fields:
\begin{equation}
	C_3 = \sum_{\alpha=1}^f A_\alpha^{(4d)} \wedge \omega_\alpha^{(1,1)},
\end{equation}
where $\omega_\alpha^{(1,1)}$ are the $(1,1)$-forms dual to the $\alpha$-th non-compact divisor $D_\alpha$ in $\tX$. Note that in four dimensions, the gauge couplings $g_\alpha$ of these $U(1)$ gauge fields tend to zero, since each divisor $D_\alpha$ has infinite volume.

However, after further dimensional reduction on $S^1_{\alpha'/R}$ to obtain the three-dimensional $\mathcal{N} = 4$ theory, we write:
\begin{equation}
	C_3 = \sum_{\alpha=1}^f A_\alpha^{(3d)} \wedge \widetilde{\omega}_\alpha^{(1,1)},
\end{equation}
where $\widetilde{\omega}_\alpha^{(1,1)}$ are now $(1,1)$-forms on $\tX \times S^1_{\alpha'/R}$. The kinetic term for $A_\alpha^{(3d)}$ arises from the dimensional reduction of the kinetic term for $C_3$ on $\tX \times S^1_{\alpha'/R}$:
\begin{equation}
	\begin{aligned}
		-\int_{\mathbb{R}^{2,1} \times \tX \times S^1} \frac{1}{2}  dC_3 \wedge \star  dC_3 
		&= -\sum_{\alpha=1}^f \int_{\mathbb{R}^{2,1}} \frac{1}{2}  \left( dA_\alpha \wedge \star  dA_\alpha \right) \int_{\tX \times S^1} \widetilde{\omega}_\alpha^{(1,1)} \wedge \star  \widetilde{\omega}_\alpha^{(1,1)} \\
		&= -\sum_{\alpha=1}^f \int_{\mathbb{R}^{2,1}} \frac{1}{2}  \left( dA_\alpha \wedge \star  dA_\alpha \right) \int_{\tX} \frac{\alpha'}{R}  \omega_\alpha^{(1,1)} \wedge \star  \omega_\alpha^{(1,1)}.
	\end{aligned}
\end{equation}
Here, the Hodge dual $\star  \widetilde{\omega}_\alpha^{(1,1)}$ incorporates the metric factor $\alpha'/R$ from the circle $S^1_{\alpha'/R}$. Therefore, by choosing a suitable scaling limit of $\alpha'/R\rightarrow 0$ in which
\begin{equation}
	\int_{\tX} \frac{\alpha'}{R}  \omega_\alpha^{(1,1)} \wedge \star  \omega_\alpha^{(1,1)}
\end{equation}
remains finite, we obtain finite gauge couplings in the resulting $\MQfour$.

Additionally, there are hypermultiplets arising from D2-branes wrapping the genus-zero curves $C$ in the resolution geometry. These hypermultiplets carry charges $q_\alpha = C \cdot D_\alpha$ under the $\alpha$-th $U(1)$ gauge group in the magnetic quiver $\MQfour$.

From a geometric perspective, the appearance of these hypermultiplets can be understood via the Gopakumar–Vafa invariant \cite{Gopakumar:1998jq,Gopakumar:1998ii}, which encodes BPS state counts in the compactification. Specifically, for a D2-brane wrapping a genus-zero curve, the genus-zero Gopakumar–Vafa invariant of degree $\textbf{d}$ enumerates the number of hypermultiplets with charge vector $\textbf{d}$:
\begin{equation}\label{GVinvariant}
	n_{\textbf{d}}^0=\text{number of hypermultiplets with charge $\mathbf{d}$ under gauge groups}.
\end{equation}
In particular, for the $r=0$ cases in this section, the degree $\textbf{d}$ would denote a vector of charges $q_\alpha$ under each non-compact divisor $D_\alpha$ $(\alpha=1,\dots,f)$.

Hence, all isolated hypersurface singularities with $r = 0$ are expected to be describable in this manner, and their magnetic quivers can be systematically derived via this approach.

\subsection{Gopakumar–Vafa Invariant from 5d SCFT}
\label{sec:GV-5d}

As discussed in Section~\ref{subsec:5d4d}, the magnetic quiver of a 4d SCFT is related to the electric quiver of its 5d counterpart via
\begin{equation}
	\mathrm{MQ}^{(4)}\equiv\mathrm{EQ}^{(5)}/U(1)^f.
\end{equation}
Here, the rank of the 5d SCFT equals the number $r$ of compact divisors in the resolution. In the cases considered in this section, $r=0$, so the resulting electric quiver ${\rm EQ}^{(5)}$ contains a flavor symmetry $U(1)^{f}$, under which certain hypermultiplets are charged. Gauging this $U(1)^{f}$ flavor symmetry yields the magnetic quiver of the corresponding 4d theory. 

This procedure is consistent with earlier results on generating Gopakumar–Vafa (GV) invariants from deformations in 5d SCFTs, as explored in various works
\cite{Collinucci:2021wty,Collinucci:2021ofd,DeMarco:2021try,Collinucci:2022rii,DeMarco:2022dgh}.

In this section, we focus on the class of \textit{compound Du Val} (cDV) singularities~\cite{wemyss2023lockdown} satisfying the $r=0$ condition, which takes the general form
\begin{equation}
	F(x,y,z,w)=x^2+P_{\mathcal{G}}(y,z)+wg(x,y,z,w)=0\,,
\end{equation}
where $P_{\mathcal{G}}(y, z)$ corresponds to an $ADE$-type singularity and $g(x,y,z,w)$ is a general polynomial,
\begin{equation}
	\begin{aligned}&P_{A_n}=y^2+z^{n+1}\\&P_{D_n}=zy^2+z^{n-1}\\&P_{E_6}=y^3+z^4\\&P_{E_7}=y^3+yz^3\\&P_{E_{8}}=y^3+z^5\,.\end{aligned}
\end{equation}
The quasi-homogeneous cases are summarized in Table~\ref{tab:cDV}.
\begin{table}[htbp]
	\centering
	\label{tab:cDV}
	\renewcommand{\arraystretch}{1.4}
	\begin{tabular}{|c|c|}
		\hline
		$(A_{N-1}, A_{k-1})$ & $x^2 + y^2 + z^k + w^N = 0$ \\ 
		\hline
		$A_{k-1}^{(k-1)}[N]$ & $x^2 + y^2 + z^k + w^N z = 0$ \\ 
		\hline
		$(A_{N-1}, D_k)$ & $x^2 + z y^2 + z^{k-1} + w^N = 0$ \\ 
		\hline
		$D_k^k[N]$ & $x^2 + z y^2 + z^{k-1} + w^N y = 0$ \\ 
		\hline
		$(A_{N-1}, E_6)$ & $x^2 + y^3 + z^4 + w^N = 0$ \\ 
		\hline
		$E_6^{(9)}[N]$ & $x^2 + y^3 + z^4 + w^N z = 0$ \\ 
		\hline
		$E_6^{(8)}[N]$ & $x^2 + y^3 + z^4 + w^N y = 0$ \\ 
		\hline
		$(A_{N-1}, E_7)$ & $x^2 + y^3 + y z^3 + w^N = 0$ \\ 
		\hline
		$E_7^{14}[N]$ & $x^2 + y^3 + y z^3 + w^N z = 0$ \\ 
		\hline
		$(A_{N-1}, E_8)$ & $x^2 + y^3 + z^5 + w^N = 0$ \\ 
		\hline
		$E_8^{(24)}[N]$ & $x^2 + y^3 + z^5 + w^N z = 0$ \\ 
		\hline
		$E_8^{(20)}[N]$ & $x^2 + y^3 + z^5 + w^N y = 0$\\ 
		\hline
	\end{tabular}
		\caption{List of quasi-homogeneous cDV singularities.}
\end{table}

\noindent
These singularities can be viewed as special deformations of the $ADE$ singularities,
\begin{equation}
	x^2+P_{\mathcal{G}}(y,z)+\sum_{i=1}^r\mu_ig_i=0,
\end{equation}
where $\mu_i$ is a function of $w$, and the monomials $g_i$ belong to the ring 
\begin{equation}
	R=\frac{\mathbb{C}[x,y,z]}{\left(F,\frac{\partial F}{\partial x},\frac{\partial F}{\partial y},\frac{\partial F}{\partial z}\right)}.
\end{equation}

This defines a fibration over the base $\mathcal{B}_\mu$ parameterized by the deformation parameters $\mu_i$. The generic fiber corresponds to a smooth deformation of the $ADE$ singularity. The base $\mathcal{B}_\mu$ is isomorphic to the space $\mathfrak{t}/\mathcal{W}$, where $\mathfrak{t}$ denotes the Cartan subalgebra of the $ADE$ Lie algebra and $\mathcal{W}$ its Weyl group.

\medskip
Let us now consider M-theory compactified on an $ADE$ singularity. The resulting 7d $\mathcal{N}=1$ theory contains vector multiplets in the adjoint representation, whose bosonic components include three real scalars $\phi_1, \phi_2, \phi_3$. After splitting the spacetime as $\mathbb{R}^7 = \mathbb{R}^5 \times \mathbb{C}_w$ and giving a holomorphic VEV to the complex scalar $\Phi = \phi_1 + i \phi_2$ which preserves half of the supersymmetry, \cite{DeMarco:2022dgh}
\begin{itemize}
	\item the zero modes of $A_\mu$ and $\phi_3$ assemble into background vector multiplets, generating a $U(1)^f$ flavor symmetry,
	\item while the zero modes of $\Phi$ localized at $w=0$ correspond to 5d massless hypermultiplets charged under the flavor symmetry.
\end{itemize}

The VEV of $\Phi$ defines a Calabi–Yau threefold $X$ as a fibration of deformed $ADE$ surfaces over $\mathbb{C}_w$. Recall that $\phi_3$ corresponds to the Kähler resolution parameters, and the D-term condition requires
\begin{equation}
	[\Phi,\phi_{3}]=0\,.
\end{equation}
Hence, the VEV of $\phi_3$ lies in the Cartan subalgebra, dual to the simple roots in the resolution process,
\begin{equation}
	\mathcal{H}=\langle\alpha_1^*,\ldots,\alpha_f^*\rangle\,.
\end{equation}
The Higgs field then resides in the subalgebra that commutes with $\phi_3$, namely the Levi subalgebra,
\begin{equation}
	\Phi(w)\in\mathcal{L}=\bigoplus_h\mathcal{L}_h\oplus\mathcal{H}\,,
\end{equation}
where $\mathcal{L}_h$ are the simple components of the Levi subalgebra.

\paragraph{Example.} 
For an $A_{n-1}$ singularity, the resolution involves a simple root $\alpha_c$ ($1 \leq c \leq n-1$). The field $\phi_3$ then takes values in the Cartan subalgebra as
\begin{equation}
	\phi_3\propto\begin{pmatrix}\frac{1}{c}\mathbb{I}_c&0\\0&-\frac{1}{n-c}\mathbb{I}_{n-c}\end{pmatrix},
\end{equation}
then
\begin{equation}
	\Phi=\begin{pmatrix}\Phi_{c\times c}&0\\0&\Phi_{(n-c)\times(n-c)}\end{pmatrix}\in\mathfrak{sl}(n)\,.
\end{equation}
The corresponding Levi subalgebra is
\begin{equation}
	\mathcal{L}=A_{c-1}\oplus A_{n-c-1}\oplus\langle\alpha_c^*\rangle\,.
\end{equation}

The admissible Higgs field configurations can be obtained from the defining threefold equation via two successive base changes, as described in \cite{DeMarco:2022dgh}. Once the Higgs field background is known, the zero modes can be computed by analyzing its fluctuations \cite{Cecotti:2010bp}:
\begin{equation}
	\Phi=\langle\Phi\rangle+\varphi\,,
\end{equation}
where $\varphi$ satisfies the BPS equations
\begin{equation}\label{Zeromodes}
	\partial\varphi=0\,,\quad\varphi\sim\varphi+[\Phi,g]\,.
\end{equation}
The Gopakumar–Vafa (GV) invariants can be obtained from the zero modes of the Higgs field. In general, choosing a vacuum expectation value (VEV) for the Higgs field corresponds to restricting to the Levi subalgebra of $G$. The 5d charged hypermultiplets then arise from the zero modes residing in the off-diagonal components of the Higgs field, and the number of such zero modes determines the GV invariants charged under the U(1) symmetries associated with the flavor or gauge symmetry of the 5d (or 3d) theory.

For example, the 5d SCFT arising from the $(A_5,D_{14})$ singularity has a block-diagonal Higgs field. Each diagonal block corresponds to a $U(1)$ flavor symmetry \cite{DeMarco:2021try}
\begin{equation}
	\begin{pmatrix}U(1)&&\\&U(1)&\\\\&&U(1)\end{pmatrix}
\end{equation}
and the corresponding zero modes are summarized as
\begin{equation}
	\begin{pmatrix}0&3&3\\\\3&12&18\\\\3&18&12\end{pmatrix}\,.
\end{equation}
Modding out by the diagonal $U(1)$ yields the effective flavor symmetry $U(1)^3/U(1)_{\text{diag}}$, and the GV invariants can then be expressed as
\begin{equation}
	n^{g=0}_{(1,0)}=3\ ,\ n^{g=0}_{(0,1)}=3\ ,\ n^{g=0}_{(1,1)}=18\,.
\end{equation}
In this way, the magnetic quiver of the associated 4d SCFT can be reconstructed via the rank-zero 5d/4d correspondence:
\begin{equation}
	\begin{tikzpicture}[x=.5cm,y=.5cm]
		\draw[ligne, black](0,0)--(4,0) ;
		\draw[ligne, black](2,-2)--(4,0) ;
		\draw[ligne, black](2,-2)--(0,0) ;
		\node[] at (-4,0) {$\mathrm{MQ}^{(4)}= $};
		\node[bd] at (0,0) [label=above:{{\scriptsize$1$}}] {};
		\node[bd] at (4,0) [label=above:{{\scriptsize$1$}}] {};
		\node[bd] at (2,-2) [label=below:{{\scriptsize$1$}}] {};
		\node[] at (2,0.5) {$18$};
		\node[] at (3.6,-1.6) {$3$};
		\node[] at (0.4,-1.6) {$3$};
	\end{tikzpicture} 
\end{equation}

\subsection{$(A_k,A_l)$ Argyres–Douglas Theory}
\label{sec:AkAl}

\subsubsection{$(A_1,A_{2N-1})$ theory}
\label{sec:A1A2N-1}

In Section~\ref{subsec:HBfromresolution}, we present the basic setup for deriving the hyperkähler metric of the generalized conifold — specifically, the singularity associated to the $(A_1, A_{2N-1})$ Argyres–Douglas theory. According to previous arguments, the corresponding magnetic quiver can be deduced from the Gopakumar–Vafa invariants, which themselves follow from the threefold equation of cDV singularities.

The singularity associated with the $(A_1, A_{2N-1})$ theory is defined by the hypersurface equation
\begin{equation}
	\mbf{X}:\ x_1^2+x_2^2+x_3^2+x_4^{2N}=0\,.
\end{equation}
From this singularity, one can compute the 4d central charges $a$, $c$ from the deformation and thus the asymptotic growth,
\begin{equation}
	a=\frac{12N^2-5N-5}{24(N+1)}\,,\ c=\frac{3N^2-N-1}{6N+6}\,,\ \mathcal{G}=2\,.
\end{equation}
It is known that this singularity has the following genus-zero GV invariant at degree one~\cite{Collinucci:2021wty}:
\begin{equation}
	\label{A1A2N-1-GV}
	n_{d=1}^{g=0}=N.
\end{equation}


We therefore conclude that the magnetic quiver associated with the $(A_1, A_{2N-1})$ theory is
\begin{equation}
	\begin{tikzpicture}[x=.5cm,y=.5cm]
		\draw[ligne, black](0,0)--(2,0);
		\node[] at (-2,0) {$\MQfour = $};
		\node[bd] at (0,0) [label=below:{{\scriptsize$1$}}] {};
		\node[bd] at (2,0) [label=below:{{\scriptsize$1$}}] {};
		\node[] at (1,-0.4) {\scriptsize$N$};
	\end{tikzpicture}
\end{equation}
Since the Coulomb branch of this magnetic quiver corresponds to the Higgs branch of the 4d SCFT, we may apply Coulomb quiver subtraction techniques~\cite{Cabrera:2018ann,Bourget:2019aer} to derive the Hasse diagram encoding its stratification:
\begin{equation}
	\begin{tikzpicture}[x=.5cm,y=.5cm]
		\node (1) [hasse] at (0,2) {};
		\node (2) [hasse] at (0,0) {};
		\draw (1) edge [] node[label=right:$A_{N-1}$] {} (2);
	\end{tikzpicture}
\end{equation}
Hence, the Higgs branch predicted from the singularity is $\mathbb{C}^2/\mathbb{Z}_N.$

The $(A_1, A_{2N-1})$ theory also admits a class S realization~\cite{Xie:2012hs}
\begin{equation}
	(A_{N-1},1,N,[N-1,1]),
\end{equation}
where we adopt the notation for Argyres--Douglas theories following~\cite{Li:2022njl}, in which a general AD theory is denoted as
\begin{equation}
	(\mathfrak{g},k,b,\mathfrak{f}).
\end{equation}
The corresponding VOA is known to be \cite{Beem:2017ooy,Song:2017oew}
\begin{equation}
	W_{-\frac{N^2}{N+1}}(\mathfrak{sl}_N, [N-1,1])\,.
\end{equation}
The associated variety of the affine vertex algebra $L_{-\frac{N^2}{N+1}}(\mathfrak{sl}_N)$ is the nilpotent cone of $\mathfrak{sl}_N$~\cite{Arakawa:2010ni}. Consequently, the associated variety of the W-algebra and thus the 4d Higgs branch is
\begin{equation}
	\mathbb{C}^2 / \mathbb{Z}_N \cong \mathcal{N}_{\mathfrak{sl}_N} \cap S_{[N-1,1]}\,.
\end{equation}
where $\mathcal{N}_{\mathfrak{sl}_N}$ denotes the nilpotent cone of $\mathfrak{sl}_N$, and $S_{[N-1,1]}$ is the Slodowy slice associated with the partition $[N-1,1]$. This result is  consistent with the expectations from the singularity analysis.

For this class of singularities, we also present an alternative approach to computing the GV invariant explicitly via an algebraic crepant resolution. We begin by rewriting the defining equation of the $(A_1, A_{2N-1})$ singularity as
\begin{equation}
	\begin{aligned}
		&x_1x_2+f_1f_2=0\,,\\ &f_1=x_3+ix_4^N\,,\\ &f_2=x_3-ix_4^N\,.
	\end{aligned}
\end{equation}
As in the case of conifold, we perform the crepant resolution $(x_1, f_1; \delta_1)$. That is, we replace $x_1\rightarrow x_1\delta_1$, $f_1\rightarrow f_1\delta_1$ and then divide the whole equation by $\delta_1$, yielding the resolved equation (note that $f_2=f_1-2ix_4^N$)
\begin{equation}
	x_1x_2+(f_1\delta_1-2ix_4^N)f_1=0\,.
\end{equation}
The resulting blow-up map $\pi:\widetilde{X} \rightarrow X$ is given explicitly by
\begin{equation}
	\pi:\mathrm{~}(x_1,x_2,f_1,f_2,\delta_1)\mathrm{~}\longmapsto\mathrm{~}(x_1\delta_1,\mathrm{~}x_2,\mathrm{~}f_1\delta_1,f_2)\,.
\end{equation}
The non-compact exceptional divisor $S_1: \delta_1 = 0$ is
\begin{equation}
	x_1x_2-2if_1x_4^N=0\,.
\end{equation}
As in the standard conifold case, to obtain the exceptional 2-cycle $C_1$, we set $x_2 = x_4 = 0$. A notable feature here is that, due to the $x_4^N$ term, the solution $x_2 = x_4 = 0$ has multiplicity $N$. In other words, there are $N$ exceptional 2-cycles of equal volume in the resolved geometry, which collectively shrink to zero volume in the singular limit, consistent with the GV invariant (\ref{A1A2N-1-GV}). 

\subsubsection{$(A_{k-1},A_{l-1})$ theory $(k\neq l)$}

Now let us generalize to the $(A_{k-1},A_{l-1})$ theory, defined by the singularity equation
\be
\label{AkAl}
\mbf{X}:\ x_1^2+x_2^2+x_3^k+x_4^l=0\,.
\ee
The four-dimensional central charges $(a,c)$ can be directly extracted from the singularity. For instance, when $k=2$, the closed-form expression for central charges is
\begin{equation}
\begin{aligned}
	a
	&=
	\frac{
		24l^2-15l-27-(5l+13)(-1)^l
	}{
		96(l+2)
	},
	\\
	c
	&=
	\frac{
		6l^2-3l-6-(l+2)(-1)^l
	}{
		24(l+2)
	},
	\\
	\mathcal{G}
	&=
	\frac{
		3(l+1)+(l+5)(-1)^l
	}{
		2(l+2)
	}.
\end{aligned}
\end{equation}

Let us now derive the GV invariants directly from the singularity equation~\eqref{AkAl}. 
The GV invariants associated with the $(A_{k-1}, A_{l-1})$ singularity were computed in~\cite{DeMarco:2021try} using the method reviewed in Section~\ref{sec:GV-5d}. 
The defining hypersurface equation can be rewritten as
\begin{equation}
	x_1^2+x_2^2+\det(x_3+ \Phi(x_4))=0\,.
\end{equation}
The determinant $\det(x_3 + \Phi)$ can be further factorized as
\begin{equation}
\det(x_3+ \Phi)=\prod_{s=1}^n \left(x_3^{k^\prime} + e^{2\pi i s/n} x_4^{l^\prime}\right)
\end{equation}
where $n = \gcd(k,l)$ and $k' = k/n$, $l' = l/n$. 
Each factor $\bigl(x_3^{k'} + e^{2\pi i s/n} x_4^{l'}\bigr)$ corresponds to a characteristic matrix of the form
\begin{equation}
	\begin{pmatrix}
		0 & * & 0 & \cdots & 0 \\
		0 & 0 & * & 0 & 0 \\
		\vdots & 0 & \ddots & \ddots & 0 \\
		0 & 0 & 0 & 0 & * \\
		-e^{2\pi i s/n} x_4 & 0 & 0 & 0 & 0
	\end{pmatrix},
\end{equation}
where $*$ denotes either constants or the variable $x_4$. 
The matrix $\Phi$ can therefore be expressed as a direct sum of these characteristic matrices.

Following~\eqref{Zeromodes}, the corresponding zero modes are organized as
\begin{equation}
	\begin{pmatrix}
		(k^\prime -1)(l^\prime -1) \text{ modes} & k^\prime l^\prime  \text{ modes} & \cdots & k^\prime l^\prime  \text{ modes} \\
		k^\prime l^\prime  \text{ modes} & \ddots &  & \vdots \\
		\vdots &  & \ddots & k^\prime l^\prime  \text{ modes} \\
		k^\prime l^\prime  \text{ modes} & \ldots & k^\prime l^\prime  \text{ modes} & (k^\prime-1)(l^\prime -1) \text{ modes}
	\end{pmatrix}.
\end{equation}
The flavor symmetry for 5d SCFT $U(1)^n / U(1)_{\mathrm{diag}}$ is realized by the adjoint action of
\begin{equation}
\varphi=	\begin{pmatrix}
		e^{i\alpha_1} \mathbb{I}_{k^\prime } &  &  &  \\
		& e^{i\alpha_2} \mathbb{I}_{k^\prime } &  &  \\
		&  & \ddots &  \\
		&  &  & e^{i\alpha_n} \mathbb{I}_{k^\prime }
	\end{pmatrix}, \quad \sum_{s=1}^n \alpha_s = \frac{2\pi n}{k^\prime }\,.
\end{equation}

The GV invariants can then be read off from the off-diagonal zero modes of $\varphi$. 
In particular, the genus-zero GV invariant is given by
\begin{equation}
	n_{(0,\cdots,0,1,0,\cdots,0,1,0)}^{0}=k^\prime l^\prime\,.
\end{equation}
This result predicts that the magnetic quiver should consist of $n$ $U(1)$ nodes, corresponding to the diagonal components of the Higgs field, with $k^\prime l^\prime$ edges connecting each pair of distinct $U(1)$ nodes. 
This structure precisely reproduces the magnetic quiver of the $(A_{k-1}, A_{l-1})$ theory.

Indeed, it is known that the magnetic quiver $\mathrm{MQ}^{(4)}$ associated with the $(A_{k-1}, A_{l-1})$ theory consists of $n \equiv \gcd(k,l)$ nodes, with $z\equiv k'l'=\frac{k l}{\gcd(k,l)^2}$ edges connecting each pair of nodes~\cite{Giacomelli:2020ryy,Xie:2021ewm}. 
As an illustrative example, the magnetic quiver corresponding to the $(A_5, A_8)$ theory is given by
\begin{equation} \begin{tikzpicture}[x=1cm,y=1cm, scale=1.5] \draw[ligne, black] (0,0)--(2,0) node[midway, above] {\scriptsize $6$}; \draw[ligne, black] (0,0)--(1,-1) node[midway, left] {\scriptsize $6$}; \draw[ligne, black] (2,0)--(1,-1) node[midway, right] {\scriptsize $6$}; \node[] at (-1,0) {$\mathrm{MQ}^{(4)} = $}; \node[bd] at (0,0) [label=left:{{\scriptsize$1$}}] {}; \node[bd] at (2,0) [label=right:{{\scriptsize$1$}}] {}; \node[bd] at (1,-1) [label=below:{{\scriptsize$1$}}] {}; \end{tikzpicture} \end{equation}

The derivation of magnetic quivers from GV invariants naturally explains the general rules governing magnetic quivers of Argyres--Douglas theories, as summarized in~\cite{Xie:2021ewm}.

The Hasse diagram can be obtained from the magnetic quiver via Coulomb branch quiver subtraction. 
Since the magnetic quiver is rather complicated, we focus on a specific subtraction pattern. 
As an illustrative example, the Hasse diagram of the Higgs branch for the case $k\neq l$ contains the following chain
\begin{equation}
	\label{AkAl-Hasse}
	\begin{tikzpicture}[x=.5cm,y=.5cm]
		\node (1) [hasse] at (0,2) {};
		\node (2) [hasse] at (0,0) {};
		\node (3) [hasse] at (0,-2) {};
		\node (dots) at (0,-2.7) {$\vdots$};
		\node (4) [hasse] at (0,-4) {};
		\node (5) [hasse] at (0,-6) {};
		\draw (1) edge [] node[label=right:$A_{z-1}$] {} (2);
		\draw (2) edge [] node[label=right:$A_{2z-1}$] {} (3);
		\draw (4) edge [] node[label=right:$A_{(n-1)z-1}$] {} (5);
	\end{tikzpicture} 
\end{equation}
Analogously to the $(A_1,A_{2N-1})$ theory, the Hasse diagram can also be derived from the crepant resolution.

Now let us try to match the Higgs branch structure with the crepant resolution $\tX$. The singularity equation (\ref{AkAl}) can be rewritten as
\be
x_1^2+x_2^2+\prod_{j=1}^n \left(x_3^{k/n}+\omega_jx_4^{l/n}\right)=0\,,
\ee
where $\omega_j$'s are numerical coefficients. Redefine coordinates
\begin{equation}
    f_j=x^{k/n}_3+\omega_jx^{l/n}_4,\quad j=1,\cdots,n\,.
\end{equation}
 We then perform the following resolution sequence to resolve (\ref{AkAl}):
\begin{equation}
	(x_1,f_j;\delta_j)\quad(j=1,\dots,n-1)\,.
\end{equation}
where $n \equiv \gcd(k,l)$. The resulting resolved geometry is described by the equation
\begin{equation}
	\begin{aligned}
		&x_1x_2+\prod_{j=1}^nf_j=0\,, \\
		&f_{j}\delta_{j}=x_{3}^{k/n}+\omega_{j}x_{4}^{l/n}\quad(j=1,\ldots,n-1)\,, \\
		&f_n=x_3^{k/n}+\omega_n x_4^{l/n}.
	\end{aligned}
\end{equation}

The layered structure of the Higgs branch can be understood as follows
\begin{itemize}
	\item Assuming $\delta_i \neq 0$ for $i=2,\dots,n-1$, the resolved equation becomes (using $x_3^{k/n}=f_1\delta_1-\omega_1 x_4^{l/n}$ and plugging into the expression of $f_n$)
	\begin{equation}
    \label{AkAl-resolved}
		x_1x_2+f_1\cdots f_{n-2} f_{n-1}(f_1\delta_1+(\omega_n-\omega_1)x_4^{l/n})=0\,.
	\end{equation}
	Here, $[x_1:f_{1}]$ located at $\delta_1 = x_2 = x_4 = 0$ forms the projective coordinates of a $\mathbb{P}^1$. Furthermore, setting $x_3 = 0$ yields $k/n$ copies of two-cycles. Thus, the final resolved geometry is of conifold type
	\begin{equation}
		x_1x_2+f_{1}x_4^{l/n}=0\,.
	\end{equation}
	Hence, since $\delta_1 = 0 = x_4$, it follows that $x_3 = 0$, and consequently the total contribution is $\frac{k}{n} \times \frac{l}{n} \equiv z$ two-cycles. According to arguments in \cite{Ooguri:1996me}, the corresponding metric describes the $A_{z-1}$ type Kleinian singularity $\sim \mb{C}^2/\mb{Z}_z$, corresponding to the top layer of (\ref{AkAl-Hasse}).
    
	\item Multiplying both sides of (\ref{AkAl-resolved}) by $\delta_{n-1}$ and plug in $f_{n-1}\delta_{n-1}=f_1\delta_1+(\omega_{n-1}-\omega_1)x_4^{l/n}$, we obtain
	\begin{equation}
		x_1x_2\delta_{n-1}+f_1\cdots f_{n-2}(f_1\delta_1+(\omega_{n-1}-\omega_1)x_4^{l/n})(f_1\delta_1+(\omega_n-\omega_1)x_4^{l/n})=0\,.
	\end{equation}
	Setting $\delta_1 = x_2 = x_4 = 0$, we find $2z$ two-cycles, corresponding to the singularity $\mathbb{C}^2/\mathbb{Z}_{2z}$, in the second to the top layer of (\ref{AkAl-Hasse}). Iterating this process eventually reconstructs the full Hasse diagram of the Higgs branch.
\end{itemize}

Although this class of theories admits a class~$\mathcal{S}$ realization of the form
\begin{equation}
	(A_{l-1},k,l,[l])\,,
\end{equation}
the associated VOA for general values of $k$ and $l$ has been much less explored. However, when $l$ and $k$ are coprime, the corresponding VOA is known to be~\cite{Song:2017oew}
\begin{equation}
	W_{-l+\frac{l}{k+l}}(\mathfrak{sl}_l,[l])\,.
\end{equation}
The associated variety of this VOA is
\begin{equation}
	\mathcal{N}_{\mathfrak{sl}_l}\cap S_{[l]}\,,
\end{equation}
which implies that the resulting $W$-algebra is lisse~\cite{Xie:2019vzr}. This is consistent with the absence of a magnetic quiver in this case.

For more general values of $(k, l)$, the magnetic quiver indicates that there exist multiple possible ways to subtract the quiver. Consequently, these theories do not appear to correspond to affine $W$-algebras as the Hasse diagrams of the Higgs branches are not of the ones of intersections between nilpotent orbits and Slodowy slices. Such Argyres–Douglas theories can instead be constructed by gauging the flavor symmetries of certain Argyres–Douglas matter theories~\cite{Xie:2017vaf,Xie:2017aqx,Beem:2023ofp}. The corresponding VOAs can then be realized through coset constructions~\cite{Xie:2019yds}. The associated varieties of these theories, however, remain to be further explored.

\subsubsection{$(A_{k-1},A_{l-1})$ theory $(k=l)$}
For the case of $k=l$, 
the central charges $a,c$ of the 4d theory  from the singularity of this series are
\begin{equation}
	\begin{aligned}
	    &c=\frac{k^3-5k+4}{24}\,,\\
        & a=\frac{k^3-6k+5}{24}\,.
	\end{aligned}
\end{equation}

The corresponding magnetic quiver $\MQfour$ has $k$ nodes with $z=\frac{kl}{\mathrm{gcd}(k,l)^2}=1$ edge between any pair of nodes. Hence in the Hasse diagram we expect minimal nilpotent orbits from the quiver subtraction, rather than Kleinian singularities. We observe that in such cases, the relation between partial resolution and the Hasse diagram also fails. 

Let us take the simplest example of $(A_2, A_2)$ with magnetic quiver:
\begin{equation}
	\begin{tikzpicture}[x=.5cm,y=.5cm]
		\draw[ligne, black](0,0)--(2,0);
		\draw[ligne, black](0,0)--(1,-1);
		\draw[ligne, black](2,0)--(1,-1);
		\node[] at (-3,0) {$\MQfour = $};
		\node[bd] at (0,0) [label=left:{{\scriptsize$1$}}] {};
		\node[bd] at (2,0) [label=right:{{\scriptsize$1$}}] {};
		\node[bd] at (1,-1) [label=below:{{\scriptsize$1$}}] {};
	\end{tikzpicture}
\end{equation}
which leads to the Hasse diagram
\begin{equation}
	\begin{tikzpicture}[x=.5cm,y=.5cm]
		\node (1) [hasse] at (0,2) {};
		\node (2) [hasse] at (0,0) {};
		\draw (1) edge [] node[label=right:$a_2$] {} (2);
	\end{tikzpicture} 
\end{equation}
The Hasse diagram contains only one layer, consistent with the Coulomb branch dimension $\widehat{r} = 1$ of the $(A_2, A_2)$ Argyres–Douglas theory. However it is not straightforward to obtain this minimal nilpotent orbit from the resolution perspective.

Nonetheless, the computation of Gopakumar-Vafa invariants still applies to these theories. For $(A_{k-1}, A_{k-1})$ Argyres–Douglas theories with $k > 1$, we can similarly rewrite the singular equation as
\begin{equation}
	xy+\prod_{i=1}^{k}f_i=0\:,\:f_i=z+\omega_iw\,,
\end{equation}
where $\omega_i$ are $k$-th roots of unity, and perform the resolution sequence
\begin{equation}
	(x,f_1;\delta_1)\:,\:\ldots\:,\:(x,f_{k-1};\delta_{k-1})\:.
\end{equation}
The resulting exceptional locus is a chain of $\mathbb{P}^1$ curves $C_1, \dots, C_{k-1}$. Each curve $C_i$ ($1 \leq i \leq k-1$) contributes a genus-0 GV invariant $n_{(0,\dots,1,\dots,0)} = 1$, where the 1 appears at the $i$-th entry. All linear combinations of the form $C_{ij} \equiv C_i + C_{i+1} + \dots + C_j$ ($1 \leq i \leq j \leq k-1$) then yield genus-0 GV invariants $n_{(0,\dots 0,1,0,\dots,1,\dots,0)} = 1$, where the 1's occupy entries $i$ through $j$. These curves precisely correspond to the positive roots of an $A_{k-1}$ Lie algebra, which is the flavor symmetry of the 4d theory. 

These GV invariants predict that the resulting magnetic quiver contains $k$ $U(1)$ gauge nodes, with $n_{(0,\dots 0,1,0,\dots,1,\dots,0)} =1$ edge between any pair of distinct gauge nodes. Hence, the GV invariants determine the magnetic quiver of the $(A_{k-1},A_{k-1})$ theory.

Using the above data, the VOA $V_X$ corresponding to the singularity of this type has central charge
\begin{equation}
    c_{2d}=-12\,c=-\frac{k^3-5k+4}{2}
\end{equation}
and 
the asymptotic growth of vacuum character is
\begin{equation}
	\mathcal{G}=-48\,(a-c)=2(k-1)\,.
\end{equation}
The associated variety of $V_X$ is $\mathrm{CB}(\mathrm{MQ}^{(4)})$ whose Hasse diagram can then be reconstructed by performing quiver subtraction.

\subsection{More general cDV type singularities}
\label{subsec:cDV}

While a general resolution method for arbitrary cDV-type singularities is not available, the connection between magnetic quivers and GV invariants is universal for cDV singularities. This universality allows one to predict the magnetic quiver for the associated 4d $\mathcal{N}=2$ SCFTs. For instance, in the $(A_{4n-1}, D_7)$ theory with $\gcd(n,3)=1$, given by the singularity equation
\be
\mbf{X}:\ x_1^2+x_2^{4n}+x_3(x_3^{5}+x_4^2)=0\,,
\ee
the magnetic quiver is given by \cite{Carta:2021whq}:
\begin{equation}\label{MQAD}
	\begin{tikzpicture}
		
		\node at (-5.0, 0) {$\MQfour = $}; 
		
		\node at (-3.5, 0) { $\mathbf{[n]_2}$};
		\node at (3.5, 0) { $\mathbf{[n]_2}$};
		
		\draw[black, thick] (-1.8, 0) -- (1.8, 0);
		\node[black] at (0, -0.35) { $3n$};
		
		\node at (0, 1.7) { $D_1$}; 
		
		\node at (-2.2, 0) { $D_1$}; 
		\node at (2.2, 0) { $D_1$};
		
		\draw[gray, thick] (-1.8, 0) -- (0, 1.35);
		\draw[gray, thick] (1.8, 0) -- (0, 1.35);
		
		\node at (-0.9, 1.15) { $n$};
		\node at (0.9, 1.15) { $n$}; 
		
		\draw[black, thick] (-2.5, 0) -- (-3.0, 0);
		\draw[black, thick] (2.5, 0) -- (3.0, 0);
		
		\node at (4.5, 0) { $\mathbf{/ \mathbb{Z}_2}$};
		
	\end{tikzpicture}
\end{equation}
Here, a diagonal $\mathbb{Z}_2$ symmetry is gauged, and $\mathbf{[n]_2}$ denotes charge-2 matter. The lines connecting the $D_1$ nodes represent half-hypermultiplets in bifundamental representations. Ignoring this $\mathbb{Z}_2$ gauging, the magnetic quiver can be reconstructed from the GV invariants \cite{DeMarco:2022dgh}:
\begin{equation}
	\begin{gathered}
		(q_{a},q_{b},q_{c})=(0,2,0):\boldsymbol{n},\\
		(q_{a},q_{b},q_{c})=(0,0,2):\boldsymbol{n},\\
		(q_{a},q_{b},q_{c})=(0,1,1):6\boldsymbol{n},\\
		(q_{a},q_{b},q_{c})=(1,1,0):2\boldsymbol{n},\\
		(q_{a},q_{b},q_{c})=(1,0,1):2\boldsymbol{n}\,.
	\end{gathered}
\end{equation}

The central charges of the corresponding VOAs for $n \leq 10$ are listed in Table~\ref{acforAD}:

\begin{table}[h]
	\centering 
	\renewcommand{\arraystretch}{1.3}
	\begin{tabular}{c|c c c c c c c}
		\hline
		$n$ & $1$ & $2$ & $4$ & $5$ & $7$ & $8$ & $10$ \\
		\hline
		$a$ & $\frac{53}{8}$ & $\frac{559}{24}$ & $\frac{3827}{56}$ & $\frac{2251}{24}$ & $\frac{1181}{8}$ & $\frac{46325}{264}$ & $\frac{24161}{104}$ \\
		$c=-\frac{1}{12}c_{2d}$ & $\frac{27}{4}$ & $\frac{703}{30}$ & $\frac{137}{2}$ & $\frac{2255}{24}$ & $\frac{739}{5}$ & $\frac{11593}{66}$ & $\frac{465}{2}$ \\
		$\mathcal{G}$ & $6$ & $\frac{34}{5}$ & $\frac{54}{7}$ & $8$ & $\frac{42}{5}$ & $\frac{94}{11}$ & $\frac{114}{13}$ \\
		\hline
	\end{tabular}
	\caption{Central charges $a$, $c$, and asymptotic growth $\mathcal{G}$ for $(A_{4n-1},D_7)$ with gcd$(n,3)=1$.}
	\label{acforAD}
\end{table}

In the following, we focus on specific cases whose Higgs branches can be extracted from the resolution of singularities. For other cDV singularities, the magnetic quiver can be directly obtained from the corresponding GV invariants.

\subsubsection{$(A_1,D_N)$ theory, odd $N$}

We express the singularity as:
\begin{equation}\label{ADeven}
	\mbf{X}:\ x_1x_2+x_3(x_3^{N-2}+x_4^2)=0\,.
\end{equation}
The central charges $a$, $c$, and the asymptotic growth are
\begin{equation}
	a=\frac{4N^2-5N+1}{16N}\,,\ c=\frac{1}{4}(N-1)\,,\ \mathcal{G}=3-\frac{3}{N}\,.
\end{equation}
The GV invariants for these theories, computed in \cite{DeMarco:2021try}, are
\begin{equation}
	n_1^0=2
\end{equation}
for odd values of $N$. Therefore, the GV invariant predicts that the magnetic quiver takes the form:
\begin{equation}
	\begin{tikzpicture}[x=.5cm,y=.5cm]
		\draw[ligne, black](0,0)--(2,0);
		\node[] at (-2,0) {$\MQfour = $};
		\node[bd] at (0,0) [label=below:{{\scriptsize$1$}}] {};
		\node[bd] at (2,0) [label=below:{{\scriptsize$1$}}] {};
		\node[] at (1,-0.4) {\scriptsize$2$};
	\end{tikzpicture}
\end{equation}
Applying the Coulomb branch subtraction then yields the Hasse diagram of the Higgs branch,
\begin{equation}\label{HasseADO}
	\begin{tikzpicture}[x=.5cm,y=.5cm]
		\node (1) [hasse] at (0,2) {};
		\node at (0,3) {};
		\node (2) [hasse] at (0,0) {};
		\draw (1) edge [] node[label=right:$A_1$] {} (2);
	\end{tikzpicture}
\end{equation}

Indeed, the magnetic quiver associated with this case is
\begin{equation}
	\mathrm{MQ}^{(4)}\cong(\mathrm{hyper})^{\otimes((N-1)/2-1)}\otimes\mathrm{SQED}[N_{f}=2]~.
\end{equation}
Note that this magnetic quiver contains a factor of $\frac{N-1}{2}-1$ free hypermultiplets, which precisely corresponds to the Higgs branch of the $(A_1, A_{N-3})$ Argyres--Douglas theory~\cite{Benvenuti:2018bav,Dedushenko:2019mnd,Closset:2020scj}. The $\mathrm{SQED}[N_f=2]$ factor reproduces the Hasse diagram in \eqref{HasseADO}, in exact agreement with the prediction from the GV invariant.

This magnetic quiver can also be understood from the crepant resolution. When $N$ is odd, the defining equation cannot be factorized further. The crepant resolution is specified by $(x_1, x_3; \delta_1)$, leading to the resolved equation
\begin{equation}
	x_1x_2+x_3(x_3^{N-2}\delta_1^{N-2}+x_4^2)=0\,.
\end{equation}
For $N>3$, a terminal Argyres--Douglas singularity of type $(A_1, A_{N-3})$ remains at $x_1 = x_2 = \delta_1 = x_4 = 0$. In addition, the exceptional divisor $\delta_1=0$ exhibits a local singularity of the form $x_1 x_2 + x_4^2 = 0$.

Restricting the resolved geometry to $\delta_1=0$ yields
\begin{equation}
	x_1x_2+x_4^2=0\,,
\end{equation}
from which we see that there are two $\mathbb{P}^1$'s of equal volume, specified by $\delta_1 = x_2 = x_4 = 0 .$ This leads to a $\mathbb{C}^2/\mathbb{Z}_2$ quotient in the Higgs branch, which precisely matches the Hasse diagram shown in \eqref{HasseADO}.

\subsubsection{$(A_1,D_N)$ theory, even $N$}
\label{sec:A1DN-even}

The singularity \eqref{ADeven} for even N has the central charges $a,c$ and asymptotic growth
\begin{equation}
	a=\frac{1}{12}(3N-5)\,,\ c=\frac{N}{4}-\frac{1}{3}\,,\ \mathcal{G}=4.
\end{equation}
The GV invariants for these theories, computed in \cite{DeMarco:2021try}, are given for even $N$ by
\begin{equation}
	n_{(1,0)}^0=1\,,\ n_{(0,1)}^0=1\,,\ n_{(1,1)}^0=\frac{N-2}{2}\,.
\end{equation}
These GV invariants predict the magnetic quiver should be 
\begin{equation}
	\begin{tikzpicture}[x=.5cm,y=.5cm]
		\draw[ligne, black](0,0)--(4,0) ;
		\draw[ligne, black](2,-2)--(4,0) ;
		\draw[ligne, black](2,-2)--(0,0) ;
		\node[] at (-4,0) {$\mathrm{MQ}^{(4)}= $};
		\node[bd] at (0,0) [label=above:{{\scriptsize$1$}}] {};
		\node[bd] at (4,0) [label=above:{{\scriptsize$1$}}] {};
		\node[bd] at (2,-2) [label=below:{{\scriptsize$1$}}] {};
		\node[] at (2,1) {$\frac{N-2}{2}$};
		\node[] at (3.6,-1.6) {$1$};
		\node[] at (0.4,-1.6) {$1$};
	\end{tikzpicture} 
\end{equation}
The expected Hasse diagram of the Higgs branch is
\begin{equation}
	\begin{tikzpicture}[x=.5cm,y=.5cm]
		\node (1) [hasse] at (0,2) {};
		\node (2) [hasse] at (0,0) {};
		\node (3) [hasse] at (0,-2) {};
		\draw (1) edge [] node[label=right:$A_{\frac{N}{2}-2}$] {} (2);
		\draw (2) edge [] node[label=right:$a_1$] {} (3);
	\end{tikzpicture} 
\end{equation}

We now reproduce the Hasse diagram from the crepant resolution of singularity \eqref{ADeven}.
The canonical threefold singularity can be rewritten as
\begin{equation}
	\begin{gathered}\begin{aligned}&x_1x_2+x_3f_1f_2=0\end{aligned}\\\begin{aligned}f_1&=x_4+ix_3^{(N-2)/2}\end{aligned}\\\begin{aligned}f_2&=x_4-ix_3^{(N-2)/2}.\end{aligned}\end{gathered}
\end{equation}
Performing the crepant resolution
\begin{equation}
	(x_1,x_3;\delta_1),(x_1,f_1;\delta_2),
\end{equation}
we obtain the resolved equation
\begin{equation}
	\begin{aligned}&\begin{aligned}x_1x_2+x_3f_1f_2=0\end{aligned}\\&\begin{aligned}f_1\delta_2&=x_4+ix_3^{(N-2)/2}\delta_1^{(N-2)/2}\end{aligned}\\&\begin{aligned}f_2&=x_4-ix_3^{(N-2)/2}\delta_1^{(N-2)/2}.\end{aligned}\end{aligned}
\end{equation}
Rewriting the first equation as
\begin{equation}
	x_1x_2+x_3f_1(f_1\delta_2-2ix_3^{(N-2)/2}\delta_1^{(N-2)/2})=0,
\end{equation}
we observe that for $N > 4$, the exceptional divisor $\delta_2 = 0$ carries a singularity of type $x_1 x_2 + \delta_1^{(N-2)/2} = 0$, while $\delta_1 = 0$ gives $x_1 x_2 + f_1^2 = 0$.

For $\delta_1 = 0$, the exceptional 2-cycle $C_1$ lies along $\delta_1 = x_2 = f_1 = 0$. Due to the quadratic term $f_1^2$, there are two copies of $C_1$ with equal volume, shrinking simultaneously in the singular limit. This corresponds to the subspace $\mathbb{C}^2 / \mathbb{Z}_2$ in the bottom layer of the Hasse diagram.

For $\delta_2 = 0$, the exceptional 2-cycle $C_2$ lies along $\delta_2 = x_2 = \delta_1 = 0$. The term $\delta_1^{(N-2)/2}$ implies $(N-2)/2$ copies of $C_2$ with equal volume, shrinking together in the singular limit, corresponding to $\mathbb{C}^2 / \mathbb{Z}_{(N-2)/2}$ in the top layer.

In this case, the $(A_1,D_N)$ theory admits the class S realization 
\begin{equation}
	(A_{k-1},1,k,[k-2,1^2])\,,
\end{equation}
for $N=2k-2$. The VOA associated to the even $N$ case is  $W_{-k + \frac{k}{k-1}}(\mathfrak{sl}_k, [k-2,1^2])$ \cite{Song:2017oew}. The associated variety predicted from the VOA is exactly
\begin{equation}
	\overline{\mathcal{O}}_{[k-1,1]}\cap S_{[k-2,1^2]}\,,
\end{equation}
consistent with the Hasse diagrams derived above.

\subsubsection{$(A_{2n-1},E_7)$ theory}

We begin with the singular geometry defined by
\begin{equation}
	\mbf{X}:\ (x_1+ix_2^n)(x_1-ix_2^n)+x_3(x_3^2+x_4^3)=0\,.
\end{equation}
 The central charges of the corresponding VOAs for $1\leq n \leq 10$ are given in Table~\ref{acAE}.
\begin{table}[h]
	\centering
	\renewcommand{\arraystretch}{1.3}
	\begin{tabular}{c|c c c c c c c c c c}
		\hline
		$n$
		& $1$ & $2$ & $3$ & $4$ & $5$
		& $6$ & $7$ & $8$ & $9$ & $10$ \\
		\hline
		$a$
		& $\frac{3}{2}$
		& $\frac{22}{3}$
		& $\frac{191}{12}$
		& $\frac{345}{13}$
		& $\frac{233}{6}$
		& $\frac{787}{15}$
		& $\frac{537}{8}$
		& $\frac{4217}{51}$
		& $\frac{2359}{24}$
		& $\frac{2202}{19}$ \\
		$c$
		& $\frac{31}{20}$
		& $\frac{244}{33}$
		& $\frac{383}{24}$
		& $\frac{346}{13}$
		& $\frac{467}{12}$
		& $\frac{788}{15}$
		& $\frac{2151}{32}$
		& $\frac{4222}{51}$
		& $\frac{1183}{12}$
		& $116$ \\
		$\mathcal{G}$
		& $\frac{12}{5}$
		& $\frac{32}{11}$
		& $2$
		& $\frac{48}{13}$
		& $4$
		& $\frac{16}{5}$
		& $\frac{9}{2}$
		& $\frac{80}{17}$
		& $14$
		& $\frac{96}{19}$ \\
		\hline
	\end{tabular}
	\caption{The values of $a$, $c$ and $\mathcal{G}$ for $(A_{2n-1},E_7)$}
    \label{acAE}
\end{table}
The genus-zero GV invariant computed from the five-dimensional SCFT yields~\cite{Collinucci:2022rii}
\begin{equation}
	n_1^0=3n.
\end{equation}
This leads to the magnetic quiver
\begin{equation}\label{MQAE}
	\begin{tikzpicture}[x=.5cm,y=.5cm]
		\draw[ligne, black](0,0)--(2,0);
		\node[] at (-2,0) {$\MQfour = $};
		\node[bd] at (0,0) [label=below:{{\scriptsize$1$}}] {};
		\node[bd] at (2,0) [label=below:{{\scriptsize$1$}}] {};
		\node[] at (1,-0.4) {\scriptsize$3n$};
	\end{tikzpicture}
\end{equation}
This result agrees with the magnetic quiver reported in \cite{Carta:2022spy}. The Coulomb branch subtraction gives the Hasse diagram 
\begin{equation}
	\begin{tikzpicture}[x=.5cm,y=.5cm]
		\node (1) [hasse] at (0,2) {};
		\node (2) [hasse] at (0,0) {};
		\draw (1) edge [] node[label=right:$A_{3n-1}$] {} (2);
	\end{tikzpicture}
\end{equation}

The same Hasse diagram can also be derived directly from the singularity. The defining equation can be rewritten by introducing
\begin{equation}
	\begin{aligned}f_1f_2+x_3(x_3^2+x_4^3)=0\\f_1=x_1+ix_2^n\\f_2=x_1-ix_2^n\,.\end{aligned}
\end{equation}
Performing the crepant resolution $(f_1, x_3; \delta_1)$ yields the resolved equations
\begin{equation}
	\begin{aligned}f_1(f_1\delta_1-2ix_2^n)+x_3(x_3^2\delta_1^2+x_4^3)&=0\\f_1\delta_1&=x_1+ix_2^n\,.\end{aligned}
\end{equation}
A terminal singularity remains at $f_1 = x_2 = \delta_1 = x_4 = 0$, and the exceptional divisor $\delta_1 = 0$ is itself singular. 

To analyze the Higgs branch structure, consider the exceptional 2-cycle $C_1$ defined by $\delta_1 = x_2 = x_4 = 0$. The term $-2i x_2^n$ contributes an order $n$ vanishing, while $x_4^3$ contributes an order 3 vanishing. This implies the existence of $3n$ two-cycles of equal volume that shrink to zero size in the singular limit, corresponding to the Higgs branch $\mathbb{C}^2 / \mathbb{Z}_{3n}$.

The free hypermultiplet sector of the magnetic quiver can also be inferred from the singularity.  Notice that the Gorenstein terminal singularity at $f_1 = x_2 = \delta_1 = x_4 = 0$ has the local form of
\begin{equation}
f_1^2\delta_1-2if_1 x_2^n+\delta_1^2+x_4^3=0\,.
\end{equation}
This singularity gives rise to $4n - 3$ free hypermultiplets on the Higgs branch.

\section{$r>0$ cases with a smooth resolution}
\label{sec:r>0resolution}
In this section, we analyze theories associated with singularities where $r > 0$ and a singular divisor is present in the resolved, smooth Calabi-Yau threefold $\tX$. Such cases include $x_1^3+x_2^3+x_3^3+x_4^4=0$ discussed in \cite{Closset:2020afy}, and for isolated hypersurface singularities with low $r$ they are classified in \cite{Closset:2021lwy}.  For configurations involving $r$ compact divisors, the type IIA analysis in the previous section proves insufficient. This is because the coupling constant associated with a compact divisor,
\begin{equation}
\frac{1}{g^2}\sim\int_{\tX} \frac{\alpha'}{R}  \omega_i^{(1,1)} \wedge \star  \omega_i^{(1,1)}
\end{equation}
vanishes in the limit $\alpha'/R\to 0$, and strongly coupled gauge dynamics are involved. Consequently, deriving the magnetic quiver for a general singularity with $r > 0$ remains challenging. Moreover, one should also be careful about the cases with non-trivial 3-cycles in $\tX$, as they would correspond to free vector multiplets in $\FTfour$~\cite{Closset:2020scj}.

\subsection{$r=1$ cases with $f=0$ and a singular divisor}
\label{subsec:r=1}

	We study singularities with a single 4-cycle ($r=1$) and low Higgs branch dimensions. We begin with the following ``smoothable'' singularities listed in \cite{Closset:2021lwy} with $f = 0$ and $b_3=0$, summarized in Table~\ref{tab:singularities}. For this specific class, we develop a method to derive the magnetic quiver and identify the corresponding VOA in certain examples.
    
\begin{table}[h]
	\centering
	\begin{tabular}{c|c|c|c|c|c|c|c|c|c}
		\hline
		sing. & $F(x)$ & $r$ & $f$ & $d_H$ & $\widehat{r}$ & $\widehat{d_H}$ & $\Delta \mathcal{A}_r$ & $b_3$ & $c$ \\
		\hline
		sing($E_8$) & $x_3^7+x_4^5 x_3+x_2^3+x_1^2$ & 1 & 0 & 29 & 29 & 1 & 0 & 0 & $\frac{6989}{186}$\\
		sing($E_7$) & $x_2^5+x_3^3 x_2+x_3 x_4^3+x_1^2$ & 1 & 0 & 17 & 17 & 1 & 0 & 0 &$\frac{1853}{114}$\\
		sing($E_6$) & $x_2^4+x_3^2 x_2+x_1^3+x_3 x_4^2$ & 1 & 0 & 11 & 11 & 1 & 0 & 0 &$\frac{671}{78}$\\
		\hline
	\end{tabular}
	\caption{Canonical singularities with $r=1$, $f=0$, $b_3=0$ and a smooth crepant resolution.}
	\label{tab:singularities}
\end{table}


We now  examine the resolution properties of these singularities. After a single weighted blow-up of the ambient space $\mathbb{C}^4$, the Calabi-Yau threefold becomes smooth. However, the resulting exceptional compact divisor $S_1:\delta_1=0$ precisely contains an $E_n$-type surface (du Val) singularity, as summarized in Table~\ref{t:SingEn}. 
\begin{table}
	\centering
\begin{tabular}{c|c|c|c}
	sing. & resolution & divisor eq. & du Val singularity \\
	\hline
	sing($E_8$) & $(x_1^{(3)},x_2^{(2)},x_3^{(1)},x_4^{(1)};\delta_1)$ & $\delta_1=x_4^5 x_3+x_2^3+x_1^2=0$ & $E_8:x_1=x_2=x_4=0$\\
	sing($E_7$) & $(x_1^{(2)},x_2^{(1)},x_3^{(1)},x_4^{(1)};\delta_1)$ & $\delta_1=x_3^3 x_2+x_3 x_4^3+x_1^2=0$ & $E_7:x_1=x_3=x_4=0$\\
	sing($E_6$) & $(x_1,x_2,x_3,x_4;\delta_1)$ & $\delta_1=x_3^2 x_2+x_1^3+x_3 x_4^2=0$ & $E_6:x_1=x_3=x_4=0$
\end{tabular}
\caption{Three examples of singularities labeled as sing($E_n$), with a smooth crepant resolution but singular divisor, 5d rank $r=1$, flavor rank $f=0$ and the number of 3-cycles $b_3=0$.}\label{t:SingEn}
\end{table}

Now we would like to provide more physical interpretations about the singular divisor on the resolved, smooth CY3s $\tX$. Resolving these surface singularities on the divisor $S_1$ of $\mathrm{sing}(E_n)$ yields a smooth generalized del Pezzo surface $gdP_n$ with a set of $(-2)$-curves forming an $E_n$ Dynkin diagram~\cite{derenthal2014singular}. This geometric operation, however, is not a resolution of the CY3 singularity and therefore does not correspond to any supersymmetric deformation of the 4d $\mathcal{N}=2$ field theory.

	Instead, we interpret the smoothing of $S_1$ as the resolution (or deformation) of the singular Higgs branch $\mc{M}_H=\mathbb{C}^2/\Gamma_{E_n}$ into the corresponding $E_n$-type ALE space, although the closed-form metric for such spaces remains unknown. Analogous to the $A_n$ case, we view the hypermultiplet moduli space metric
\begin{equation}
	ds^2=\tau_2^{-2}[V^{-1}(\mathrm{d}d_1-\vec{A}\cdot\mathrm{d}\vec{y})^2+V|\mathrm{d}\vec{y}|^2]
\end{equation}
	as the limit where the ALE ``centers'' coincide.

    In other words, the singular limit $\mathbb{C}^2/\Gamma_{E_n}$ of the $E_n$-type ALE space exactly corresponds to the case of the singular  $S_1$, where all the $(-2)$-curves on the $gdP_n$ are contracted to zero-volume, i.e. the case of the resolved $\tX$. Due to the shrinking of all $(-2)$-curves, $\tX$ only has a single compact 2-cycle, that is the Poincar\'e dual of the compact 4-cycle $S_1$. Thus we have the flavor rank $f=0$ of $\FTfour$.

Now we provide another physical argument for the Higgs branch of $\FTfour$ being $\mathbb{C}^2/\Gamma_{E_n}$. It is instructive to compare the 5d and 4d theories from $\mathrm{sing}(E_n)$ with that from a local $dP_n$ singularity. Denote $\mathbf{X} = \mathrm{sing}(E_n)$ and $\mathbf{X}' = \text{local } dP_n$. Thus M-theory on $\mathbf{X}'$ gives the rank-1 Seiberg $E_n$ theory, denoted by $\mathcal{T}^{5\mathrm{d}}_{\mathbf{X}'}$. From the crepant resolution perspective, $\tX$ can be viewed as a special point in the complex structure moduli space of $\widetilde{\mathbf{X}'}$. From the 5d Coulomb branch perspective, $\FT$ and $\mathcal{T}^{5\mathrm{d}}_{\mathbf{X}'}$ have the same Coulomb branch structure, but different extended Coulomb branch, as $\mathcal{T}^{5\mathrm{d}}_{\mathbf{X}'}$ is the rank-1 Seiberg $E_n$ theory whose flavor rank equals to $n$.  

Moreover, the 5d SCFTs $\FT$ from M-theory on $\mathbf{X}$ and $\mathbf{X}'$ share the same Higgs branch dimension $d_H=h^\vee-1$. Along with the relation between the resolved $\tX$ and $\widetilde{\mathbf{X}'}$, we would expect $\FT$ and $\mathcal{T}^{5\mathrm{d}}_{\mathbf{X}'}$ to have the same Higgs branch---the minimal nilpotent orbit of $E_n$ for $\mathbf{X} = \mathrm{sing}(E_n)$.

Note that we have the flavor rank $f=0$ for $\mathbf{X} = \mathrm{sing}(E_n)$, which can be confirmed by both the Coulomb branch spectrum of $\FTfour$ and the lack of additional 2-cycles in the resolution geometry. This is a bit puzzling, because if $f=0$, how can there be a $d_H=h^\vee-1$-dimensional Higgs branch that is the minimal nilpotent orbit of $E_n$? The interpretation should be that $f=0$ is really meant for the 4d non-Lagrangian SCFT $\FTfour$. But for the 5d SCFT $\FT$, its Coulomb branch is a special locus of the extended Coulomb branch of the rank-1 Seiberg $E_n$ theory, whose actual flavor rank equals to $n$, which is different from $f$.

\begin{figure}
\centering
\includegraphics[height=7cm]{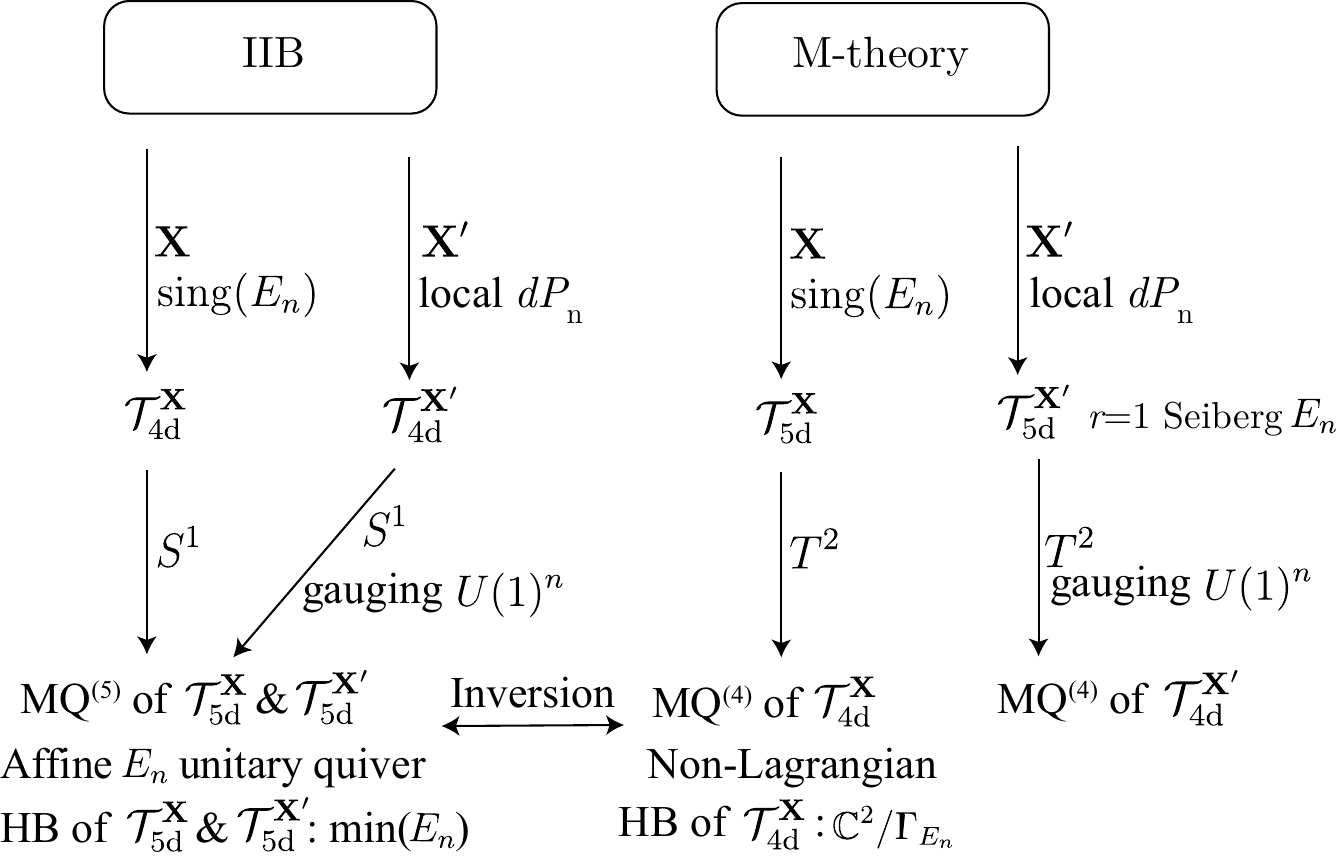}\label{f:En-Inversion}
\caption{Consider two related singularities $\mathbf{X}=\mathrm{sing}(E_n)$ listed in Table~\ref{t:SingEn} and $\mathbf{X}'=$ local $dP_n$ singularity, we can define 4d $\mc{N}=2$ SCFTs $\FTfour$, $\mathcal{T}^{4\mathrm{d}}_{\mathbf{X}'}$ and 5d SCFTs $\FT$, $\mathcal{T}^{5\mathrm{d}}_{\mathbf{X}'}$. Here we give the relation between their Higgs branches, via dimensional reduction, inversion and gauging. For the Higgs branch of $\mathcal{T}^{4\mathrm{d}}_{\mathbf{X}'}$, we provide a discussion of examples such as $\mathbf{X}': x_1^3+x_2^3+x_3^3+x_4^3=0$ in section~\ref{sec:D2nD2n}.}
\end{figure}

Nonetheless, since $f=0$ for $\mathbf{X} = \mathrm{sing}(E_n)$, we have $r = \widehat{d}_H$ and $\widehat{r} = d_H$. From the relations between $\FT$ and $\FTfour$ \cite{Closset:2020scj}, dimensional reduction to 3d implies that $D_{T^2}\FT$ and $D_{S^1}\FTfour$ should be 3d mirror to each other. In this sense,
\begin{equation}
	\text{HB of }\FTfour=\text{HB of }D_{S^1}\FTfour=\text{CB of }D_{T^2}\FT.
\end{equation}
Since the Higgs branch of $\FT$ for $\mathbf{X} = \mathrm{sing}(E_n)$ is the minimal nilpotent orbit of $E_n$:
\begin{equation}
	\text{HB of }D_{T^2}\FT=\text{HB of }\FT=\text{HB of }\mathcal{T}^{5\mathrm{d}}_{\mathbf{X}'}=\overline{\mathcal{O}}_{\min}(E_n).
\end{equation}
	Using the inversion between 3d Coulomb and Higgs branches~\cite{Grimminger:2020dmg}, we conclude that
\begin{equation}
	\text{HB of }D_{S^1}\FTfour=\text{Inv(CB of }D_{S^1}\FTfour)=\mathbb{C}^2/\Gamma_{E_n}.
\end{equation}
	Hence, the Higgs branch of $\FTfour$ is identified as the Kleinian singularity $\mathbb{C}^2/\Gamma_{E_n}$. We summarize this chain of relations in Figure~\ref{f:En-Inversion}.

From the above argument, we are led to a proposal that after the $S^1$ reduction of $\FTfour$ for $\mathbf{X}=\mathrm{sing}(E_n)$, it flows under the RG to the following three-dimensional $\mc{N}=4$ electric quivers with $E_6$, $E_7$ and $E_8$ shapes, respectively:
\begin{equation}
	\begin{tikzpicture}[x=.5cm,y=.5cm]
		\node at (-3.0, 0) {$\EQfour (E_6) = $}; 
		\draw[ligne, black](0,0)--(2,0);
		\draw[ligne, black](2,0)--(4,0);
		\draw[ligne, black](4,0)--(8,0);
		\node[bd] at (0,0) [label=below:{{\scriptsize$1$}}] {};
		\node[bd] at (2,0) [label=below:{{\scriptsize$2$}}] {};
		\node[bd] at (4,0) [label=below:{{\scriptsize$3$}}] {};
		\node[bd] at (6,0) [label=below:{{\scriptsize$2$}}] {};
		\node[bd] at (8,0) [label=below:{{\scriptsize$1$}}] {};
		\node[bd] at (4,2) [label=left:{{\scriptsize$2$}}] {};
		\node[bd] at (4,4) [label=left:{{\scriptsize$1$}}] {};
		\draw[ligne, black](4,0)--(4,2);
		\draw[ligne, black](4,2)--(4,4);
	\end{tikzpicture}
\end{equation}
\begin{equation}
	\begin{tikzpicture}[x=.5cm,y=.5cm]
		\node at (-3.0, 0) {$\EQfour (E_7) = $}; 
		\draw[ligne, black](0,0)--(2,0);
		\draw[ligne, black](2,0)--(4,0);
		\draw[ligne, black](4,0)--(8,0);
		\draw[ligne, black](8,0)--(10,0);
		\draw[ligne, black](10,0)--(12,0);
		\draw[ligne, black](6,0)--(6,2);
		\node[bd] at (0,0) [label=below:{{\scriptsize$1$}}] {};
		\node[bd] at (2,0) [label=below:{{\scriptsize$2$}}] {};
		\node[bd] at (4,0) [label=below:{{\scriptsize$3$}}] {};
		\node[bd] at (6,0) [label=below:{{\scriptsize$4$}}] {};
		\node[bd] at (8,0) [label=below:{{\scriptsize$3$}}] {};
		\node[bd] at (10,0) [label=below:{{\scriptsize$2$}}] {};
		\node[bd] at (12,0) [label=below:{{\scriptsize$1$}}] {};
		\node[bd] at (6,2) [label=left:{{\scriptsize$2$}}] {};
	\end{tikzpicture}
\end{equation}
\begin{equation}
	\begin{tikzpicture}[x=.5cm,y=.5cm]
		\node at (-3.0, 0) {$\EQfour (E_8)= $}; 
		\draw[ligne, black](0,0)--(2,0);
		\draw[ligne, black](2,0)--(4,0);
		\draw[ligne, black](4,0)--(8,0);
		\draw[ligne, black](8,0)--(10,0);
		\draw[ligne, black](10,0)--(12,0);
		\draw[ligne, black](12,0)--(14,0);
		\node[bd] at (0,0) [label=below:{{\scriptsize$2$}}] {};
		\node[bd] at (2,0) [label=below:{{\scriptsize$4$}}] {};
		\node[bd] at (4,0) [label=below:{{\scriptsize$6$}}] {};
		\node[bd] at (6,0) [label=below:{{\scriptsize$5$}}] {};
		\node[bd] at (8,0) [label=below:{{\scriptsize$4$}}] {};
		\node[bd] at (10,0) [label=below:{{\scriptsize$3$}}] {};
		\node[bd] at (12,0) [label=below:{{\scriptsize$2$}}] {};
		\node[bd] at (14,0) [label=below:{{\scriptsize$1$}}] {};
		\node[bd] at (4,2) [label=left:{{\scriptsize$3$}}] {};
		\draw[ligne, black](4,0)--(4,2);
	\end{tikzpicture}
\end{equation}
The Coulomb and Higgs branches of these quivers are
\begin{equation}
	\text{CB}=\bar{\mathcal{O}}_{\text{min}}(E_i)\ ,\ \text{HB}=\mathbb{C}^2/\Gamma_{E_i}\,.
\ee

The relation between the Higgs branch of the 5d SCFT  $X^\prime$ and that of 4d SCFT can be extended to general $\mathbf{X}$ with $r>1$, $f = b_3 = 0$, and smooth crepant resolutions, assuming the validity of the inversion procedure on the magnetic quiver of $\FT$ conjectured in Section~\ref{sec:inversion}. Practically, for a given $\mathbf{X}$ with a singular divisor after crepant resolution, we identify a 5d SCFT from M-theory on $\mathbf{X}'$ whose resolution yields the same 5d Coulomb branch data as that of $\mathbf{X}$\footnote{Note that the corresponding $\mathbf{X}'$ may not be a hypersurface singularity in general.}. If the Hasse diagram of $\FT$ can be computed, we can derive the Higgs branch of $\FT$ and, via the inversion procedure (if valid), deduce the Higgs branch for the 4d SCFT $\FTfour$. The two Higgs branches are related by symplectic duality.

For these singularities, we can identify their corresponding VOAs with known affine W-algebras. The spectrum of the Coulomb branch operators for the theory from each singularity is
\begin{itemize}
	\item sing$(E_8)\ :	x_{3}^{7}+x_{4}^{5}x_{3}+x_{2}^{3}+x_{1}^{2}:$
	\begin{equation}
    \label{eq:CBspectrumE6}
		\begin{aligned}
			\Biggl\{&\frac{32}{31},\frac{38}{31},\frac{42}{31},\frac{44}{31},\frac{48}{31},\frac{50}{31},\frac{54}{31},\frac{60}{31},\frac{66}{31},\frac{68}{31},\frac{72}{31},\frac{74}{31},\frac{78}{31},\frac{80}{31},\frac{84}{31},\frac{90}{31},\frac{102}{31},\frac{104}{31},\frac{108}{31},\frac{110}{31},\\
			& \frac{114}{31},\frac{120}{31},\frac{138}{31},\frac{140}{31},\frac{144}{31},\frac{150}{31},\frac{174}{31},\frac{180}{31},\frac{210}{31}\Biggr\}
		\end{aligned}
	\end{equation}
	\item sing$(E_7)\ :x_{2}^{5}+x_{3}^{3}x_{2}+x_{3}x_{4}^{3}+x_{1}^{2}:$
	\begin{equation}
    \label{eq:CBspectrumE7}
		\left\{\frac{20}{19},\frac{24}{19},\frac{26}{19},\frac{28}{19},\frac{30}{19},\frac{32}{19},\frac{36}{19},\frac{42}{19},\frac{44}{19},\frac{46}{19},\frac{48}{19},\frac{50}{19},\frac{54}{19},\frac{66}{19},\frac{68}{19},\frac{72}{19},\frac{90}{19}\right\}
	\end{equation}
	\item sing$(E_6)\ :x_2^4+x_3^2x_2+x_1^3+x_3x_4^2:$
	\begin{equation}
    \label{eq:CBspectrumE8}
		\left\{\frac{14}{13},\frac{17}{13},\frac{18}{13},\frac{20}{13},\frac{21}{13},\frac{24}{13},\frac{30}{13},\frac{32}{13},\frac{33}{13},\frac{36}{13},\frac{48}{13}\right\}
	\end{equation}
\end{itemize}
	Notice that these Coulomb branch spectra have the universal denominator $h^\vee(E_i)+1$ for $i=6,7,8$. We therefore propose that these 4d theories are also Argyres--Douglas theories of the type
\begin{equation}\label{singular1data}
	(E_i,1,h^\vee,\mathfrak{f}),~~i=6,7,8.
\end{equation}
Now we need to find the correct $\mathfrak{f}$ which is the Nahm label of the regular puncture in the class-S construction of AD theories.

For this class of Argyres--Douglas theories, the rank is given by \cite{Li:2022njl,Shan:2023xtw}
\begin{equation}
    \begin{aligned}\frac{1}{2}\left[(h_j\frac{k}{b}+h_j-1)\mathrm{rank} (\mathfrak{j})-\dim\mathcal{O}_{prim}+\dim\mathcal{O}_{\mathfrak{f}}^{Hitchin}\right]\,,\end{aligned}
\end{equation}
where $\mathcal{O}_{\mathfrak{f}}^{Hitchin}$ denotes the Spaltenstein dual of the nilpotent
orbit $\mathfrak{f}.$
 Comparing with the number of CB operators in \ref{eq:CBspectrumE6}, \ref{eq:CBspectrumE7} and \ref{eq:CBspectrumE8}, we find that
\begin{equation}
	\dim\mathcal{O}_\mathfrak{f}^{Hitchin}=2(h^\vee-1)\,,
\end{equation}
which is the dimension of the minimal nilpotent orbit of $E_i$. Because of the duality between the Hitchin label and the Nahm label, we conclude that
\begin{equation}
	\mathfrak{f}=\mathcal{O}_{\text{subregular}}\,.
\end{equation}
One can see that central charges and CB spectrum of $(E_k, 1,h^\vee,\mathcal{O}_{\text{subregular}})$ match the ones of $\FTfour$, hence these two different constructions should be dual to each other in the IR.

The VOA corresponding to $(E_k, 1,h^\vee,\mathcal{O}_{\text{subregular}})$ is the affine $W$-algebra  \cite{Song:2017oew,Xie:2019yds}:
\begin{equation}
    \label{eq:affineWalgSingE}
	\mathcal{W}_{k_{2d}}(E_i,\mathcal{O}_{\text{subregular}}),~~k_{2d}=-h^\vee+\frac{h^\vee}{h^\vee+1}\,.
\end{equation}
The central charge of the W-algebra can be calculated through 
\begin{equation}
    \begin{aligned}c_{4d}=-\frac{c_{2d}}{12}=-\frac{1}{12}\left(\mathrm{dim}\mathfrak{g}^0-\frac{1}{2}\mathrm{dim}\mathfrak{g}^{\frac{1}{2}}-\frac{12}{k_{2d}+h_{\mathfrak{g}}^\vee}\left|\rho-(k_{2d}+h_{\mathfrak{g}}^\vee)\frac{h}{2}\right|^2\right)\,.\end{aligned}
\end{equation}
 Substituting the relevant data, we find agreement with the geometric engineering results in Table~\ref{tab:singularities}
\begin{equation}
	c_{2d}=-\frac{1342}{13},-\frac{3706}{19},-\frac{13978}{31}\,.
\end{equation}

Another property of 4d theories corresponding to these singularities is that their Higgs branches are the Kleinian singularities $\mathbb{C}^2/\Gamma_{E_i}$ with 
\begin{equation}
	\dim_{\mathbb{H}}\text{HB}=1\,.
\end{equation}
The associated variety of the affine $W$-algebra $\mathcal{W}_{-h^\vee+\frac{h^\vee}{h^\vee+1}}(E_i,\mathcal{O}_{\text{subregular}})$ is given by the intersection \cite{Arakawa:2010ni}
\begin{equation}
	X=\mathcal{N}\cap S_{\text{subregular}}=\mathbb{C}^2/\Gamma_{E_i}\,,
\end{equation}
with dimension
\begin{equation}
	\dim_{\mathbb{C}}\text{HB}=\dim_{\mathbb{C}}\mathcal{O}_{\text{prin}}-\dim_{\mathbb{C}}\mathcal{O}_{\text{subregular}}=2\,.
\end{equation}
Here  $\mathcal{N}$ is the nilpotent cone, and $S_f$ is the Slodowy slice.  The match between Higgs branch and associated variety is another evidence that the affine W-algebras \ref{eq:affineWalgSingE} are VOAs corresponding to $\FTfour$ discussed in this section.

\subsection{$r=2$ cases with $f=0$ and a singular divisor}
\label{subsec:r=2}

Similar to the cases before, we now try to find singularities which have a smooth crepant resolution, but $r$ singular 4-cycles. As before we also require that $f=0$ and $b_3=0$. All such cases of IHS with $r=2$, $f=0$ and $b_3=0$ are already listed in \cite{Closset:2021lwy}, which we summarize in Table~\ref{t:rank-2}.

\begin{table}
    \centering
    \begin{tabular}{|c|c|c|c|c|c|c|c|c|c|}
    \hline
        $F(x)$ & $r$ & $f$ & $d_H$ & $\hat r$ & $\hat d_H$ & $\Delta \mathcal{A}_r$ & $b_3$ & $c$ \\
        \hline
         $x_1^2+x_2^5+x_3^{11}+x_3 x_4^3$ & 2 & 0 & 46 & 46 & 2 & 0 & 0 & $\frac{8303}{93}$ \\
         \hline
         $x_1^2+x_2^3+x_2 x_3^5+x_3 x_4^7$ & 2 & 0 & 46 & 46 & 2 & 0 & 0 & $\frac{5267}{57}$ \\
         \hline
         $x_1^5+x_2^3+x_2 x_3^3+x_3 x_4^2$ & 2 & 0 & 22 & 22 & 2 & 0 & 0 & $\frac{1133}{39}$ \\
         \hline
         $x_1^2+x_2 x_4^4+x_2^3 x_3+x_3^5 x_4$ & 2 & 0 & 30 & 30 & 2 & 0 & 0 & $\frac{137}{3}$ \\
         \hline
         $x_1^4+x_2 x_4^3+x_2^2 x_3+x_3^2 x_4$ & 2 & 0 & 18 & 18 & 2 & 0 & 0 & $\frac{61}{3}$ \\
         \hline
    \end{tabular}
    \label{t:rank-2}
    \caption{The list of IHS singularities with a smooth crepant resolution but singular divisors, for the cases of $r=2$, $f=b_3=0$.}
\end{table}

\subsubsection{$x_1^2+x_2^5+x_3^{11}+x_3x_4^3=0$}

The first singularity
\begin{equation}
	\mathbf{X}:\ x_1^2+x_2^5+x_3^{11}+x_3x_4^3=0 
\end{equation}
can be resolved by the sequence
\begin{equation}
(x_1^{(2)},x_2^{(1)},x_3^{(1)},x_4^{(1)};\delta_1),(x_1^{(3)},x_2^{(1)},x_4^{(2)},\delta_1^{(1)};\delta_2)\,.
\end{equation}
The resolved equation is
\begin{equation}
	x_1^2+x_2^5\delta_1+x_3^{11}\delta_1^7\delta_2+x_3x_4^3=0 \,.
\end{equation}
The exceptional divisor $\delta_2=0$ has an $E_8$ du Val singularity at $x_1=x_2=x_4=0$.

$\FTfour$ has the Coulomb branch spectrum
\begin{equation}
\begin{aligned}
	\Big\{ & \tfrac{32}{31},\; \tfrac{38}{31},\; \tfrac{42}{31},\; \tfrac{44}{31},\; \tfrac{48}{31},\; \tfrac{50}{31},\; \tfrac{54}{31}, \tfrac{60}{31},\; \tfrac{64}{31},\; \tfrac{68}{31},\; \tfrac{72}{31},\; \tfrac{74}{31},\; \tfrac{78}{31},\; \tfrac{80}{31}, 
	\tfrac{84}{31},\; \tfrac{90}{31},\; \tfrac{98}{31},\; \tfrac{102}{31},\; \tfrac{104}{31},\; \tfrac{108}{31},\; \tfrac{110}{31}, \\[6pt]
	& \tfrac{114}{31},\; \tfrac{120}{31},\; \tfrac{130}{31},\; \tfrac{132}{31},\; \tfrac{134}{31},\; \tfrac{138}{31},\; \tfrac{140}{31}, \tfrac{144}{31},\; \tfrac{150}{31},\; \tfrac{164}{31},\; \tfrac{168}{31},\; \tfrac{170}{31},\; \tfrac{174}{31},\; \tfrac{180}{31}, \tfrac{198}{31},\; \tfrac{200}{31},\; \tfrac{204}{31},  \\[6pt]
	& \tfrac{210}{31},\; \tfrac{230}{31},\; \tfrac{234}{31},\; \tfrac{240}{31}, \tfrac{264}{31},\; \tfrac{270}{31},\; \tfrac{300}{31},\; \tfrac{330}{31} \Big\}\,.
\end{aligned}
\end{equation}
From the CB spectrum and central charges, we conjecture that this theory is also the Argyres--Douglas theory 
\begin{equation}
	(E_8^{30}[1],\mathfrak{f}=E_8(a_2))
\end{equation}
with associated VOA
\begin{equation}
	W_{-30+\frac{30}{31}}(\mathfrak{e}_8,E_8(a_2))\,.
\end{equation}
Here, the type of regular puncture is determined from the rank formula. By
substituting the values of $k,b$, and the rank of the 4d theory into the formula, one obtains the dimension of the corresponding dual orbit
\begin{equation}
	\dim \mathcal{O}_\mathfrak{f}^H=2\widehat{r}\,.
\end{equation}
In this case, the only relevant orbit is the sub-subregular $\mathfrak{f}=E_8(a_2)$. The associated variety of the corresponding VOA is precisely
\begin{equation}\label{Hasser2}
	\mathcal{N}\cap \mathcal{O}_\mathfrak{f}
\end{equation}
with complex dimension $240-236=4$.

Furthermore, one can compute the central charge
\begin{equation}
	c_{2d}=-\frac{33212}{31}
\end{equation}
which agrees with the result $c_{4d}=8303/93$ from geometric engineering.

Using the result of \eqref{Hasser2}, the Higgs branch Hasse diagram of $\FTfour$ exhibits the following layered structure:
\begin{equation}
\label{r=2-1-4d-MQ}
	\begin{tikzpicture}[x=.5cm,y=.5cm]
		\node[bd] at (4,0) [label=left:{{$E_8(a_2)$}}] {};
		\node[bd] at (4,2) [label=left:{{$E_8(a_1)$}}] {};
		\node[bd] at (4,4) [label=left:{{$E_8$}}] {};
		\draw[ligne, black](4,0)--(4,2) node[midway,left] {$E_7$};
		\draw[ligne, black](4,2)--(4,4) node[midway,left] {$E_8$};
	\end{tikzpicture}
\end{equation}
Now we claim that the singularity $\mathbf{X}'$ that share the same resolution sequence with $\mathbf{X}$ is 
\begin{equation}
	x_1^2+x_2^5+x_3^{10}+x_3x_4^3=0\,,
\end{equation}
which has $r=2$, $f=8$, $d_H=46$, $\widehat{r}=38$. The 5d SCFT $\mathcal{T}^{5\mathrm{d}}_{\mathbf{X}'}$ has the flavor symmetry $G_F=E_8$ and IR gauge theory description $SU(2)_0-SU(2)-5\mbf{F}$~\cite{Closset:2020scj}. The Lagrangian description of the 4d $\mc{N}=2$ SCFT $\FTfour$ can be found in Fig. 2 of \cite{Closset:2020scj}, which is an $SU$ type quiver with the same shape as (\ref{r=2-1-MQ5}). A remarkable property of this singularity is that it admits the same resolution sequence as $\mathbf{X}$
\begin{equation}
(x_1^{(2)},x_2^{(1)},x_3^{(1)},x_4^{(1)};\delta_1),(x_1^{(3)},x_2^{(1)},x_4^{(2)},\delta_1^{(1)};\delta_2)\,.
\end{equation}
Nonetheless, the crepant resolution $\widetilde{\mathbf{X}'}$ and the exceptional divisors on $\widetilde{\mathbf{X}'}$ are all smooth. The resulting 5d SCFT $\mathcal{T}^{5\mathrm{d}}_{\mathbf{X}'}$ has the magnetic quiver
\begin{equation}
\label{r=2-1-MQ5}
	\begin{tikzpicture}[x=.5cm,y=.5cm]
    \node at (-2.0, 0) {$MQ^{(5)} = $}; 
		\draw[ligne, black](0,0)--(2,0);
		\draw[ligne, black](2,0)--(4,0);
		\draw[ligne, black](4,0)--(8,0);
		\draw[ligne, black](8,0)--(10,0);
		\draw[ligne, black](10,0)--(12,0);
			\draw[ligne, black](12,0)--(14,0);
		\draw[ligne, black](8,0)--(8,2);
		\node[bd] at (0,0) [label=below:{{\scriptsize$2$}}] {};
		\node[bd] at (2,0) [label=below:{{\scriptsize$4$}}] {};
		\node[bd] at (4,0) [label=below:{{\scriptsize$6$}}] {};
		\node[bd] at (6,0) [label=below:{{\scriptsize$8$}}] {};
		\node[bd] at (8,0) [label=below:{{\scriptsize$10$}}] {};
		\node[bd] at (10,0) [label=below:{{\scriptsize$7$}}] {};
		\node[bd] at (12,0) [label=below:{{\scriptsize$4$}}] {};
	\node[bd] at (14,0) [label=below:{{\scriptsize$1$}}] {};
		\node[bd] at (8,2) [label=left:{{\scriptsize$5$}}] {};
	\end{tikzpicture}
\end{equation}
The associated Higgs branch Hasse diagram for $\mathcal{T}^{5\mathrm{d}}_{\mathbf{X}'}$ can then be extracted via quiver subtraction
\begin{equation}
\label{r=2-1-5d-MQ}
	\begin{tikzpicture}[x=.5cm,y=.5cm]
		\node[bd] at (4,0) [] {};
		\node[bd] at (4,2) [] {};
		\node[bd] at (4,4) [] {};
		\draw[ligne, black](4,0)--(4,2) node[midway,left] {$\mathfrak{e}_8$};
		\draw[ligne, black](4,2)--(4,4) node[midway,left] {$\mathfrak{e}_7$};
	\end{tikzpicture}
\end{equation}
We therefore observe that the two Hasse diagrams (\ref{r=2-1-4d-MQ}) and (\ref{r=2-1-5d-MQ}) associated with $\FTfour$ and $\mathcal{T}^{5\mathrm{d}}_{\mathbf{X}'}$ are related by the inversion.

\subsubsection{$x_1^5+x_2^3+x_2 x_3^3+x_3 x_4^2=0$}

Following the same procedure, we consider the singularity $\mathbf{X}: x_1^5+x_2^3+x_2 x_3^3+x_3 x_4^2=0$. The crepant resolution of $\mathbf{X}$ is achieved by the sequence
\begin{equation}
	(x_1^{(1)},x_2^{(1)},x_3^{(1)},x_4^{(1)};\delta_1),(x_2^{(1)},x_4^{(1)},\delta_1^{(1)};\delta_2)\,,
\end{equation}
giving
\begin{equation}
	x_1^5\delta_1^2+x_2^3\delta_2+x_2 x_3^3\delta_1+x_3 x_4^2=0\,.
\end{equation}
The exceptional divisor $\delta_2=0$ has a $D_7$ du Val singularity at $\delta_1=x_3=x_4=0$.

The corresponding singularity $\mathbf{X}'$ is
\begin{equation}
	\mathbf{X}':\ x_1^5+x_2^2x_1+x_3^5+x_3x_4^2=0\,,
\end{equation}
with the associated quantities given by
\begin{equation}
	r=2\ ,\ f=8\ ,\ d_H=22\ ,\ \widehat{r}=14\,.
\end{equation}
The resolution sequence is the same as that of $\mathbf{X}$
\begin{equation}
	(x_1^{(1)},x_2^{(1)},x_3^{(1)},x_4^{(1)};\delta_1),(x_2^{(1)},x_4^{(1)},\delta_1^{(1)};\delta_2)\,,
\end{equation}
which matches the blow-up sequence for the singularity $x_1^5+x_2^3+x_2 x_3^3+x_3 x_4^2$. The resulting 5d SCFT $\mathcal{T}^{5\mathrm{d}}_{\mathbf{X}'}$ is the one with $G_F=SO(14)\times U(1)$ and IR gauge theory description $SU(3)_{3/2}+7\mbf{F}$, $Sp(2)+7\mbf{F}$ and $[1]-SU(2)-SU(2)-[4]$, with the magnetic quiver 
\begin{equation}
	\begin{tikzpicture}[x=.5cm,y=.5cm]
    \node at (-2.0, 0) {$MQ^{(5)} = $}; 
		\draw[ligne, black](0,0)--(2,0);
		\draw[ligne, black](2,0)--(4,0);
		\draw[ligne, black](4,0)--(8,0);
		\draw[ligne, black](8,0)--(10,0);
		\draw[ligne, black](10,0)--(12,0);
		\draw[ligne, black](8,0)--(8,2);
		\draw[ligne, black](8,2)--(8,4);
		\node[bd] at (0,0) [label=below:{{\scriptsize$1$}}] {};
		\node[bd] at (2,0) [label=below:{{\scriptsize$2$}}] {};
		\node[bd] at (4,0) [label=below:{{\scriptsize$3$}}] {};
		\node[bd] at (6,0) [label=below:{{\scriptsize$4$}}] {};
		\node[bd] at (8,0) [label=below:{{\scriptsize$5$}}] {};
		\node[bd] at (10,0) [label=below:{{\scriptsize$3$}}] {};
		\node[bd] at (12,0) [label=below:{{\scriptsize$1$}}] {};
		\node[bd] at (8,2) [label=left:{{\scriptsize$3$}}] {};
		\node[bd] at (8,4) [label=left:{{\scriptsize$1$}}] {};
	\end{tikzpicture}
\end{equation}
By applying the quiver subtraction, we obtain the Hasse diagram for HB$(\mathcal{T}^{5\mathrm{d}}_{\mathbf{X}'})$
\begin{equation}
	\begin{tikzpicture}[x=.5cm,y=.5cm]
		\node[bd] at (4,0) [] {};
		\node[bd] at (4,2) [] {};
		\node[bd] at (4,4) [] {};
		\draw[ligne, black](4,0)--(4,2) node[midway,left] {$\mathfrak{d}_7$};
		\draw[ligne, black](4,2)--(4,4) node[midway,left] {$\mathfrak{e}_6$};
	\end{tikzpicture}
\end{equation}
The inversion yields the following Hasse diagram
\begin{equation}\label{hasse1}
	\begin{tikzpicture}[x=.5cm,y=.5cm]
		\node[bd] at (4,0) [] {};
		\node[bd] at (4,2) [] {};
		\node[bd] at (4,4) [] {};
		\draw[ligne, black](4,0)--(4,2) node[midway,left] {$E_6$};
		\draw[ligne, black](4,2)--(4,4) node[midway,left] {$D_7$};
	\end{tikzpicture}
\end{equation}
which can also be obtained via the Higgs branch subtraction procedure \cite{Bennett:2024loi}. We conjecture that this should be the Hasse diagram of the Higgs branch of the $\FTfour$ for the singularity $\mathbf{X}:x_1^5+x_2^3+x_2 x_3^3+x_3 x_4^2$.

It can be  checked that the Hasse diagram \eqref{hasse1} can not be embedded into the Hasse diagram of any nilpotent orbits. Moreover, $E_6$ is not a subalgebra of $D_7$. Hence, the Higgs branch is not of the form
\begin{equation}
	\bar{\mathbb{O}}\cap S_{\mathfrak{f}}.
\end{equation}
Therefore, the VOA corresponding to this theory is neither an affine Kac–Moody algebra nor an affine W-algebra.


\subsubsection{$x_1^4+x_2x_4^3+x_2^2x_3+x_3^2x_4=0$}

For the singularity
\begin{equation}
	\mathbf{X}:\ x_1^4+x_2x_4^3+x_2^2x_3+x_3^2x_4=0\,.
\end{equation}
Its resolution sequence reads 
\begin{equation}
\label{r=2-3-resolution}
	(x_1^{(1)},x_2^{(1)},x_3^{(1)},x_4^{(1)};\delta_1),(x_3^{(1)},\delta_1^{(1)};\delta_2)\,.
\end{equation}
After the resolution, the resolved equation
\begin{equation}
	x_1^4\delta_1+x_2x_4^3\delta_1+x_2^2x_3+x_3^2x_4\delta_2=0
\end{equation}
has an $A_7$ du Val singularity on the exceptional divisor $\delta_2=0$ at $x_1=x_2=\delta_1=0$.

For a singularity $\mathbf{X}'$ with the same resolution sequence and smooth divisors, we did not find an IHS representation, but we do know from the resolution sequence (\ref{r=2-3-resolution}), that the intersection numbers among the compact divisors in $\widetilde{\mathbf{X}'}$ can be summarized in the following diagram:
\be
\includegraphics[height=2cm]{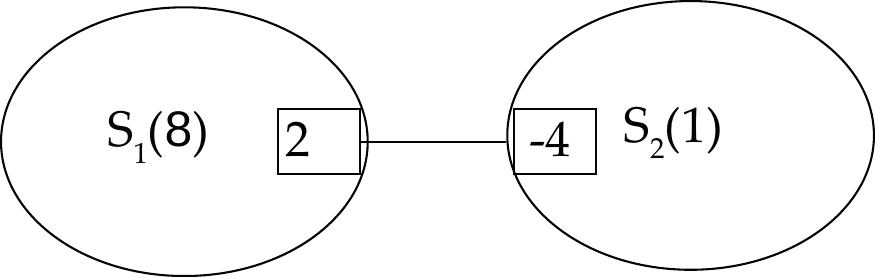}
\ee

The convention of such diagram is the same as in \cite{Closset:2021lwy}, where the computation methods for triple intersection numbers are presented in e.g. ~\cite{Tian:2021cif,Closset:2021lwy}:
\begin{itemize}
	\item Each oval represents an irreducible compact divisor $S_i$. The number in parentheses attached to $S_i$ is the triple self-intersection $S_i^3$.
	\item A line connecting two divisors $S_i$ and $S_j$ carries the numbers $S_i \cdot S_j^2$ (written next to $S_i$) and $S_j \cdot S_i^2$ (written next to $S_j$).
	\item A number placed at the junction of three divisors $S_i$, $S_j$, and $S_k$ denotes the mixed intersection $S_i \cdot S_j \cdot S_k$.
\end{itemize}

From the triple intersection numbers, it was known that~\cite{Jefferson:2018irk,Apruzzi:2019opn} the corresponding 5d rank-two SCFT $\mathcal{T}^{5\mathrm{d}}_{\mathbf{X}'}$ is the one with $G_F=SU(8)\times SU(2)$ and IR gauge theory descriptions $SU(3)_{1/2}+7\mbf{F}$ and $[2]-SU(2)-SU(2)-[3]$, with the UV magnetic quiver~\cite{VanBeest:2020kxw}
\begin{equation}
	\begin{tikzpicture}[x=.5cm,y=.5cm]
    \node at (-2.0, 0) {$MQ^{(5)} = $}; 
		\draw[ligne, black](0,0)--(2,0);
		\draw[ligne, black](2,0)--(4,0);
		\draw[ligne, black](4,0)--(8,0);
		\draw[ligne, black](8,0)--(10,0);
		\draw[ligne, black](10,0)--(12,0);
		\draw[ligne, black](6,0)--(6,2);
		\draw[ligne, black](6,2)--(6,4);
		\node[bd] at (0,0) [label=below:{{\scriptsize$1$}}] {};
		\node[bd] at (2,0) [label=below:{{\scriptsize$2$}}] {};
		\node[bd] at (4,0) [label=below:{{\scriptsize$3$}}] {};
		\node[bd] at (6,0) [label=below:{{\scriptsize$4$}}] {};
		\node[bd] at (8,0) [label=below:{{\scriptsize$3$}}] {};
		\node[bd] at (10,0) [label=below:{{\scriptsize$2$}}] {};
		\node[bd] at (12,0) [label=below:{{\scriptsize$1$}}] {};
		\node[bd] at (6,2) [label=left:{{\scriptsize$2$}}] {};
		\node[bd] at (6,4) [label=left:{{\scriptsize$1$}}] {};
	\end{tikzpicture}
\end{equation}
Performing quiver subtraction yields the Hasse diagram for the Higgs branch of $\mathcal{T}^{5\mathrm{d}}_{\mathbf{X}'}$
\begin{equation}
	\begin{tikzpicture}[x=.5cm,y=.5cm]
		\draw(0,0)--(-2,2) node[midway, left] {$\mathfrak{a}_1$};
		\draw(0,0)--(2,2) node[midway, right] {$\mathfrak{a}_7$};
		\draw(-2,2)--(0,4)  node[midway, left] {$\mathfrak{e}_7$};
		\draw(2,2)--(0,4)  node[midway, right] {$\mathfrak{e}_6$};
		\node[bd] at (0,0) [] {};
		\node[bd] at (-2,2) [] {};
		\node[bd] at (2,2) [] {};
		\node[bd] at (0,4) [] {};
	\end{tikzpicture}
\end{equation}

We conjecture that after an inversion, we get the Higgs branch Hasse diagram for $\FTfour$:
\begin{equation}\label{MQr22}
	\begin{tikzpicture}[x=.5cm,y=.5cm]
		\draw(0,0)--(-2,2) node[midway, left] {$E_7$};
		\draw(0,0)--(2,2) node[midway, right] {$E_6$};
		\draw(-2,2)--(0,4)  node[midway, left] {$A_1$};
		\draw(2,2)--(0,4)  node[midway, right] {$A_7$};
		\node[bd] at (0,0) [] {};
		\node[bd] at (-2,2) [] {};
		\node[bd] at (2,2) [] {};
		\node[bd] at (0,4) [] {};
	\end{tikzpicture}
\end{equation}

Similar to the previous case, the Higgs branch (or the associated variety) cannot be written as the intersection of a nilpotent orbit; consequently, the corresponding VOA is neither an affine Kac–Moody algebra nor a W-algebra.

\subsubsection{$x_{1}^{2}+x_{2}x_{4}^{4}+x_{2}^{3}x_{3}+x_{3}^{5}x_{4}=0$}

The singularity
\begin{equation}
\mathbf{X}:\ x_{1}^{2}+x_{2}x_{4}^{4}+x_{2}^{3}x_{3}+x_{3}^{5}x_{4}=0
\end{equation}
has the resolution sequence 
\begin{equation}
	(x_1^{(2)},x_2^{(1)},x_3^{(1)},x_4^{(1)};\delta_1),(x_1^{(1)},x_2^{(1)},\delta_1^{(1)};\delta_2)\,.
\end{equation}
The resolved equation 
\begin{equation}
x_{1}^{2}+x_{2}x_{4}^{4}\delta_1+x_{2}^{3}x_{3}\delta_2+x_{3}^{5}x_{4}\delta_1^2=0
\end{equation}
has a $D_8$ du Val singularity on the exceptional divisor $\delta_2=0$ at $x_1=x_4=\delta_1=0$.

For a singularity $\mathbf{X}'$ with the same resolution sequence and smooth divisors, triple intersection numbers give rise to the diagram 
\be
\includegraphics[height=2cm]{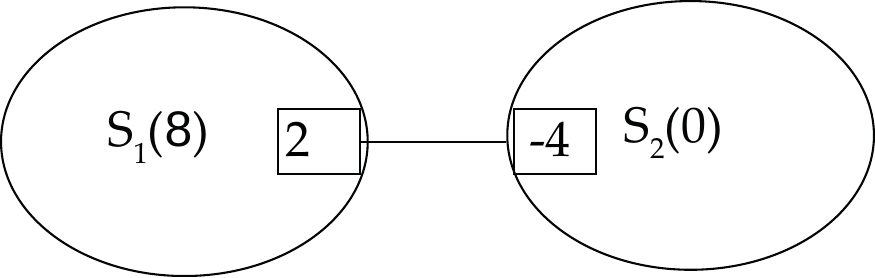}
\ee

Its corresponding 5d SCFT $\mathcal{T}^{5\mathrm{d}}_{\mathbf{X}'}$ is the one with $G_F=SO(16)\times SU(2)$ and IR gauge theory descriptions $SU(3)_{1}+8\mbf{F}$, $Sp(2)+8\mbf{F}$ and $[2]-SU(2)-SU(2)-[4]$, with magnetic quiver
\begin{equation}
	\begin{tikzpicture}[x=.5cm,y=.5cm]
    \node at (-4.0, 0) {$MQ^{(5)} = $}; 
		\draw[ligne, black](0,0)--(-2,0);
		\draw[ligne, black](0,0)--(2,0);
		\draw[ligne, black](2,0)--(4,0);
		\draw[ligne, black](4,0)--(8,0);
		\draw[ligne, black](8,0)--(10,0);
		\draw[ligne, black](10,0)--(12,0);
		\draw[ligne, black](12,0)--(14,0);
			\node[bd] at (-2,0) [label=below:{{\scriptsize$1$}}] {};
		\node[bd] at (0,0) [label=below:{{\scriptsize$2$}}] {};
		\node[bd] at (2,0) [label=below:{{\scriptsize$4$}}] {};
		\node[bd] at (4,0) [label=below:{{\scriptsize$6$}}] {};
		\node[bd] at (6,0) [label=below:{{\scriptsize$5$}}] {};
		\node[bd] at (8,0) [label=below:{{\scriptsize$4$}}] {};
		\node[bd] at (10,0) [label=below:{{\scriptsize$3$}}] {};
		\node[bd] at (12,0) [label=below:{{\scriptsize$2$}}] {};
		\node[bd] at (14,0) [label=below:{{\scriptsize$1$}}] {};
		\node[bd] at (4,2) [label=left:{{\scriptsize$3$}}] {};
		\draw[ligne, black](4,0)--(4,2);
	\end{tikzpicture}
\end{equation}
leading to the Higgs branch Hasse diagram
\begin{equation}
	\begin{tikzpicture}[x=.5cm,y=.5cm]
		\draw(0,0)--(-2,2) node[midway, left] {$\mathfrak{a}_1$};
		\draw(0,0)--(2,2) node[midway, right] {$\mathfrak{d}_8$};
		\draw(-2,2)--(0,4)  node[midway, left] {$\mathfrak{e}_8$};
		\draw(2,2)--(0,4)  node[midway, right] {$\mathfrak{e}_7$};
		\node[bd] at (0,0) [] {};
		\node[bd] at (-2,2) [] {};
		\node[bd] at (2,2) [] {};
		\node[bd] at (0,4) [] {};
	\end{tikzpicture}
\end{equation}
We conjecture that after an inversion, we get the Higgs branch Hasse diagram for $\FTfour$:
\begin{equation}\label{MQr23}
	\begin{tikzpicture}[x=.5cm,y=.5cm]
		\draw(0,0)--(-2,2) node[midway, left] {$E_8$};
		\draw(0,0)--(2,2) node[midway, right] {$E_7$};
		\draw(-2,2)--(0,4)  node[midway, left] {$A_1$};
		\draw(2,2)--(0,4)  node[midway, right] {$D_8$};
		\node[bd] at (0,0) [] {};
		\node[bd] at (-2,2) [] {};
		\node[bd] at (2,2) [] {};
		\node[bd] at (0,4) [] {};
	\end{tikzpicture}
\end{equation}
Obviously, this VOA is also neither an affine Kac–Moody algebra nor a W-algebra.

\subsubsection{$x_{1}^{2}+x_{2}^{3}+x_{2}x_{3}^{5}+x_{3}x_{4}^{7}=0$}

For the singularity
\begin{equation}
	\mathbf{X}:\ x_{1}^{2}+x_{2}^{3}+x_{2}x_{3}^{5}+x_{3}x_{4}^{7}=0
\end{equation}
with resolution sequence
\begin{equation}
	(x_1^{(3)},x_2^{(2)},x_3^{(1)},x_4^{(1)};\delta_1)\,,(x_1^{(1)},x_2^{(1)},\delta_1^{(1)};\delta_2)\,.
\end{equation}
The resolved equation reads
\be
x_{1}^{2}+x_{2}^{3}\delta_2+x_{2}x_{3}^{5}\delta_1+x_{3}x_{4}^{7}\delta_1^2=0\,.
\ee
There is a $D_{10}$ du Val singularity on the exceptional divisor $\delta_2=0$ at $x_1=x_3=\delta_1=0$.

For a singularity $\mathbf{X}'$ with the same resolution sequence and smooth divisors, the triple intersection numbers give rise to the diagram
\be
\includegraphics[height=2cm]{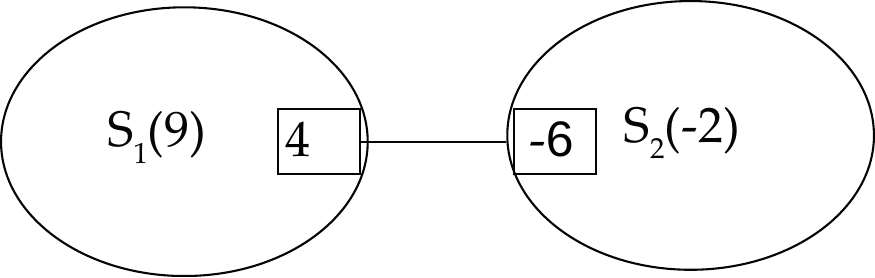}
\ee

The 5d SCFT $\mathcal{T}^{5\mathrm{d}}_{\mathbf{X}'}$ is the one with $G_F=SO(20)$ and IR gauge theory descriptions $SU(3)_{1/2}+9\mbf{F}$, $Sp(2)+9\mbf{F}$ and $[3]-SU(2)-SU(2)-[4]$, with the magnetic quiver 
\begin{equation}
	\begin{tikzpicture}[x=.5cm,y=.5cm]
    \node at (-4.0, 0) {$MQ^{(5)} = $}; 
		\draw[ligne, black](0,0)--(-2,0);
		\draw[ligne, black](0,0)--(2,0);
		\draw[ligne, black](2,0)--(4,0);
		\draw[ligne, black](4,0)--(8,0);
		\draw[ligne, black](8,0)--(10,0);
		\draw[ligne, black](10,0)--(12,0);
		\draw[ligne, black](12,0)--(14,0);
		\draw[ligne, black](14,0)--(16,0);
		\node[bd] at (-2,0) [label=below:{{\scriptsize$2$}}] {};
		\node[bd] at (0,0) [label=below:{{\scriptsize$5$}}] {};
		\node[bd] at (2,0) [label=below:{{\scriptsize$8$}}] {};
		\node[bd] at (4,0) [label=below:{{\scriptsize$7$}}] {};
		\node[bd] at (6,0) [label=below:{{\scriptsize$6$}}] {};
		\node[bd] at (8,0) [label=below:{{\scriptsize$5$}}] {};
		\node[bd] at (10,0) [label=below:{{\scriptsize$4$}}] {};
		\node[bd] at (12,0) [label=below:{{\scriptsize$3$}}] {};
		\node[bd] at (14,0) [label=below:{{\scriptsize$2$}}] {};
			\node[bd] at (16,0) [label=below:{{\scriptsize$1$}}] {};
		\node[bd] at (2,2) [label=left:{{\scriptsize$4$}}] {};
		\draw[ligne, black](2,0)--(2,2);
	\end{tikzpicture}
\end{equation}
leading to the Hasse diagram
\begin{equation}
	\begin{tikzpicture}[x=.5cm,y=.5cm]
		\node[bd] at (4,0) [] {};
		\node[bd] at (4,2) [] {};
		\node[bd] at (4,4) [] {};
		\draw[ligne, black](4,0)--(4,2) node[midway,left] {$\mathfrak{d}_{10}$};
		\draw[ligne, black](4,2)--(4,4) node[midway,left] {$\mathfrak{e}_8$};
	\end{tikzpicture}
	\end{equation}
We conjecture that after an inversion, we get the Higgs branch Hasse diagram for $\FTfour$:
\begin{equation}\label{MQr24}
	\begin{tikzpicture}[x=.5cm,y=.5cm]
		\node[bd] at (4,0) [] {};
		\node[bd] at (4,2) [] {};
		\node[bd] at (4,4) [] {};
		\draw[ligne, black](4,0)--(4,2) node[midway,left] {$E_8$};
		\draw[ligne, black](4,2)--(4,4) node[midway,left] {$D_{10}$};
	\end{tikzpicture}
\end{equation}
This VOA is a new quasi-lisse VOA and does not admit a class S realization.

For all these examples, we see that the type of du Val singularity on the exceptional divisor $\delta_2=0$ always appears on the top layer of the conjectured Hasse diagram of the Higgs branch of $\FTfour$, from the inversion.

\subsection{Examples with smooth resolution and smooth divisors}
\label{subsec:smoothresolution}

Besides the previous examples of $\FTfour$ with small Higgs branch dimensions $\h d_H=1,2$, in this section we discuss some examples of $\mathbf{X}$ with fully smooth crepant resolution and smooth divisors.

\subsubsection{$(D_{2n},D_{2n})$ Argyres-Douglas theory}
\label{sec:D2nD2n}

The first class of examples is the $(D_{2n},D_{2n})$ Argyres-Douglas theory, given by IIB on the following $\mathbf{X}$:
\be
x_1^{2n-1}+x_2^{2n-1}+x_1 x_3^2+x_2x_4^2=0\,.
\ee
The resolution sequence is
\be
(x_1,x_2,x_3,x_4;\delta_1)\ ,\ (x_3,x_4,\delta_1;\delta_2)\ ,\dots,\ (x_3,x_4,\delta_{n-2};\delta_{n-1})\,.
\ee

It was known that the corresponding $\FTfour$ is a Lagrangian quiver gauge theory \cite{Closset:2020afy}
\be
\begin{array}{rcl}
& [1] &\\
& \vert & \\
& SU(n) & \\
& \vert & \\
 {[1]}-SU(2)-\dots-SU(2n-2)- & SU(2n-1) & -SU(n)-{[1]}
\end{array}\,.
\ee

For example when $n=2$, $\mathbf{X}$ is the local $dP_6$ singularity, and 
\be
\begin{array}{rcl}
& [1] &\\
& \vert & \\
& SU(2) & \\
& \vert & \\
\FTfour= {[1]}-SU(2)- & SU(3) & -SU(2)-{[1]}
\end{array}\,.
\ee
From the quiver description, we observe the following Higgsing patterns:
\be
\label{D4D4-Higgsing}
\begin{array}{rcl}
& [1] &\\
& \vert & \\
& SU(2) & \\
& \vert & \\
 {[1]}-SU(2)- & SU(3) & -SU(2)-{[1]}\\
& \downarrow & \\
{[2]}-SU(2) & - & SU(2)-{[2]}\\
& \downarrow & \\
& SU(2) & -{[4]}\\
& \downarrow & \\
& \cdot & 
 \end{array}\,.
\ee

Thus the Hasse diagram reads (we only keep one branch)
\begin{equation}\label{Hasse from Lagrangian}
		\begin{tikzpicture}[x=1cm,y=1cm]
			\node (1) [hasse] at (0,1) {};
\node (2) [hasse] at (0,0) {};
\node (3) [hasse] at (0,-1) {};
\node (4) [hasse] at (0,-2) {};

\draw (1) edge [] node[label=left:$d_4$] {} (2);
\draw (2) edge [] node[label=left:$a_1$] {} (3);
\draw (3) edge [] node[label=left:$A_3$] {} (4);
		\end{tikzpicture} 
	\end{equation}   
The structure can be derived by the Higgs branch dimension of each layer, with the top node corresponds to the fully Higgsed theory and the bottom node corresponds to the un-Higgsed quiver. For example, the top leaf corresponds to the fully Higgsing of $SU(2)-[4]$, which gives the $d_4$ minimal nilpotent orbit~\cite{Bourget:2019aer}. The bottom layer $A_3$ is due to that the gauge rank of the quiver at the top of (\ref{D4D4-Higgsing}) exceeds the gauge rank of the second to the top of (\ref{D4D4-Higgsing}) by three. The middle leaf can be argued similarly.

The magnetic quivers for $(D_m,D_n)$ theories are derived in \cite{Carta:2021dyx}. For the case of $(D_{2n},D_{2n})$, the quiver has a $C_n$ in the middle, connecting to $2n+1$ nodes of $D_1$ via bifundamental hypermultiplet, and then quotient out a diagonal $\mb{Z}_2$.

For our example, taking $n=2$ yields the magnetic quiver
\begin{equation}
	\begin{tikzpicture}[scale=1.2, baseline=-0.5ex]
    \node at (-2.0, 0) {$\MQfour = $}; 
		\draw (0,0) -- (1,0);
		\draw (0,0) -- (0.3,0.9);
		\draw (0,0) -- (0.3,-0.9);
		\draw (0,0) -- (-0.8,0.6);
		\draw (0,0) -- (-0.8,-0.6);
		
		\node[draw,circle,fill=black,inner sep=1.5pt] at (0,0) {};
		\node[draw,circle,fill=black,inner sep=1.5pt] at (1,0) {};
		\node[draw,circle,fill=black,inner sep=1.5pt] at (0.3,0.9) {};
		\node[draw,circle,fill=black,inner sep=1.5pt] at (-0.8,0.6) {};
		\node[draw,circle,fill=black,inner sep=1.5pt] at (0.3,-0.9) {};
		\node[draw,circle,fill=black,inner sep=1.5pt] at (-0.8,-0.6) {};
		
		\node at (-0.1,-0.3) {$~C_2$};
		\node at (0.5,0.9) {$~D_1$};
		\node at (0.5,-0.9) {$~D_1$};
		\node at (-1,0.6) {$D_1~$};
		\node at (-1,-0.6) {$D_1~$};
		\node at (1.2,0) {$~D_1$};
	\end{tikzpicture}
	\ / \ \mathbb{Z}_2
\end{equation}
Applying the Decay and Fission procedure \cite{Bourget:2023dkj,Bourget:2024mgn,Lawrie:2024wan}, we obtain the Hasse diagram for this magnetic quiver, which is identical to (\ref{Hasse from Lagrangian}).

In general, orthosymplectic quivers involve several subtleties. In the present analysis, however, we focus only on the allowed decay for the magnetic quiver. The decay algorithm applies to good quivers. For balanced nodes in a magnetic quiver, we subtract the quiver corresponding to the minimal nilpotent orbit or to the $A_k$, $D_k$ entries listed in Table 3 of \cite{Lawrie:2024wan}. The transverse slice is then determined by rebalancing the transverse quiver. For nodes with $b > 0$, one may further decay the gauge node to a lower rank. For example, in the first step the balanced node is $C_2$, which may decay to $C_1$. Subtracting the corresponding $C_1$ quiver and rebalancing determines that the flavor attached to $C_1$ should be $N_f = 3$, and hence the transverse slice is $A_3$.

For higher $n$, decay and fission still work, although the subtraction procedure becomes more involved. Here we only list the Hasse diagram of $n=3$ which has the magnetic quiver
\begin{equation}
	\begin{tikzpicture}[scale=1.2, baseline=-0.5ex]
    \node at (-2.0, 0) {$\MQfour = $}; 
		\draw (0,0) -- (1,0);
		\draw (0,0) -- (0.3,0.9);
		\draw (0,0) -- (0.3,-0.9);
		\draw (0,0) -- (-0.8,0.6);
		\draw (0,0) -- (-0.8,-0.6);
		\draw (0,0) -- (0.8,-0.4);
		\draw (0,0) -- (0.8,0.4);
		
		\node[draw,circle,fill=black,inner sep=1.5pt] at (0,0) {};
		\node[draw,circle,fill=black,inner sep=1.5pt] at (1,0) {};
		\node[draw,circle,fill=black,inner sep=1.5pt] at (0.3,0.9) {};
		\node[draw,circle,fill=black,inner sep=1.5pt] at (-0.8,0.6) {};
		\node[draw,circle,fill=black,inner sep=1.5pt] at (0.3,-0.9) {};
		\node[draw,circle,fill=black,inner sep=1.5pt] at (-0.8,-0.6) {};
			\node[draw,circle,fill=black,inner sep=1.5pt] at (0.8,-0.4) {};
			\node[draw,circle,fill=black,inner sep=1.5pt] at (0.8,0.4) {};
		
		\node at (-0.1,-0.3) {$~C_3$};
		\node at (0.5,0.9) {$~D_1$};
		\node at (0.5,-0.9) {$~D_1$};
		\node at (-1,0.6) {$D_1~$};
		\node at (-1,-0.6) {$D_1~$};
		\node at (1.2,0) {$~D_1$};
		\node at (1.2,-0.4) {$~D_1$};
		\node at (1.2,0.4) {$~D_1$};
	\end{tikzpicture}
	\ / \ \mathbb{Z}_2
\end{equation}
Applying the decay procedure yields the Hasse diagram
\begin{equation}\label{MQDD2}
	\begin{tikzpicture}[scale=1.3, baseline=-0.5ex,
		every node/.style={font=\small},
		dot/.style={draw,circle,fill=black,inner sep=2pt}
		]
		
		\node[dot] (z) at (0,4) {};
		\node[dot] (t) at (0,3) {};      
		\node[dot] (a) at (0,2) {};
		\node[dot] (b1) at (-1,1) {};
		\node[dot] (b2) at (1,1) {};
		\node[dot] (c) at (0,0) {};
		\node[dot] (g) at (2,0) {};
		\node[dot] (d2) at (1,-1) {};
		\node[dot] (e) at (1,-2) {};
		
		\draw[-] (t) -- (a) node[midway,right] {$A_1$};
		
		\draw[-] (a) -- (b1) node[midway,left] {$D_3$};
		\draw[-] (a) -- (b2) node[midway,right] {$A_1$};
		
		\draw[-] (b1) -- (c) node[midway,left] {$A_3$};
		\draw[-] (b2) -- (g) node[midway,right] {$A_1$};
		\draw[-] (b2) -- (c) node[midway,right] {$D_4$};
		\draw[-] (g) -- (d2) node[midway,right] {$D_5$};
		
		\draw[-] (c) -- (d2) node[midway,right] {$A_3$};
		
			\draw[-] (d2) -- (e) node[midway,right] {$D_3$};
		
		\draw[-] (z) -- (t) node[midway,right] {$d_4$};
		
	\end{tikzpicture}
\end{equation}

\subsubsection{$(D_{2n},D_{2m})$ $(m>n)$ Argyres-Douglas theory}

We generalize the discussions to the case of $(D_{2n},D_{2m})$ $(m>n)$ Argyres-Douglas theory, given by the singularity
\be
x_1^{2n-1}+x_2^{2m-1}+x_1 x_3^2+x_2x_4^2=0\,.
\ee
For the resolution, we first partially resolve it using the same resolution sequence as the $(D_{2n},D_{2n})$ case:
\be
(x_1,x_2,x_3,x_4;\delta_1)\ ,\ (x_3,x_4,\delta_1;\delta_2)\ ,\dots,\ (x_3,x_4,\delta_{n-2};\delta_{n-1})\,.
\ee
The singularity equation now becomes
\be
x_1 x_3^2+x_2x_4^2+x_1^{2n-1}\prod_{j=1}^{n-2}\delta_j^{2n-2j-2}+x_2^{2m-1}\prod_{j=1}^{n-1}\delta_j^{2m-2j-2}=0\,.
\ee
In the above equation, there is a remaining terminal singularity at $\delta_{n-1}=x_1=x_3=x_4=0$, locally in the form of
\be
x_1 x_3^2+x_4^2+x_1^{2n-1}+\delta_{n-1}^2=0\,,
\ee
which is the canonical threefold singularity for $(A_1,D_{2n})$ Argyres-Douglas theory. The crepant resolution and Higgs branch Hasse diagram of this singularity were already discussed in section~\ref{sec:A1DN-even}.

Let us review the basic magnetic quivers associated with the $(D_n,D_m)$ theories \cite{Carta:2021dyx}.
The magnetic quivers for the known cases can be summarized as follows.
\begin{itemize}
	\item For $n$ and $m$ both odd, and such that either
	$\frac{n-1}{\gcd(n-1,m-1)}$ or $\frac{m-1}{\gcd(n-1,m-1)}$ is even,
	the magnetic quiver is given by a $C_{(n-1)/2}$ SQCD with $n$ flavors,
	\begin{equation}
		C_{(n-1)/2}-[D_n],
	\end{equation}
	together with 
	\begin{equation}
		H_{\mathrm{free}}=\frac{1}{2}(mn-n^2+n-1)
	\end{equation}
	free hypermultiplets.
	\item For $(D_{4n-1},D_{4m+4n-3})$ with $m,n\geq 1$, the magnetic quiver is
	\begin{equation}
		C_{2n-1}-[D_{4n-1}]
	\end{equation}
	together with $2(4mn-m-n)$ free hypermultiplets.
	\item For $(D_n,D_m)$ theories admitting two mass parameters, the required condition is that either $n$ is even and $m$ is odd, or $n$ is odd and $m$ is even.
	When $n$ is even and $m$ is odd, the magnetic quiver is
	\begin{equation}
		\begin{bmatrix}D_n\end{bmatrix}-C_{n/2}-D_1
	\end{equation}
	with 
	\begin{equation}
		H_\mathrm{free}=\frac{1}{2}n(m-n-1)
	\end{equation}
	free hypermultiplets. For the case of $n$ odd and $m$ even, the magnetic quiver takes the form
	\begin{equation}
		[C_t]-D_1-C_{(n-1)/2}-[D_{n-1}]
	\end{equation}
where
	\begin{equation}
		t=\frac{1}{2}(m-n+1),
	\end{equation}
and the number of free hypermultiplets is 
	\begin{equation}
		\begin{aligned}H_{\mathrm{free}}=(n-2)t=\frac{1}{2}(n-2)(m-n+1).\end{aligned}
	\end{equation}
	\item For theories with $2M+1$ mass parameters, $M\geq 1$, let
	\begin{equation}
		\begin{aligned}&n=4M\mathfrak{n}-(2M-1),\quad m=n+4M\mathfrak{m}\\&\mathrm{gcd}(2\mathfrak{n}-1,2\mathfrak{m})=1.\end{aligned}
	\end{equation}
where $\mathfrak{n},\mathfrak{m}\geq 1$ or $\mathfrak{n}=1,\mathfrak{m}=0$.
The magnetic quiver contains a balanced central gauge node
$C_{(n-1)/2}=C_{M(2\mathfrak{n}-1)}$,
which is connected to one flavor node $[D_1]$ and to $2M$ gauge nodes $D_1$.
	
	More precisely, the central node is connected to each $D_1$ gauge node by an edge with multiplicity $2\mathfrak{n}-1$.
Each pair of $D_1$ gauge nodes is connected by an edge with multiplicity $\mathfrak{m}(2\mathfrak{n}-1)$.
Each $D_1$ gauge node carries 
	\begin{equation}
		t=\mathfrak{m}(\mathfrak{n}+1)+(2\mathfrak{n}-1)
	\end{equation}
	charge-one hypermultiplets.
In addition, the theory contains
	\begin{equation}
		H_\mathrm{free}=2M(\mathfrak{m}-1)(\mathfrak{n}-1)
	\end{equation}
	free hypermultiplets. Here, the multiplicity denotes the number of half-hypermultiplets connecting the two gauge nodes.
	\item  For theories with $2M+2$ mass parameters, $M\geq 1$, let
	\begin{equation}
		\begin{aligned}&n=(4M-2)\mathfrak{n}-(2M-2),\quad m=n+(4M-2)\mathfrak{m},\\&\mathrm{gcd}(2\mathfrak{n}-1,2\mathfrak{m})=1\end{aligned}
	\end{equation}
	where $\mathfrak{n},\mathfrak{m}\geq 1$ or $\mathfrak{n}=1,\mathfrak{m}=0$.
	The magnetic quiver contains a balanced central gauge node $C_{n/2}$ connected to $2M+1$ gauge nodes of type $D_1$.
	
	Choose two of the $D_1$ nodes, denoted by $B_1$ and $B_2$, and connect them to the central node by edges of multiplicity one.
The remaining nodes, denoted by $A_1,\ldots,A_{2M-1}$, are connected to the central node by edges of multiplicity $2\mathfrak{n}-1$.
Any two $A_j$ nodes are connected by an edge of multiplicity $\mathfrak{m}(2\mathfrak{n}-1)$.
Each $A_i$ carries $\mathfrak{m}(\mathfrak{n}-1)$ charge-two hypermultiplets, while the $B_i$ nodes carry no charged matter.
Furthermore, $B_1$ is connected to each $A_i$ by an edge of multiplicity $\mathfrak{m}$.

After modding out by a $\mathbb{Z}_2$ symmetry, the theory contains
	\begin{equation}
		H_\mathrm{free}=(2M-1)(\mathfrak{m}-1)(\mathfrak{n}-1)
	\end{equation}
	free hypermultiplets.
\end{itemize}

We now apply the decay procedure to determine the Hasse diagram associated with these magnetic quivers.
As a simple example, consider the first case mentioned above, with magnetic quiver
\begin{equation}
	C_{(n-1)/2}-[D_n]\,.
\end{equation}
At each step, we remove a $C_1$ gauge node, which eventually yields the magnetic quiver
\begin{equation}
	C_{1}-[D_n]\,,
\end{equation}
whose Coulomb branch geometry is the type $D_n$ Kleinian singularity~\cite{Cabrera:2017njm}.
Rebalancing the magnetic quiver requires that, at each step,
\begin{equation}
	b=0,2,\cdots, n-4-1.
\end{equation}
These steps correspond to Coulomb branch geometries
\begin{equation}
	D_3,D_5,\cdots, D_{n-2}\,.
\end{equation}
The resulting Hasse diagram is therefore
\begin{equation}
	\begin{tikzpicture}[x=1cm,y=1cm]
		\node (1) [hasse] at (0,1) {};
		\node (2) [hasse] at (0,0) {};
		\node (3) [hasse] at (0,-1) {};
		\node (4) [hasse] at (0,-2) {};
		\node (5) [hasse] at (0,-3) {};
		
		\draw (1) edge [] (2);
		\draw [dashed] (2) edge [] (3);
		
		\draw  (3) edge [] (4);
		
		\draw (4) edge [] (5);

		\node [left] at (0,0.5) {$D_n$};
		\node [left] at (0,-0.5) {};
		\node [left] at (0,-1.5) {$D_{5}$};
		\node [left] at (0,-2.5) {$D_{3}$};
	\end{tikzpicture} 
\end{equation}

We may also consider the magnetic quiver
\begin{equation}
	\begin{bmatrix}D_n\end{bmatrix}-C_{n/2}-D_1
\end{equation}
One may first decay the central gauge node $C_{n/2}$ and then continue the decay, or alternatively decay the $D_1$ gauge node, which reduces to the previous case.
Decaying the central $C_{n/2}$ node yields a Coulomb branch geometry $D_3$.
Further decaying the $D_1$ node gives the Coulomb branch geometry $A_{n-5}$ \cite{Cabrera:2017njm}.
Other transverse slices can be determined in a similar manner.

\subsection{ Higgs branch Hilbert series for electric quivers}
\label{sec:HB-Hilbert}

In this subsection, we examine the generator-relation structure of the Higgs branch using the Higgs branch Hilbert series. In general, computing Hilbert series requires an associated Lagrangian quiver description, and thus the standard techniques are not directly applicable to non-Lagrangian theories.

However, as argued earlier, the $S^1$ reduction of a four-dimensional $\mathcal{N}=2$ SCFT sometimes flows to a Lagrangian quiver theory or to a theory with Lagrangian mirror. This observation enables us to analyze the Higgs branch by studying the Hilbert series of the corresponding three-dimensional quiver.

Suppose the quiver theory has gauge group $G$ and $N_h$ hypermultiplets, each charged under representations $R_i$ of $G$ and $R_i^\prime$ of the flavor group. The generic form of the Higgs branch Hilbert series is
\begin{equation}
	\mathrm{HS}(t)=\int_Gd\mu_G\mathrm{~Pfc}(w,t)\mathrm{PE}\left[\sum_{i=1}^{N_h}\left(\chi_{R_i}(w)\chi_{R_i^{\prime}}(x)+\chi_{\overline{R_i}}(w)\chi_{\overline{R_i^{\prime}}}(x)\right)t^{1/2}\right].
\end{equation}
Here the plethystic exponential encodes the symmetric products of scalars in hypermultiplets, while the prefactor $\mathrm{~Pfc}(w,t)$ captures the relations among operators 
\begin{equation}
	\mathrm{Pfc}(w,t)=\mathrm{PE}\left[\sum_{i=1}^{N_r}\mathrm{char}_{R_i^{\prime\prime}}(w)t^{d_i}\right]^{-1}.
\end{equation}
Typically, the relations appear at quadratic order ($d_i=1$) and transform in the adjoint representation of the gauge group.  

To ensure gauge invariance, we integrate over the Haar measure
\begin{equation}
	\int_Gd\mu_G=\frac{1}{(2\pi i)^r}\oint_{|w|_1=1}\cdots\oint_{|w|_r=1}\frac{dw_1}{w_1}\cdots\frac{dw_r}{w_r}\prod_{\alpha\in\Delta^+}\left(1-\prod_{k=1}^rw_k^{\alpha_k}\right).
\end{equation}
where $r=\mathrm{rank}(G)$ and $\Delta^+$ denotes the set of positive roots.

\paragraph{Example: SQCD.}  
For a $U(n)$ gauge group with $U(m)$ flavor symmetry, the Hilbert series takes the form
\begin{equation}
\int_{U(n)}\mathrm{d}\mu_{U(n)}\frac{\mathrm{PE}\left[\chi_f^{U(n)}(z)\chi_{\bar{f}}^{U(m)}(w)t+\chi_{\bar{f}}^{U(n)}(z)\chi_{f}^{U(m)}(w)t\right]}{\mathrm{PE}\left[\chi_{\text{adj}}^{U(n)}t^2\right]},
\end{equation}
where we have rescaled $t \mapsto t^2$ for convenience.

The Hilbert series also admits an equivalent ``gluing'' description \cite{Hanany:2011db,Benvenuti:2010pq,Hanany:2010qu}. In this approach, both the Coulomb and Higgs branches can be obtained by splitting a general quiver into smaller pieces, evaluating each component, and gluing them together while projecting onto gauge-invariant operators.

As an example, consider the quiver tail
\begin{equation}
		\begin{tikzpicture}[x=.5cm,y=.5cm]
		\node[draw,circle,fill=black,inner sep=1.5pt,label=below:{{\scriptsize$1$}}] at (0,0) {};
		\node[draw,circle,fill=black,inner sep=1.5pt,label=below:{{\scriptsize$2$}}] at (2,0) {};
		\node[draw,circle,fill=black,inner sep=1.5pt,label=below:{{\scriptsize$3$}}] at (4,0) {};
		\node[draw,circle,fill=black,inner sep=1.5pt,label=below:{{\scriptsize$4$}}] at (6,0) {};
		\node[draw,rectangle,inner sep=1.5pt,minimum size=2mm,label=below:{{\scriptsize$5$}}] at (8,0) {};
		\draw[ligne, black] (0,0)--(8,0);
	\end{tikzpicture}
\end{equation}
which can be decomposed into smaller quiver theories
\begin{equation}
		\begin{tikzpicture}[x=.5cm,y=.5cm]
		\node[draw,circle,fill=black,inner sep=1.5pt,label=below:{{\scriptsize$1$}}] at (0,0) {};
		\node[draw, rectangle, inner sep=1.5pt, minimum size=2mm, label=below:{{\scriptsize$2$}}] at (2,0) {};

		\draw[ligne, black] (0,0)--(2,0);
			\draw[ligne, black] (4,0)--(6,0);
		\node[draw, rectangle, inner sep=1.5pt, minimum size=2mm, label=below:{{\scriptsize$2$}}] at (4,0)  {};
		\node[draw, rectangle, inner sep=1.5pt, minimum size=2mm, label=below:{{\scriptsize$3$}}] at (6,0) {};
		
	\draw[ligne, black] (8,0)--(10,0);
	\node[draw, rectangle, inner sep=1.5pt, minimum size=2mm, label=below:{{\scriptsize$3$}}] at (8,0)  {};
	\node[draw, rectangle, inner sep=1.5pt, minimum size=2mm, label=below:{{\scriptsize$4$}}] at (10,0) {};

		\draw[ligne, black] (12,0)--(14,0);
		\node[draw, rectangle, inner sep=1.5pt, minimum size=2mm, label=below:{{\scriptsize$4$}}] at (12,0)  {};
		\node[draw, rectangle, inner sep=1.5pt, minimum size=2mm, label=below:{{\scriptsize$5$}}] at (14,0) {};
	\end{tikzpicture}
\end{equation}
The Hilbert series is then given by
\begin{equation}
	\begin{aligned}\int\mathrm{d}\mu_{U(2)}\int\mathrm{d}\mu_{U(3)}\int\mathrm{d}\mu_{U(4)}\frac{H_{[2]-[3]}H_{[3]-[4]}H_{[4]-[5]}H_{(1)-[2]}}{\mathrm{PE}\left[\chi_{\text{adj}}^{U(2)}t^2\right]\mathrm{PE}\left[\chi_{\text{adj}}^{U(3)}t^2\right]\mathrm{PE}\left[\chi_{\text{adj}}^{U(4)}t^2\right]}\end{aligned}
\end{equation}
Here the contribution from a bifundamental hypermultiplet is
\begin{equation}
	H_{[k]-[l]}=\mathrm{PE}\left[\chi_f^{SU(k)}\chi_{\bar{f}}^{SU(l)}\frac{q_x}{q_y}t+\chi_{\bar{f}}^{SU(k)}\chi_{f}^{SU(l)}\frac{q_y}{q_x}t\right],
\end{equation}
where we have factorized $U(N)\cong SU(N)\times U(1)$, and $q_i$ denotes the fugacity associated with the $U(1)$.  

For example, the Hilbert series of the quiver $(1)-[2]$ is
\begin{equation}
\int\mathrm{d}\mu_{U(1)}(1-t^2)\mathrm{PE}\left[\frac{1}{a} \chi_{f}^{U(2)}t+\chi_{\bar{f}}^{U(2)}a\right],
\end{equation}
which evaluates to
\begin{equation}
	\prod_{i=1}^2(1-t^{2i})\mathrm{PE}\begin{bmatrix}\chi_{\text{adj}}^{U(2)}t^2\end{bmatrix}
\end{equation}
More generally, for a quiver tail of the form
\begin{equation}
	(1)-(2)-\cdots-[n]
\end{equation}
the Higgs branch Hilbert series is given by
\begin{equation}
	\mathrm{PE}\left[\chi_{\text{adj}}^{U(n)}t^2\right]\prod_{i=1}^{n}(1-t^{2i}).
\end{equation}
Hence, starting from the building blocks corresponding to bifundamental hypermultiplets and SQCD, one can systematically construct the Hilbert series of general $U(N)$ quiver gauge theories through the gluing procedure.

Now consider the quiver theory corresponding to singularity $x_1^5+x_2^3+x_2x_3^3+x_3x_4^2=0$. For the quiver
\begin{equation}
	\begin{tikzpicture}[x=.5cm,y=.5cm]
		\draw[ligne, black](0,0)--(2,0);
		\draw[ligne, black](2,0)--(4,0);
		\draw[ligne, black](4,0)--(8,0);
		\draw[ligne, black](8,0)--(10,0);
		\draw[ligne, black](10,0)--(12,0);
		\draw[ligne, black](8,0)--(8,2);
		\draw[ligne, black](8,2)--(8,4);
		\node[bd] at (0,0) [label=below:{{\scriptsize$1$}}] {};
		\node[bd] at (2,0) [label=below:{{\scriptsize$2$}}] {};
		\node[bd] at (4,0) [label=below:{{\scriptsize$3$}}] {};
		\node[bd] at (6,0) [label=below:{{\scriptsize$4$}}] {};
		\node[bd] at (8,0) [label=below:{{\scriptsize$5$}}] {};
		\node[bd] at (10,0) [label=below:{{\scriptsize$3$}}] {};
		\node[bd] at (12,0) [label=below:{{\scriptsize$1$}}] {};
		\node[bd] at (8,2) [label=left:{{\scriptsize$3$}}] {};
		\node[bd] at (8,4) [label=left:{{\scriptsize$1$}}] {};
	\end{tikzpicture}
\end{equation}
the quiver can be decomposed into two components.

The first component is
\begin{equation}
	\begin{tikzpicture}[x=.5cm,y=.5cm]
		\node[draw,circle,fill=black,inner sep=1.5pt,label=below:{{\scriptsize$1$}}] at (0,0) {};
		\node[draw,circle,fill=black,inner sep=1.5pt,label=below:{{\scriptsize$2$}}] at (2,0) {};
		\node[draw,circle,fill=black,inner sep=1.5pt,label=below:{{\scriptsize$3$}}] at (4,0) {};
		\node[draw,circle,fill=black,inner sep=1.5pt,label=below:{{\scriptsize$4$}}] at (6,0) {};
		\node[draw,rectangle,inner sep=1.5pt,minimum size=2mm,label=below:{{\scriptsize$5$}}] at (8,0) {};
		\draw[ligne, black] (0,0)--(8,0);
	\end{tikzpicture}
\end{equation}
whose Hilbert series is
\begin{equation}
	\mathrm{PE}\left[\chi_{\mathrm{adj}}^{U(5)}t^2\right]\prod_{i=1}^5(1-t^{2i}).
\end{equation}

The second component is
\begin{equation}
	\begin{tikzpicture}[x=.5cm,y=.5cm]
		\node[draw,circle,fill=black,inner sep=1.5pt,label=below:{{\scriptsize$1$}}] at (0,0) {};
		\node[draw,circle,fill=black,inner sep=1.5pt,label=below:{{\scriptsize$3$}}] at (2,0) {};
		\node[draw,rectangle,inner sep=1.5pt,minimum size=2mm,label=below:{{\scriptsize$5$}}] at (4,0) {};
		\draw[ligne, black] (0,0)--(4,0);
	\end{tikzpicture}
\end{equation}
whose Hilbert series takes the form
\begin{equation}
	\begin{aligned}
		&H_{(1)-(3)-[5]}
		=\int \mathrm{d}\mu_{U(3)}\,
		H_{\mathrm{glue}}(t,z_1,z_2,z_3)\,
		H_{[3]-[5]}(t,z_i,w_i)\,
		H_{(1)-[3]}(t,z_1,z_2,z_3).
	\end{aligned}
\end{equation}
The gluing factor for $U(5)$ is
\begin{equation}
H_{\text{glue}}(t, z_i) = \frac{1}{\mathrm{PE}\left[\chi_{\mathrm{adj}}^{U(5)}t^2\right]}.
\end{equation}
Consequently, the Higgs branch Hilbert series of the full quiver reads
\begin{equation}
\begin{aligned}
	& 	\int\mathrm{d}\mu_{U(5)}(z_i)\frac{1}{1-t^2}\mathrm{PE}\left[\chi_{\mathrm{adj}}^{U(5)}(z_i)t^2\right]\prod_{i=1}^5(1-t^{2i})\frac{1}{\mathrm{PE}\left[\chi_{\mathrm{adj}}^{U(5)}t^2\right]}H_{[1]-(3)-[5]}^2(z_i)\\
	=& \int\mathrm{d}\mu_{U(5)}\frac{\prod_{i=1}^5(1-t^{2i})}{1-t^2}H_{[1]-(3)-[5]}^2(z_i),\\
    =& 1+t^6+t^8+t^{10}+3t^{12}+\mathcal{O}(t^{13}).
\end{aligned}
\end{equation}
where the prefactor $\frac{1}{1-t^2}$ accounts for the decoupling of the overall $U(1)$ gauge group.

After obtaining the Higgs branch Hilbert series, we can identify the generators and relations of the Higgs branch chiral ring by applying the inverse of the plethystic exponential, known as the plethystic logarithm (PLog):
\begin{equation}
	\mathrm{PLog}[g(t_1,\ldots,t_n)]=\mathrm{PE}^{-1}[g(t_1,\ldots,t_n)]=\sum_{k=1}^\infty\frac{\mu(k)}{k}\log(g(t_1^k,\ldots,t_n^k)),
\end{equation}
where $\mu(k)$ is the M$\ddot{o}$bius function
\begin{equation}
	\mu(k):=\left\{\begin{array}{ll}0&\quad k\text{ has repeated prime factors}\\1&\quad k=1\\(-1)^n&\quad k\text{ is a product of }n\text{ distinct primes}\end{array}\right.
\end{equation}
For this case, the plethystic logarithm expands as
\begin{equation}
	t^6+t^8+t^{10}+2 t^{12}-t^{14}-2 t^{16}-4 t^{18}-3 t^{20}+\mathcal{O}(t^{21}).
\end{equation}
Here, the terms with positive coefficients in the plethystic logarithm correspond to the generators of the Higgs branch, while those with negative coefficients correspond to the relations among them. This Hilbert series also indicates that the moduli space is not a complete intersection; that is, its dimension does not equal the number of generators minus the number of relations.

Following the same procedure, we can compute the Higgs branch Hilbert series for various cases. For instance, the quiver corresponds to $\mathbf{X}:$ $x_1^4+x_2x_4^3+x_2^2x_3+x_3^2x_4=0$
\begin{equation}
	\begin{tikzpicture}[x=.5cm,y=.5cm]
		\draw[ligne, black](0,0)--(2,0);
		\draw[ligne, black](2,0)--(4,0);
		\draw[ligne, black](4,0)--(8,0);
		\draw[ligne, black](8,0)--(10,0);
		\draw[ligne, black](10,0)--(12,0);
		\draw[ligne, black](6,0)--(6,2);
		\draw[ligne, black](6,2)--(6,4);
		\node[bd] at (0,0) [label=below:{{\scriptsize$1$}}] {};
		\node[bd] at (2,0) [label=below:{{\scriptsize$2$}}] {};
		\node[bd] at (4,0) [label=below:{{\scriptsize$3$}}] {};
		\node[bd] at (6,0) [label=below:{{\scriptsize$4$}}] {};
		\node[bd] at (8,0) [label=below:{{\scriptsize$3$}}] {};
		\node[bd] at (10,0) [label=below:{{\scriptsize$2$}}] {};
		\node[bd] at (12,0) [label=below:{{\scriptsize$1$}}] {};
		\node[bd] at (6,2) [label=left:{{\scriptsize$2$}}] {};
		\node[bd] at (6,4) [label=left:{{\scriptsize$1$}}] {};
	\end{tikzpicture}
\end{equation}
can be decomposed into
\begin{equation}
	\begin{tikzpicture}[x=.5cm,y=.5cm]
		\node[draw,circle,fill=black,inner sep=1.5pt,label=below:{{\scriptsize$1$}}] at (0,0) {};
		\node[draw,circle,fill=black,inner sep=1.5pt,label=below:{{\scriptsize$2$}}] at (2,0) {};
		\node[draw,circle,fill=black,inner sep=1.5pt,label=below:{{\scriptsize$3$}}] at (4,0) {};
		\node[draw,rectangle,inner sep=1.5pt,minimum size=2mm,label=below:{{\scriptsize$4$}}] at (6,0) {};
		\draw[ligne, black] (0,0)--(6,0);
	\end{tikzpicture}
\end{equation}
and 
\begin{equation}
	\begin{tikzpicture}[x=.5cm,y=.5cm]
		\node[draw,circle,fill=black,inner sep=1.5pt,label=below:{{\scriptsize$1$}}] at (0,0) {};
		\node[draw,circle,fill=black,inner sep=1.5pt,label=below:{{\scriptsize$2$}}] at (2,0) {};
		\node[draw,rectangle,inner sep=1.5pt,minimum size=2mm,label=below:{{\scriptsize$4$}}] at (4,0) {};
		\draw[ligne, black] (0,0)--(4,0);
	\end{tikzpicture}
\end{equation}

The Hilbert series of these components are given by
\begin{equation}
	\begin{aligned}
		&H_{(1)-(2)-[4]}=\int\mathrm{d}\mu_{U(2)}H_{\text{glue}}(z_i)H_{[2]-[4]}(t,z_i,w_i)H_{(1)-[2]}(t,z_1,z_2,z_3)\\
		& H_{(1)-(2)-(3)-[4]}=\mathrm{PE}\begin{bmatrix}\chi_{\mathrm{adj}}^{U(4)}t^2\end{bmatrix}\prod_{i=1}^4(1-t^{2i}).
	\end{aligned}
\end{equation}
Combining the contributions from both components, the Higgs branch Hilbert series expands as
\begin{equation}
	1+t^6+2t^8+t^{10}+3t^{12}+\mathcal{O}(t^{13})
\end{equation}
with the plethystic logarithm
\begin{equation}
t^6+2t^8+t^{10}+2t^{12}-2t^{14}-4t^{16}-5t^{18}-5t^{20}+\mathcal{O}(t^{21}).
\end{equation}
The Hilbert series for other quivers can be computed following a similar procedure.

\section{VOA character from BPS quiver}
\label{sec:VOA-BPS}
\subsection{Intersection forms and BPS quivers}
\label{subsec:IntersectionBPSquiver}
According to the SCFT/VOA correspondence, the Schur index of a four-dimensional SCFT is identified with the vacuum character of its associated VOA. Operationally, the Schur index can be computed from the BPS quiver on the Coulomb branch \cite{Cordova:2015nma}. Within the framework of geometric engineering, this BPS quiver can be extracted from the deformation data of the Calabi–Yau threefold \cite{Klemm:1996bj,Aspinwall:2004jr,Alim:2011ae,Alim:2011kw}.


As we discuss in section \ref{sec:VOAfromsingularity}, the BPS quiver is interpreted as the intersection of vanishing cycles as 
\begin{equation}
	n_{ij}=\Sigma_i\cdot\Sigma_j=-n_{ji}\,.
\end{equation}
Thus, understanding the BPS quiver requires analyzing the intersection form of the vanishing cycles arising from the singularity deformation. Relevant results are summarized in \cite{arnold1998singularity}. 

For instance, the intersection form associated with the $(A_1,G)$ singularity is summarized in Table~\ref{intersectionAG}.
\begin{table}[h!]
	\centering
\renewcommand{\arraystretch}{1}
	\begin{tabular}{c c}
		\hline
		$(A_1,G)$ & Intersection form $n_{ij}$ \\
		\hline
		$(A_1,A_N)$ &
		$\begin{pmatrix}
			0 & 1 & 0 & \dots & 0 & 0\\
			-1 & 0 & 1 & \dots & 0 & 0\\
			0 & -1 & 0 & \dots & 0 & 0\\
			\vdots & \vdots & \vdots & \ddots & \vdots & \vdots\\
			0 & 0 & 0 & \dots & 0 & 1\\
			0 & 0 & 0 & \dots & -1 & 0
		\end{pmatrix}$ \\[1em]
		
		$(A_1,D_N)$ &
		$\begin{pmatrix}
			0 & 1 & 0 & \dots & 0 & 0 & 0\\
			-1 & 0 & 1 & \dots & 0 & 0 & 0\\
			0 & -1 & 0 & \dots & 0 & 0 & 0\\
			\vdots & \vdots & \vdots & \ddots & \vdots & \vdots & \vdots\\
			0 & 0 & 0 & \dots & 0 & 1 & 1\\
			0 & 0 & 0 & \dots & -1 & 0 & 0\\
			0 & 0 & 0 & \dots & -1 & 0 & 0
		\end{pmatrix}$ \\[1em]
		
		$(A_1,E_6)$ &
		$\begin{pmatrix}
			0 & 1 & 0 & 0 & 0 & 0\\
			-1 & 0 & 1 & 0 & 0 & 0\\
			0 & -1 & 0 & 1 & 0 & 1\\
			0 & 0 & -1 & 0 & 1 & 0\\
			0 & 0 & 0 & -1 & 0 & 0\\
			0 & 0 & -1 & 0 & 0 & 0
		\end{pmatrix}$ \\[1em]
		
		$(A_1,E_7)$ &
		$\begin{pmatrix}
			0 & 1 & 0 & 0 & 0 & 0 & 0\\
			-1 & 0 & 1 & 0 & 0 & 0 & 0\\
			0 & -1 & 0 & 1 & 0 & 0 & 0\\
			0 & 0 & -1 & 0 & 1 & 0 & 1\\
			0 & 0 & 0 & -1 & 0 & 1 & 0\\
			0 & 0 & 0 & 0 & -1 & 0 & 0\\
			0 & 0 & 0 & -1 & 0 & 0 & 0
		\end{pmatrix}$ \\[1em]
		
		$(A_1,E_8)$ &
		$\begin{pmatrix}
			0 & 1 & 0 & 0 & 0 & 0 & 0 & 0\\
			-1 & 0 & 1 & 0 & 0 & 0 & 0 & 0\\
			0 & -1 & 0 & 1 & 0 & 0 & 0 & 0\\
			0 & 0 & -1 & 0 & 1 & 0 & 0 & 0\\
			0 & 0 & 0 & -1 & 0 & 1 & 0 & 1\\
			0 & 0 & 0 & 0 & -1 & 0 & 1 & 0\\
			0 & 0 & 0 & 0 & 0 & -1 & 0 & 0\\
			0 & 0 & 0 & 0 & -1 & 0 & 0 & 0
		\end{pmatrix}$ \\
		\hline
	\end{tabular}
	\caption{Intersection forms for $(A_1,G)$ theories}
	\label{intersectionAG}
\end{table}

In general, the intersection form can be reconstructed from the simple singularity. Now consider two isolated singularities 
$F: (\mathbb{C}^n, 0) \to (\mathbb{C}, 0)$ with $F = F(z)$ and 
$F': (\mathbb{C}^m, 0) \to (\mathbb{C}, 0)$ with $F' = F'(w)$. Their direct sum $F \oplus F'$ is defined by
\begin{equation}
	F(z)+F'(w)\ {:}(\mathbb{C}^{n+m},0)\to(\mathbb{C},0)\,.
\end{equation}
The intersection matrix of $F \oplus F'$ can be expressed in terms of those of $F$ and $F'$ as \cite{gabrielov1973intersection}
\begin{equation}
	\begin{aligned}(\Delta_{ij_1}\circ\Delta_{ij_2})&=(\Sigma_{j_1}^{\prime}\cdot\Sigma_{j_2}^{\prime})\,,\\(\Delta_{i_1j}\circ\Delta_{i_2j})&=(\Sigma_{i_1}\cdot\Sigma_{i_2})\,,\\(\Delta_{i_1j_1}\circ\Delta_{i_2j_2})&=-(\Sigma_{i_1}\cdot\Sigma_{i_2})(\Sigma_{j_1}^{\prime}\cdot\Sigma_{j_2}^{\prime})\quad~\text{for}~(i_2-i_1)(j_2-j_1)>0&\mathrm{,}\end{aligned}
\end{equation}
and 
\begin{equation}
	(\Delta_{i_1j_1}\circ\Delta_{i_2j_2})=0\quad~\text{for}~(i_2-i_1)(j_2-j_1)<0\,,
\end{equation}
where $\Sigma$ and $\Sigma^\prime$ denote the cycles of $F$ and $F'$ respectively. Here $(\Delta_{ij}\circ\Delta_{kl})$ denotes the intersection pairing of $F\oplus F'$, where the subscripts label the rows and columns of the intersection matrix corresponding to the cycles. For example, $(\Delta_{ij}\circ\Delta_{ik})$ represents the intersection pairing of two cycles labeled by $j$ and $k$ within the same column.

For example, consider $F \oplus F' = x^2 + y^2 + z^4 + w^3$.  $x^2 + y^2$ does not contribute to the intersection form. The nontrivial contributions come from $z^4$ and $w^3$, corresponding to the $A_3$ and $A_2$ Dynkin diagrams \cite{arnold1998singularity}, respectively. The resulting intersection form is depicted as 
\begin{figure}[htbp]
	\centering  
\begin{tikzpicture}[scale=1.2, every node/.style={circle, fill=black, inner sep=1.5pt}]
	
	\node (n1) at (0,0) {};
	\node (n2) at (2,0) {};
	\node (n3) at (4,0) {};
	
	\node (n4) at (0,2) {};
	\node (n5) at (2,2) {};
	\node (n6) at (4,2) {};
	
	\draw (n1) -- (n2) -- (n3);
	\draw (n4) -- (n5) -- (n6);
	\draw (n1) -- (n4);
	\draw (n2) -- (n5);
	\draw (n3) -- (n6);
	
	\draw[dashed] (n1) -- (n5);
	\draw[dashed] (n2) -- (n6);
	
\end{tikzpicture}
\end{figure}
which lies in the mutation class of the $E_6$-type intersection form. This direct sum approach enables the determination of the BPS quiver for the $(G,G')$ Argyres-Douglas theories proposed in \cite{Cecotti:2010fi}.

Besides the direct sum method, there is also the polar curve method \cite{gabrielov1979polar}, which proves useful in determining the intersection forms of Arnold's 14 exceptional unimodal singularities \cite{DelZotto:2011an,Cecotti:2011gu}. This method allows us to extract the intersection form for an isolated singularity $F$ depending on a linear function $z$. The final intersection form of $F$ can be expressed in terms of the restriction $F|_{z=0}$ and the polar curve of $F$ relative to $z$, i.e., the critical points of the family $F - \lambda z$. Applications to the BPS quiver of $D_p(G)$ theories can be found in \cite{Hosseini:2021ged}. 

In our paper, in addition to the singularities corresponding to $(G,G^\prime)$ theories, 
we are mainly interested in the following class of weighted homogeneous polynomials:
\begin{equation}\label{OR singularity}
	\begin{aligned}F(z_0,\ldots,z_n)=z_0^{a_0}+z_0z_1^{a_1}+\ldots+z_{n-1}z_n^{a_n},\quad n\geq1.\end{aligned}
\end{equation}
Define $r_k=a_0a_1\cdots a_k$ and $r_{-1}=1$. The integers $c_0,\ldots,c_\mu$ are defined via the coefficients in the expansion
\begin{equation}
	\begin{aligned}\prod_{i=-1}^n(t^{r_i}-1)^{(-1)^{n-i}}&=c_\mu t^\mu+\cdots+c_1t+c_0.\end{aligned}
\end{equation}
P.~Orlik and R.~Randell~\cite{orlik1977monodromy} conjectured that there exists a distinguished basis of vanishing cycles of $f$. This conjecture was
subsequently proved in~\cite{varolgunes2023seifert}. In the conventions
used here, the corresponding intersection matrix takes the form
\begin{equation}
	S_{ij}=-L^{-1}-(-1)^n (L^{-1})^t
\end{equation}
where the Seifert matrix $L$ of $f$ is given by 
\begin{equation}
	L=-(-1)^{n(n+1)/2}\begin{pmatrix}c_0&0&\cdots&\cdots&0&0&0\\c_1&c_0&0&\cdots&\cdots&0&0\\c_2&c_1&c_0&0&\cdots&\cdots&0\\\vdots&\vdots&\vdots&\ddots&\ddots&\vdots&\vdots\\c_{\mu-3}&c_{\mu-4}&c_{\mu-5}&\cdots&c_0&0&0\\c_{\mu-2}&c_{\mu-3}&c_{\mu-4}&\cdots&c_1&c_0&0\\c_{\mu-1}&c_{\mu-2}&c_{\mu-3}&\cdots&c_2&c_1&c_0\end{pmatrix}.
\end{equation}

In addition to these mathematical techniques, other approaches exist to determine the BPS quivers of certain class S theories (although many of these theories lack known geometric engineering realizations) \cite{Alim:2011ae,Alim:2011kw,Xie:2012gd,Xie:2012jd,Cecotti:2013lda}.

\subsection{Schur index from the BPS quiver}

\label{subsec:Schurindex}
The massive particles on the Coulomb branch of a 4d
$\mathcal{N}=2$ theory carry charges valued in a lattice $\Gamma$,
equipped with a Dirac pairing $\langle \cdot , \cdot \rangle$.
The mass of a single-particle state with charge $\gamma \in \Gamma$
satisfies the BPS bound
\begin{equation}
	M\geq|\mathcal{Z}(\gamma)|
\end{equation}
where $\mathcal{Z}(\gamma)$ denotes the central charge.
States that saturate this bound are referred to as BPS particles.
The spectrum of BPS particles may exhibit discontinuities, described by the wall crossing formula.

For each $\gamma \in \Gamma$, we introduce a variable $X_\gamma$ obeying
the quantum torus algebra
\begin{equation}
	X_\gamma X_{\gamma^{\prime}}=q^{\frac{\langle\gamma,\gamma^{\prime}\rangle}{2}}X_{\gamma+\gamma^{\prime}}=q^{\langle\gamma,\gamma^{\prime}\rangle}X_{\gamma^{\prime}}X_\gamma\,.
\end{equation}
The $q$-exponential is defined as
\begin{equation}
	E_q(z) = \prod_{i=0}^\infty \left(1 + q^{i+\frac{1}{2}} z\right)^{-1} 
	= \sum_{n=0}^\infty \frac{\left(-q^{\frac{1}{2}} z\right)^n}{(q)_n}\,.
\end{equation}
where the $q$-Pochhammer symbol is  
\begin{equation}
	(q)_n\equiv\begin{cases}1&\quad\mathrm{if}~n=0\,,\\ \prod_{k=1}^n(1-q^k)&\quad\mathrm{if}~n\geq1\,.\end{cases}
\end{equation}
The $q$-exponential satisfies a number of useful identities. For instance,
when $\langle \gamma_1, \gamma_2 \rangle = 1$, one has
\begin{equation}
	E_q(X_{\gamma_1})E_q(X_{\gamma_2})=E_q(X_{\gamma_2})E_q(X_{\gamma_1+\gamma_2})E_q(X_{\gamma_1})\,.
\end{equation}

The Kontsevich--Soibelman (KS) operator $\mathcal{O}(q)$ is defined as an
ordered product of $q$-exponentials, arranged according to the phase of
the central charge $\mathcal{Z}(\gamma)$, with smaller phases placed to
the left.
In practice, the KS operator can be determined via quiver mutation of the
BPS quiver~\cite{Cecotti:2010fi,Xie:2012gd}, which encodes the BPS spectrum together with the
ordering by phase.
For example, if the BPS chamber consists of the charges
\begin{equation}
	\gamma_1,\gamma_2,\gamma_3\,,
\end{equation}
the corresponding KS operator is
\begin{equation}
	\mathcal{O}(q)=E_q(X_{-\gamma_1})E_q(X_{-\gamma_2})E_q(X_{-\gamma_3})E_q(X_{\gamma_1})E_q(X_{\gamma_2})E_q(X_{\gamma_3})\,.
\end{equation}
The Schur index is conjectured to be given by \cite{Cordova:2015nma}
\begin{equation}
\label{Schur-conj}
	\mathcal{I}(q,z_1,\cdots,z_n)=(q)_\infty^{2\widehat{r}}\mathrm{Tr}\left[\mathcal{O}(q)\right]\left(\mathrm{Tr}[X_{\gamma_{f_1}}],\cdots,\mathrm{Tr}[X_{\gamma_{f_i}}]\right)\,,
\end{equation}
where $\widehat{r}$ denotes the rank of the Coulomb branch.
The flavor fugacities are associated with an integral basis
$\{ \gamma_{f_i} \}$ of the flavor lattice.
The trace is defined by
\begin{equation}
	\mathrm{Tr}[X_\gamma]=\begin{cases}\prod_i\mathrm{Tr}[X_{\gamma_{f_i}}]^{f_i(\gamma)}&\langle\gamma,\gamma^{\prime}\rangle=0~\forall\gamma^{\prime}\in\Gamma,\\0&\mathrm{else},\end{cases}
\end{equation}
where $f_i(\gamma)$ denotes the flavor charge of $\gamma$. The trace can then be expressed as a function of the flavor fugacities $z_i$,
\begin{equation}
	\mathrm{Tr}[X_{\gamma_{f_i}}]=h_i(z_1,\cdots,z_{n_f})\,.
\end{equation}
The explicit form of these functions can be determined by examining the lowest-order
terms in the index and matching them with the vacuum module of the corresponding VOAs.

However, when applying this construction to more complicated BPS quivers,
one typically encounters infinitely many negative powers of $q$,
making it difficult to extract the Schur index explicitly.
As explained in~\cite{Cordova:2016uwk}, the product over the entire phase plane is not
well-defined. A method proposed in~\cite{Cordova:2016uwk} is to introduce quantum spectrum
operators, which separately capture the contributions of BPS particles
and antiparticles,
\begin{equation}
	\mathcal{S}(q)=E_q(X_{\gamma_i})\cdots 
\end{equation}
and $\bar{\mathcal{S}}(q)$ including the anti-particles
\begin{equation}
	\bar{\mathcal{S}}(q)=E_q(X_{-\gamma_i})\cdots\,.
\end{equation}
These operators are well defined on the phase intervals
$[0,\pi)$ and $[\pi,2\pi)$, respectively.
The Schur index can then be written more conveniently as
\begin{equation}
	\mathcal{I}(q)=(q)_\infty^{2\widehat{r}}\mathrm{~Tr~}\left[\bar{\mathcal{S}}(q)\mathcal{S}(q)\right].
\end{equation}
To avoid convergence issues, one introduces an integer cutoff $N$ such that
\begin{equation}
	\begin{aligned}X_{a_i\gamma_i}&\mapsto0\quad&\mathrm{if}\quad\sum_ia_i&>N\,.\end{aligned}
\end{equation}
The value of $N$ is theory-dependent and must be chosen sufficiently large
compared with the highest power of $q$ retained in the Schur index.

As an illustration, we consider the $(A_2,A_2)$ theory, whose BPS quiver is
\begin{equation}\label{box2}
	\begin{tikzpicture}[x=1cm, y=1cm]
		\node [circle, draw] (1) at (0, 0) {};
		\node [circle, draw] (2) at (2, 0) {};
		
		\node [circle, draw] (3) at (0,-2) {};
		\node [circle, draw] (4) at (2,-2) {};
		
		\node[above=0.3cm, draw=none] at (1) {$\gamma_1$};
		\node[above=0.3cm, draw=none] at (2) {$\gamma_2$};
		\node[below=0.3cm, draw=none] at (3) {$\gamma_4$};
		\node[below=0.3cm, draw=none] at (4) {$\gamma_{3}$};
		
		\draw[->] (1) -- (2);
		\draw[->] (2) -- (4);
		\draw[<-] (1) -- (3);
		\draw[<-] (3) -- (4);
	\end{tikzpicture}
\end{equation}
The flavor charges are
\begin{equation}
	\gamma_1+\gamma_3,\gamma_2+\gamma_4\,.
\end{equation}
We choose the BPS chamber~\cite{Cecotti:2010fi,Xie:2012gd}
\begin{equation}
	\gamma_1,\gamma_3,\gamma_2+\gamma_3,\gamma_1+\gamma_4,\gamma_4,\gamma_2\,.
\end{equation}
Upon imposing the cutoff $N=9$, the quantum spectrum operator
$\mathcal{S}(q)$ takes the form
\begin{equation}
	\begin{aligned}
		1
		&+ \sqrt{q}\Bigl(
		- X_{\gamma_1}-X_{\gamma_2}-X_{\gamma_3}-X_{\gamma_4}
		\Bigr)+ q\Bigl(
		X_{2\gamma_1}+X_{2\gamma_2}+X_{2\gamma_3}+X_{2\gamma_4}
		+X_{\gamma_1+\gamma_3}+X_{\gamma_2+\gamma_4}
		+X_{\gamma_1+\gamma_2+\gamma_3+\gamma_4}
		\Bigr) \\[4pt]
		&+ q^{3/2}\Bigl(
		- X_{\gamma_i}-X_{3\gamma_i}
		+X_{\gamma_1+\gamma_2}+X_{\gamma_2+\gamma_3}
		+X_{\gamma_1+\gamma_4}+X_{\gamma_3+\gamma_4} 
		- X_{2\gamma_1+\gamma_3}-X_{\gamma_1+2\gamma_3}
		- X_{2\gamma_2+\gamma_4}-X_{\gamma_2+2\gamma_4} \\
		&\qquad- X_{2\gamma_1+\gamma_2+\gamma_3+\gamma_4}
		- X_{\gamma_1+2\gamma_2+\gamma_3+\gamma_4} 
		- X_{\gamma_1+\gamma_2+2\gamma_3+\gamma_4}
		- X_{\gamma_1+\gamma_2+\gamma_3+2\gamma_4}
		\Bigr) \\[4pt]
		&+ q^{2}\Bigl(
		X_{2\gamma_i}+X_{4\gamma_i}
		+2X_{\gamma_1+\gamma_3}+2X_{\gamma_2+\gamma_4} 
		+X_{3\gamma_1+\gamma_3}+X_{2\gamma_1+2\gamma_3}
		+X_{\gamma_1+3\gamma_3}
		+X_{3\gamma_2+\gamma_4} \\
		&\qquad
		+X_{\gamma_1+\gamma_2+\gamma_3+\gamma_4}
		+X_{3\gamma_1+\gamma_2+\gamma_3+\gamma_4}
		+X_{\gamma_1+3\gamma_2+\gamma_3+\gamma_4} 
		+X_{2\gamma_1+\gamma_2+2\gamma_3+\gamma_4}
		+X_{\gamma_1+\gamma_2+3\gamma_3+\gamma_4}
		+X_{2\gamma_2+2\gamma_4} \\
		&\qquad
		+X_{\gamma_1+2\gamma_2+\gamma_3+2\gamma_4}
		+X_{2\gamma_1+2\gamma_2+2\gamma_3+2\gamma_4} 
		+X_{\gamma_2+3\gamma_4}
		+X_{\gamma_1+\gamma_2+\gamma_3+3\gamma_4}
		\Bigr)+\mathcal{O}(q^{3/2})\,.
	\end{aligned}
\end{equation}
Setting $\mathrm{Tr}X_{\gamma_1+\gamma_3}=z_1$ and $\mathrm{Tr}X_{\gamma_2+\gamma_4}=z_2$,
the Schur index becomes
\begin{equation}
	\begin{aligned}
	&(q)_\infty^{2}\mathrm{Tr}\left[\mathcal{O}(q)\right]=1
		+ \Bigl(
		2+\frac{1}{z_1}+z_1+\frac{1}{z_2}
		+\frac{1}{z_1 z_2}+z_2+z_1 z_2
		\Bigr) q \\[4pt]
		&+ \Bigl(
		6
		+ \frac{1}{z_1^2} + \frac{3}{z_1} + 3 z_1 + z_1^2
		+ \frac{1}{z_2^2} + \frac{3}{z_2} + 3 z_2 + z_2^2 
		+ \frac{1}{z_1^2 z_2^2}
		+ \frac{1}{z_1 z_2^2}
		+ \frac{1}{z_1^2 z_2}
		+ \frac{3}{z_1 z_2}
		+ \frac{z_1}{z_2}
		+ \frac{z_2}{z_1} \\[2pt]
		&\qquad
		+ 3 z_1 z_2
		+ z_1^2 z_2
		+ z_1 z_2^2
		+ z_1^2 z_2^2
		\Bigr) q^2
		+ \mathcal{O}(q^3)\,  \\[2pt]
		&\qquad
		=1+\chi_8^{SU(3)}q+\left(\chi_1^{SU(3)}+\chi_8^{SU(3)}+\chi_{27}^{SU(3)}\right)q^2+ \mathcal{O}(q^3)\,.
	\end{aligned}
\end{equation}
This matches the vacuum module of the affine Kac--Moody algebra
$\widehat{SU(3)}_{-\frac{3}{2}}$, as expected from the equivalence
$(A_2,A_2) = (A_1,D_4)$.

Following this procedure, the Schur indices of the $(A_2,A_{N-1})$ theory with $\gcd(3,N)=1$ and  the $(A_3,A_{N-1})$ theory with $\gcd (4,N)=1$ are presented in \cite{Nishinaka:2025ytu}.

There is a more convenient way to introduce the cutoff $\gamma$, following the approach of \cite{Gaiotto:2024ioj}.  We assume that the minimal chamber of the BPS particles takes the form
\begin{equation}
	\gamma_1^\prime,\gamma_2^\prime,\cdots,\gamma_N^\prime\,,
\end{equation}
where each $\gamma_i^\prime$ is a linear combination of the original charges $\gamma_i$. 
The spectrum operator can then be expanded as
\begin{equation}
	S_\gamma(q)=\sum_{n_i\geq 0}^{\sum_in_i\gamma^\prime_i=\gamma}\frac{(-q)^{\frac{1}{2}\sum_in_i}q^{\frac{1}{2}\sum_{i<j}\langle\gamma^\prime_i,\gamma^\prime_j\rangle n_in_j}}{\prod_i(q)_{n_i}} X_{\sum_in_i\gamma^\prime_i}\,.
\end{equation}
Replacing $\gamma_i^\prime$ by $-\gamma_i^\prime$ gives the expansion of $S_{-\gamma}(q)$, which differs from $S_{\gamma}(q)$ only by the fugacity.
Notice that the trace selects the terms proportional to the flavor fugacity; therefore, the final expression for the Schur index is
\begin{equation}\label{generalindex}
	\mathcal{I}(\mathbf{b},q)=(q)_{\infty}^{2\widehat{r}}\sum_{\gamma,\gamma^{\prime}}^{\gamma-\gamma^{\prime}\in\Gamma_{f}}S_{\gamma}(q)S_{\gamma^{\prime}}(q)X_{\gamma-\gamma^\prime}\,.
\end{equation}
If we focus on the unflavored Schur index, the result further simplifies to
\begin{equation}\label{unflavor}
	\mathcal{I}(q)=(q)_{\infty}^{2\widehat{r}}\sum_{\gamma}S_{\gamma}(q)^2\,.
\end{equation}
Using this formula, the unflavored Schur index can be computed much more efficiently.
\subsection{Schur index of singularities associated with lisse VOAs}
\label{subsec:lisseVOA}
In this subsection, we analyze the BPS quivers and Schur indices associated with Orlik--Randell type singularities~\eqref{OR singularity}, as well as the dependence of the Schur index on the intersection matrix, the minimal chamber, and flavor charge lattice.

Recall that a holomorphic function
\begin{equation}
	\tilde F \colon (\mathbb{C}^{n+k},0) \to (\mathbb{C},0)
\end{equation}
is called a \emph{stabilization} of
\begin{equation}
	F \colon (\mathbb{C}^n,0) \to (\mathbb{C},0)
\end{equation}
if it is of the form
\begin{equation}
	\tilde{F}=F+z_{n+1}^2+\cdots+z_{n+k}^2\,.
\end{equation}
Two singularities related by stabilization have the same intersection form of vanishing cycles, up to the symmetry property of the intersection pairing, which is symmetric or antisymmetric depending on the parity of $k$~\cite{arnold1998singularity}.

For the $D_N^N[k]$ singularity, the defining equation is (see Table~\ref{tab:cDV})
\begin{equation}\label{generalsingularity}
	x_{1}^{2}+x_{2}^{N-1}+x_{2}x_{3}^{2}+z^{k}x_{3}=0
\end{equation}
which is precisely the stabilization of~\eqref{OR singularity} with
$a_0=N-1$, $a_1=2$, and $a_2=k$.
Choosing the intersection pairing to be antisymmetric reproduces the intersection matrix associated with the $D_N^N[k]$ singularity.

As an illustrative example, consider the $D_4^4[2]$ singularity. The intersection matrix associated with the hypersurface
\begin{equation}\label{example}
	x_{2}^{3}+x_{2}x_{3}^{2}+z^{2}x_{3}=0
\end{equation}
is given by
\begin{equation}
\label{symexample}
	\left(
	\begin{array}{rrrrrrrr}
		-2 &  1 &  0 & -1 &  1 &  0 &  0 &  0 \\
		 1 & -2 &  1 &  0 & -1 &  1 &  0 &  0 \\
		 0 &  1 & -2 &  1 &  0 & -1 &  1 &  0 \\
		-1 &  0 &  1 & -2 &  1 &  0 & -1 &  1 \\
		 1 & -1 &  0 &  1 & -2 &  1 &  0 & -1 \\
		 0 &  1 & -1 &  0 &  1 & -2 &  1 &  0 \\
		 0 &  0 &  1 & -1 &  0 &  1 & -2 &  1 \\
		 0 &  0 &  0 &  1 & -1 &  0 &  1 & -2
	\end{array}
	\right).
\end{equation}
Adding a quadratic term $x_1^2$ to~\eqref{example} requires antisymmetrizing the intersection matrix~\eqref{symexample}. After this modification, the resulting intersection matrix becomes
\begin{equation}
    \left(
	\begin{array}{rrrrrrrr}
		 0 &  1 &  0 & -1 &  1 &  0 &  0 &  0 \\
		-1 &  0 &  1 &  0 & -1 &  1 &  0 &  0 \\
		 0 & -1 &  0 &  1 &  0 & -1 &  1 &  0 \\
		 1 &  0 & -1 &  0 &  1 &  0 & -1 &  1 \\
		-1 &  1 &  0 & -1 &  0 &  1 &  0 & -1 \\
		 0 & -1 &  1 &  0 & -1 &  0 &  1 &  0 \\
		 0 &  0 & -1 &  1 &  0 & -1 &  0 &  1 \\
		 0 &  0 &  0 & -1 &  1 &  0 & -1 &  0
	\end{array}
	\right),
\end{equation}
which is precisely the intersection matrix of the $D_4^4[2]$ singularity.
Similarly, the $E_7^{14}[k]$ singularity can be realized as a stabilization of~\eqref{OR singularity} with
$a_0=3$, $a_1=3$, and $a_2=k$.

Let us now consider the $D_N^N[1]$ singularity,
\begin{equation}
	x_1^2+x_2^{N-1}+x_2x_3^2+zx_3=0.
	\label{eq:DNN1}
\end{equation}
After removing the quadratic stabilization, the corresponding
chain exponents are $(a_0,a_1,a_2)=(N-1,2,1)$.
Thus, writing $m=N-2$, one obtains
\begin{equation}
	L^{-1}
	=
	\begin{pmatrix}
		1      & 0      & 0      & \cdots & 0 \\
		-1     & 1      & 0      & \ddots & \vdots \\
		0      & -1     & 1      & \ddots & 0 \\
		\vdots & \ddots & \ddots & \ddots & 0 \\
		0      & \cdots & 0      & -1     & 1
	\end{pmatrix}_{m\times m}.
\end{equation}
Consequently,
\begin{equation}
	B_{ij}
	=\delta_{i,j-1}-\delta_{i-1,j},
\end{equation}
which is the exchange matrix of an oriented $A_{N-2}$ quiver,
up to reversing all arrows.

This result is also manifest directly from the singularity. Under
the holomorphic coordinate transformation
\begin{equation}
	z'=z+x_2x_3,
\end{equation}
one finds
\begin{equation}
	x_1^2+x_2^{N-1}+x_2x_3^2+zx_3
	=
	x_1^2+x_2^{N-1}+z'x_3.
\end{equation}
The latter is the suspension of the $A_{N-2}$ singularity. Therefore
\begin{equation}
	D_N^N[1]\simeq(A_1,A_{N-2}).
	\label{eq:DNN1-equivalence}
\end{equation}
Accordingly, its BPS quiver and Schur index coincide with the
standard results for the $(A_1,A_{N-2})$ theory.

In general, a four-dimensional $\mathcal{N}=2$ SCFT corresponds to a quasi-lisse VOA~\cite{Beem:2017ooy}, and its Schur index involves flavor fugacities $z_i$. In this work, we first focus on singularities corresponding to lisse VOAs. A lisse VOA is defined as a VOA with a trivial associated variety, which equivalently implies that its Zhu $C_2$ algebra is finite-dimensional~\cite{Arakawa:2010dtu}. Physically, this corresponds to a four-dimensional SCFT whose Higgs branch is zero-dimensional.

The singularities associated with lisse VOAs are summarized in Table~\ref{singularitylisse}~\cite{Xie:2019vzr}.
\begin{table}[h]
	\centering
	\begin{tabular}{|c|c|c|c|}
		\hline
		$\mathfrak{j}$  & $b$ & Singularity & condition\\ \hline
		$A_{N-1}$ & $N$ &
		$x_1^2+x_2^2+x_3^N+z^k=0$ & $\gcd(k,N)=1$\\ \hline
		$D_N$ & $2N-2$ &
		$x_1^2+x_2^{N-1}+x_2 x_3^2+z^k=0$ & $\frac{2N-2}{\mathrm{gcd}(k,2N-2)}\text{ is even }\&\mathrm{~gcd}(k,2N-2)\mathrm{~is~odd}$ \\ \hline
		& $N$ &
		$x_1^2+x_2^{N-1}+x_2 x_3^2+z^k x_3=0$ & $\frac{N}{\mathrm{gcd}(k,N)}\mathrm{~is~even}$\\ \hline
		$E_6$ & $12$ &
		$x_1^2+x_2^3+x_3^4+z^k=0$ & $k\neq 3n$\\ \hline
		& $9$ &
		$x_1^2+x_2^3+x_3^4+z^k x_3=0$ & $k\neq 9n$\\ \hline
		$E_7$ & $18$ &
		$x_1^2+x_2^3+x_2 x_3^3+z^k=0$ &$k\neq 2n$\\ \hline
		& $14$ &
		$x_1^2+x_2^3+x_2 x_3^3+z^k x_3=0$ &$k\neq 2n$ \\ \hline
		$E_8$ & $30$ &
		$x_1^2+x_2^3+x_3^5+z^k=0$ &$k\neq 30n$  \\ \hline
		& $24$ &
		$x_1^2+x_2^3+x_3^5+z^k x_3=0$ &$k\neq 24n$ \\ \hline
		& $20$ &
		$x_1^2+x_2^3+x_3^5+z^k x_2=0$ & $k\neq 20n$ \\ \hline
	\end{tabular}
	\caption{Singularities corresponding to lisse VOA}
	\label{singularitylisse}
\end{table}

When $b = h^\vee$, the corresponding singularities are referred to as $(A_{k-1},J)$ singularities. 
For $b \neq h^\vee$, they are usually denoted as $J^b[k]$ singularities.

From this classification, we find that Orlik--Randell type singularities correspond to lisse VOAs when
\begin{equation}
	D_N^N[k]~ \text{with $\frac{N}{\gcd(k,N)}$ is even},~E_7^{14}[k]~ \text{with $k\neq 2n$}.
\end{equation}
For a lisse VOA, the Schur index admits a particularly simple expression,
$$
	\mathcal{I}(q)=(q)_{\infty}^{2\widehat{r}}\sum_{\gamma}S_{\gamma}(q)^2.
$$
Given the BPS quiver, the Schur index is completely determined by the intersection matrix together with a choice of minimal chamber. 
In particular, for a lisse VOA the Schur index can be computed from the corresponding spectrum operator
\begin{equation}
	S_\gamma(q)=\sum_{n_i\geq 0}^{\sum_in_i\gamma^\prime_i=\gamma}\frac{(-q)^{\frac{1}{2}\sum_in_i}q^{\frac{1}{2}\sum_{i<j}\langle\gamma^\prime_i,\gamma^\prime_j\rangle n_in_j}}{\prod_i(q)_{n_i}}.
\end{equation}
Here the data specifying the theory include:
\begin{itemize}
	\item the set of BPS particles in the minimal chamber;
	\item the intersection pairing between their charges, encoded in
	\(\sum_{i<j}\langle \gamma_i',\gamma_j' \rangle\).
\end{itemize}

For other types of lisse VOAs, the associated singularities can be divided into $(G,G^\prime)$-type singularities and certain $J^b[k]$-type singularities. The lisse VOAs associated with the $(A,A)$ type have already been studied in \cite{Nishinaka:2025ytu}. The BPS quiver of the $(G,G')$ type is given by the square tensor product of the $G$- and $G^{\prime}$-type quivers \cite{Cecotti:2010fi}, 
\begin{equation}
	G\square G^{\prime}\,.
\end{equation}
The BPS quiver for $E_6^b[k]$ and $E_8^b[k]$ has the following form \cite{Hosseini:2021ged}, see Table~\ref{tab:square_tensor_examples}.
\begin{table}[htbp]
	\centering
	\begin{tabular}{c c}
		\hline
		Theory & BPS quiver \\
		\hline
		$E_6^{(8)}[k]$  & $A_3 \boxtimes A_2^{2}[k]$ \\
		$E_6^{(9)}[k]$  & $A_2 \boxtimes A_3^{3}[k]$ \\
		$E_8^{(20)}[k]$ & $A_4 \boxtimes A_2^{2}[k]$ \\
		$E_8^{(24)}[k]$ & $A_2 \boxtimes A_4^{4}[k]$ \\
		\hline
	\end{tabular}
	\caption{BPS quiver for $E_6^b[k]$ and $E_8^b[k]$}
	\label{tab:square_tensor_examples}
\end{table}
The BPS quiver of the theory $A_{N}^N[k]$ corresponds to the planar singularity
\begin{equation}
	y^{N+1}+yz^{k}=0\,,
\end{equation}
which generically gives rise to a BPS quiver of the form
\begin{equation}
	\begin{tikzpicture}[x=1.5cm,y=1.5cm]
		\node[circle, draw] (1) at (0,0) {};
		\node[circle, draw] (2) at (0,1) {};
		\node[circle, draw] (3) at (1,0) {};
		\node[circle, draw] (4) at (1,1) {};
		\node[circle, draw] (5) at (2,0) {};
		\node[circle, draw] (6) at (2,1) {};
		\node[circle, draw] (7) at (3,0) {};
		\node[circle, draw] (8) at (3,1) {};
		\node[circle, draw] (9) at (0,2) {};
		\node[circle, draw] (10) at (1,2) {};
		\draw[->] (9) -- (10);
		\draw[->] (1) -- (3);
		\draw[->] (10) -- (4);
		\draw[->] (4) -- (9);
		\draw[->] (9) -- (2);
		\draw[->] (2) -- (1);
		\draw[->] (1) -- (3);
		\draw[->] (3) -- (2);
		\draw[->] (4) -- (3);
		\draw[->] (2) -- (4);
		\draw[->] (6) -- (10);
		\draw[->] (4) -- (6);
		\draw[->] (6) -- (5);
		\draw[->] (5) -- (4);
		\draw[->] (3) -- (5);
		\draw[->] (5) -- (7);
		\draw[->] (7) -- (6);
		\draw[->] (6) -- (8);
		\draw[->] (8) -- (7);
	\end{tikzpicture}
\end{equation}
For lisse VOAs of this type, the minimal chamber no longer enjoys the simple property of being generated by a purely clockwise mutation sequence. Consequently, the corresponding Schur index takes a more intricate form. In general, the Schur index for a lisse VOA is given by \eqref{unflavor}. Since these theories do not possess flavor symmetry, the sum over $\gamma$ corresponds to a sum over charge vectors of the form
\begin{equation}
	\sum_i m_i\gamma_i
\end{equation}
with all $m_i\geq 0$. The structure of $S_{\gamma}(q)$ is governed by the extended intersection matrix
\begin{equation}
	q^{\frac{1}{2}\sum_{i<j}\langle\gamma^\prime_i,\gamma^\prime_j\rangle n_in_j}\,.
\end{equation}
This factor differs from that of the singularity \eqref{generalsingularity}, as the minimal chamber now contains composite charge vectors of the form
\begin{equation}
	\sum_i n_i\gamma_i\,.
\end{equation}

Let us illustrate this mechanism by considering the $(G,G^\prime)$ singularity, whose minimal chamber can be constructed following the prescription of \cite{Cecotti:2010fi}. 
As a concrete example, we study the $(A_{2},D_{4})$ singularity, whose BPS quiver is given by
\begin{equation}
	\begin{tikzpicture}[x=1.5cm,y=1.5cm]
		\node[circle, draw] (1) at (0,0) {};
		\node[circle, draw] (2) at (0,2) {};
		\node[circle, draw] (3) at (2,0) {};
		\node[circle, draw] (4) at (2,2) {};
		\node[circle, draw] (5) at (3.5,0.5) {};
		\node[circle, draw] (6) at (4,-0.5) {};
		\node[circle, draw] (7) at (3.5,2.5) {};
		\node[circle, draw] (8) at (4,1.5) {};
		
		\node[left=0.2cm]  at (1) {1};
		\node[left=0.2cm]  at (2) {2};
		\node[right=0.2cm] at (3) {3};
		\node[right=0.2cm] at (4) {4};
		\node[right=0.2cm] at (5) {5};
		\node[right=0.2cm] at (6) {6};
		\node[left=0.2cm]  at (7) {7};
		\node[left=0.2cm]  at (8) {8};
		\draw[->] (2) -- (4);
		\draw[->] (4) -- (3);
		\draw[->] (3) -- (1);
		\draw[->] (1) -- (2);
		\draw[->] (3) -- (5);
		\draw[->] (5) -- (7);
		\draw[->] (7) -- (4);
		\draw[->] (3) -- (6);
		\draw[->] (6) -- (8);
		\draw[->] (8) -- (4);
		
	\end{tikzpicture}
\end{equation}
This singularity corresponds to a lisse VOA. 
The minimal chamber consists of the following set of BPS charges:
\begin{equation}
	\gamma^\prime=\{\gamma_2,\ \gamma_3,\ \gamma_8,\ \gamma_7,\ 
		\gamma_1+\gamma_2,\ \gamma_3+\gamma_4,\ 
		\gamma_6+\gamma_8,\ \gamma_5+\gamma_7,\ 
		\gamma_6,\ \gamma_4,\ \gamma_1,\ \gamma_5\}\,.
\end{equation}
The corresponding contribution $S_{\gamma}(q)$ takes the form
\begin{equation}
	\sum_{n_i\geq 0}^{\sum_in_i\gamma_i^{\prime}=\gamma}\frac{(-q)^{\frac{1}{2}\sum_in_i}q^{\frac{1}{2}\sum_{i<j}\langle\gamma_i^{\prime},\gamma_j^{\prime}\rangle n_in_j}}{\prod_i(q)_{n_i}}\,.
\end{equation}
The quadratic factor $q^{\frac{1}{2}\sum_{i<j}\langle\gamma^\prime_i,\gamma^\prime_j\rangle n_in_j}=q^{n\cdot A\cdot n^T}$ is determined by the symmetric matrix $A$, which in this case is given explicitly by
\begin{equation}
\left(
\begin{array}{cccccccccccc}
	0 & 0 & 0 & 0 & -\frac{1}{2} & \frac{1}{2} & 0 & 0 & 0 & \frac{1}{2} & -\frac{1}{2} & 0 \\
	0 & 0 & 0 & 0 & \frac{1}{2} & -\frac{1}{2} & \frac{1}{2} & \frac{1}{2} & \frac{1}{2} & -\frac{1}{2} & \frac{1}{2} & \frac{1}{2} \\
	0 & 0 & 0 & 0 & 0 & \frac{1}{2} & -\frac{1}{2} & 0 & -\frac{1}{2} & \frac{1}{2} & 0 & 0 \\
	0 & 0 & 0 & 0 & 0 & \frac{1}{2} & 0 & -\frac{1}{2} & 0 & \frac{1}{2} & 0 & -\frac{1}{2} \\
	-\frac{1}{2} & \frac{1}{2} & 0 & 0 & 0 & 0 & 0 & 0 & 0 & \frac{1}{2} & -\frac{1}{2} & 0 \\
	\frac{1}{2} & -\frac{1}{2} & \frac{1}{2} & \frac{1}{2} & 0 & 0 & 0 & 0 & \frac{1}{2} & -\frac{1}{2} & \frac{1}{2} & \frac{1}{2} \\
	0 & \frac{1}{2} & -\frac{1}{2} & 0 & 0 & 0 & 0 & 0 & -\frac{1}{2} & \frac{1}{2} & 0 & 0 \\
	0 & \frac{1}{2} & 0 & -\frac{1}{2} & 0 & 0 & 0 & 0 & 0 & \frac{1}{2} & 0 & -\frac{1}{2} \\
	0 & \frac{1}{2} & -\frac{1}{2} & 0 & 0 & \frac{1}{2} & -\frac{1}{2} & 0 & 0 & 0 & 0 & 0 \\
	\frac{1}{2} & -\frac{1}{2} & \frac{1}{2} & \frac{1}{2} & \frac{1}{2} & -\frac{1}{2} & \frac{1}{2} & \frac{1}{2} & 0 & 0 & 0 & 0 \\
	-\frac{1}{2} & \frac{1}{2} & 0 & 0 & -\frac{1}{2} & \frac{1}{2} & 0 & 0 & 0 & 0 & 0 & 0 \\
	0 & \frac{1}{2} & 0 & -\frac{1}{2} & 0 & \frac{1}{2} & 0 & -\frac{1}{2} & 0 & 0 & 0 & 0 \\
\end{array}
\right).
\end{equation}
This matrix is obtained by extending the original intersection matrix to include the composite charge vectors
\begin{equation}
	\{ 
	\gamma_1+\gamma_2,\ \gamma_3+\gamma_4,\ 
	\gamma_6+\gamma_8,\ \gamma_5+\gamma_7\}\,.
\end{equation}
In this way, the intersection matrix of the singularity together with the structure of the minimal chamber uniquely determines the extended intersection matrix.
Collecting all contributions and applying the Schur index formula for lisse VOAs \eqref{unflavor}, we obtain the character of the vacuum module of the corresponding VOA.

For singularities corresponding to quasi-lisse VOAs, the Schur index takes the general form
\begin{equation}
	(q)_{\infty}^{2\widehat{r}}\sum_{\gamma,\gamma^{\prime}}^{\gamma-\gamma^{\prime}\in\Gamma_{f}}S_{\gamma}(q)S_{\gamma^{\prime}}(q)X_{\gamma-\gamma^\prime}\,.
\end{equation}
Here the factors $S_{\gamma}(q)$ and $S_{\gamma^{\prime}}(q)$ depend on the intersection matrix and the minimal chamber of the associated BPS quiver, while the condition $\gamma-\gamma^{\prime}\in\Gamma_f$ selects the allowed pairs of charges $(\gamma,\gamma^{\prime})$ contributing to the index.

As an example, for the $(A_2,A_2)$ theory \eqref{box2}, the flavor lattice is generated by
\begin{equation}
	\gamma_1+\gamma_3,\gamma_2+\gamma_4\,.
\end{equation}
These generators impose the constraint
\begin{equation}
	\gamma-\gamma^\prime\in\mathbb{Z}(\gamma_1+\gamma_3)+\mathbb{Z}(\gamma_2+\gamma_4)\,.
\end{equation}
Once these constraints are solved, the Schur index is obtained by evaluating the corresponding $S_{\gamma}(q)$ and $S_{\gamma^{\prime}}(q)$.

In summary, the Schur index is controlled by the functions $S_{\gamma}(q)$, which depend on a symmetric matrix obtained by extending the intersection matrix to include composite BPS charges appearing in the minimal chamber. 
The flavor factor $X_{\gamma-\gamma^{\prime}}$ further constrains the allowed charge pairs contributing to the index.

As discussed in Section~\ref{sec:VOAfromsingularity}, the modular data of the associated VOA can also be extracted from the BPS quiver and is therefore determined by the intersection matrix of the corresponding singularity. 
However, this approach does not appear to work straightforwardly for theories whose Coulomb branches contain operators of integer scaling dimension. 
We leave a systematic study of the modular properties of lisse VOAs, as well as possible extensions to singularities associated with theories possessing only fractional Coulomb branch dimensions, for future work.

\section{Discussions}
	In this paper, we discuss various aspects of the correspondence between quasi-homogeneous isolated hypersurface singularities and the resulting four-dimensional $\mathcal{N}=2$ SCFTs and their associated VOAs, including central charges, Coulomb branch operators, Higgs branches/associated varieties, and the Schur index/vacuum module character. For a large class of cases, these quantities can be determined from singularity theory.
	
	The geometric engineering contains many examples that do not admit a class S realization, and for which the associated VOA data is typically unknown. Through this work, we aim to build a bridge between geometric engineering and the theory of VOAs.

There remain several interesting directions for further exploration:
\begin{itemize}
	\item In this paper, we propose a general method to derive magnetic quivers for singularities with $r=0$, as well as for singularities with $r>0$, $f=b_3=0$, and admitting smooth crepant resolutions. However, the case $r>0$ relies on results from five-dimensional SCFTs, and the method to derive magnetic quivers for general five-dimensional SCFTs remains an open problem. For more general $r>0$ singularities, a systematic derivation of magnetic quivers from singularity theory is still lacking.
	\item  We find that the Schur index is determined by the intersection matrix of the vanishing cycles associated with the deformation of the singularity. However, for a general quasi-homogeneous isolated hypersurface singularity, it remains an open question how to explicitly compute the Seifert matrix or intersection matrix.
	\item In this work, the singularity is realized as a hypersurface in $\mathbb{C}^4$. After performing a transformation on one variable, $z \to e^z$, one obtains a hypersurface in $\mathbb{C}^3 \times \mathbb{C}^*$ \cite{Giacomelli:2017ckh}. Many class $\mathcal{S}$ theories admit realizations in this setting \cite{Giacomelli:2020ryy,Carta:2021whq}. The geometric engineering of theories arising from such hypersurfaces remains to be explored.
	\item As mentioned in Section~\ref{sec:VOAfromsingularity}, techniques based on three-dimensional rank-zero SCFTs can be used to construct modular matrices for the associated VOAs. It would be interesting to apply these methods to a large class of singularities that possess BPS quivers but lack integer scaling dimension Coulomb branch operators. We plan to investigate the modular properties of the VOAs associated with these singularities, in particular the lisse VOAs, in future work.
	
	For theories with integer scaling dimension Coulomb branch operators, the three-dimensional approach typically does not apply straightforwardly. Generalizing these results to encompass general quasi-lisse VOAs remains an open problem.
	\item In Section~\ref{subsec:r=1}, we derive electric quivers for three $r=1$ singularities and identify their class $\mathcal{S}$ realizations and corresponding W-algebras. It would be interesting to derive the associated VOAs directly from these electric quivers.
\end{itemize}

\acknowledgments We would like to thank Andrei Gabrielov, Chiung Hwang, Chunhao Li, Peng Shan and Zhenghao Zhong for helpful discussions. 
WY is supported by National Key R\&D Program of China (No. 2025YFA1017400).
YNW and PY are supported by National Natural Science
Foundation of China under Grant No. 12422503. PY is also supported by National Natural Science Foundation of China under Grant No. 12447142.

\bibliographystyle{JHEP}     
\bibliography{FM}

@article{Collinucci:2021ofd,
    author = "Collinucci, Andr\'es and De Marco, Mario and Sangiovanni, Andrea and Valandro, Roberto",
    title = "{Higgs branches of 5d rank-zero theories from geometry}",
    eprint = "2105.12177",
    archivePrefix = "arXiv",
    primaryClass = "hep-th",
    doi = "10.1007/JHEP10(2021)018",
    journal = "JHEP",
    volume = "10",
    number = "18",
    pages = "018",
    year = "2021"
}

@article{DeMarco:2021try,
    author = "De Marco, Mario and Sangiovanni, Andrea",
    title = "{Higgs Branches of rank-0 5d theories from M-theory on $(A_j,A_l)$ and $(A_k,D_n)$ singularities}",
    eprint = "2111.05875",
    archivePrefix = "arXiv",
    primaryClass = "hep-th",
    month = "11",
    year = "2021"
}

@article{VanBeest:2020kxw,
    author = {Van Beest, Marieke and Bourget, Antoine and Eckhard, Julius and Sch\"afer-Nameki, Sakura},
    title = "{(5d RG-flow) Trees in the Tropical Rain Forest}",
    eprint = "2011.07033",
    archivePrefix = "arXiv",
    primaryClass = "hep-th",
    doi = "10.1007/JHEP03(2021)241",
    journal = "JHEP",
    volume = "03",
    pages = "241",
    year = "2021"
}

@article{Giacomelli:2020ryy,
    author = "Giacomelli, Simone and Mekareeya, Noppadol and Sacchi, Matteo",
    title = "{New aspects of Argyres--Douglas theories and their dimensional reduction}",
    eprint = "2012.12852",
    archivePrefix = "arXiv",
    primaryClass = "hep-th",
    doi = "10.1007/JHEP03(2021)242",
    journal = "JHEP",
    volume = "03",
    pages = "242",
    year = "2021"
}

@article{Closset:2020scj,
    author = "Closset, Cyril and Schafer-Nameki, Sakura and Wang, Yi-Nan",
    title = "{Coulomb and Higgs Branches from Canonical Singularities: Part 0}",
    eprint = "2007.15600",
    archivePrefix = "arXiv",
    primaryClass = "hep-th",
    doi = "10.1007/JHEP02(2021)003",
    journal = "JHEP",
    volume = "02",
    pages = "003",
    year = "2021"
}

@article{Closset:2020afy,
    author = "Closset, Cyril and Giacomelli, Simone and Schafer-Nameki, Sakura and Wang, Yi-Nan",
    title = "{5d and 4d SCFTs: Canonical Singularities, Trinions and S-Dualities}",
    eprint = "2012.12827",
    archivePrefix = "arXiv",
    primaryClass = "hep-th",
    doi = "10.1007/JHEP05(2021)274",
    journal = "JHEP",
    volume = "05",
    pages = "274",
    year = "2021"
}

@Article{Wang:2016yha,
  author        = {Wang, Yifan and Xie, Dan and Yau, Stephen S.T. and Yau, Shing-Tung},
  title         = {{$4d$ $\mathcal{N} = 2$ SCFT from complete intersection singularity}},
  journal       = {Adv. Theor. Math. Phys.},
  year          = {2017},
  volume        = {21},
  pages         = {801--855},
  archiveprefix = {arXiv},
  doi           = {10.4310/ATMP.2017.v21.n3.a6},
  eprint        = {1606.06306},
  primaryclass  = {hep-th},
}

@article{Grimminger:2020dmg,
    author = "Grimminger, Julius F. and Hanany, Amihay",
    title = "{Hasse Diagrams for $\mathbf{3d}$ $\mathbf{\mathcal{N}=4}$ Quiver Gauge Theories -- Inversion and the full Moduli Space}",
    eprint = "2004.01675",
    archivePrefix = "arXiv",
    primaryClass = "hep-th",
    month = "4",
    year = "2020"
}

@article{Gaiotto:2009we,
      author         = "Gaiotto, Davide",
      title          = "{N=2 dualities}",
      journal        = "JHEP",
      volume         = "08",
      year           = "2012",
      pages          = "034",
      doi            = "10.1007/JHEP08(2012)034",
      eprint         = "0904.2715",
      archivePrefix  = "arXiv",
      primaryClass   = "hep-th",
      SLACcitation   = "%%CITATION = ARXIV:0904.2715;%%"
}

@article{Apruzzi:2019opn,
      author         = "Apruzzi, Fabio and Lawrie, Craig and Lin, Ling and
                        Schafer-Nameki, Sakura and Wang, Yi-Nan",
      title          = "{Fibers add Flavor, Part I: Classification of 5d SCFTs,
                        Flavor Symmetries and BPS States}",
      journal        = "JHEP",
      volume         = "11",
      year           = "2019",
      pages          = "068",
      doi            = "10.1007/JHEP11(2019)068",
      eprint         = "1907.05404",
      archivePrefix  = "arXiv",
      primaryClass   = "hep-th",
      SLACcitation   = "%%CITATION = ARXIV:1907.05404;%%"
}

@Article{Cabrera:2018jxt,
  author        = {Cabrera, Santiago and Hanany, Amihay and Yagi, Futoshi},
  title         = {{Tropical Geometry and Five Dimensional Higgs Branches at Infinite Coupling}},
  journal       = {JHEP},
  year          = {2019},
  volume        = {01},
  pages         = {068},
  archiveprefix = {arXiv},
  doi           = {10.1007/JHEP01(2019)068},
  eprint        = {1810.01379},
  primaryclass  = {hep-th},
  reportnumber  = {Imperial/TP/18/AH/10},
  slaccitation  = {%%CITATION = ARXIV:1810.01379;%%},
}

@Article{Morrison:1996xf,
  author =        {Morrison, David R. and Seiberg, Nathan},
  title =         {{Extremal transitions and five-dimensional supersymmetric field theories}},
  journal =       {Nucl. Phys.},
  year =          {1997},
  volume =        {B483},
  pages =         {229-247},
  archiveprefix = {arXiv},
  doi =           {10.1016/S0550-3213(96)00592-5},
  eprint =        {hep-th/9609070},
  primaryclass =  {hep-th},
  reportnumber =  {DUKE-TH-96-130, RU-96-80},
  slaccitation =  {%%CITATION = HEP-TH/9609070;%%}
}

@article{Ferlito:2017xdq,
      author         = "Ferlito, Giulia and Hanany, Amihay and Mekareeya,
                        Noppadol and Zafrir, Gabi",
      title          = "{3d Coulomb branch and 5d Higgs branch at infinite
                        coupling}",
      journal        = "JHEP",
      volume         = "07",
      year           = "2018",
      pages          = "061",
      doi            = "10.1007/JHEP07(2018)061",
      eprint         = "1712.06604",
      archivePrefix  = "arXiv",
      primaryClass   = "hep-th",
      reportNumber   = "IMPERIAL-TP-17-AH-08",
      SLACcitation   = "%%CITATION = ARXIV:1712.06604;%%"
}

@article{Lawrie:2012gg,
      author         = "Lawrie, Craig and Schafer-Nameki, Sakura",
      title          = "{The Tate Form on Steroids: Resolution and Higher
                        Codimension Fibers}",
      journal        = "JHEP",
      volume         = "04",
      year           = "2013",
      pages          = "061",
      doi            = "10.1007/JHEP04(2013)061",
      eprint         = "1212.2949",
      archivePrefix  = "arXiv",
      primaryClass   = "hep-th",
      reportNumber   = "KCL-MTH-12-14",
      SLACcitation   = "%%CITATION = ARXIV:1212.2949;%%"
}

@Article{Jefferson:2018irk,
  author        = {Jefferson, Patrick and Katz, Sheldon and Kim, Hee-Cheol and Vafa, Cumrun},
  title         = {{On Geometric Classification of 5d SCFTs}},
  journal       = {JHEP},
  year          = {2018},
  volume        = {04},
  pages         = {103},
  archiveprefix = {arXiv},
  doi           = {10.1007/JHEP04(2018)103},
  eprint        = {1801.04036},
  primaryclass  = {hep-th},
  slaccitation  = {%%CITATION = ARXIV:1801.04036;%%},
}

@Article{Intriligator:1997pq,
  author        = {Intriligator, Kenneth A. and Morrison, David R. and Seiberg, Nathan},
  title         = {{Five-dimensional supersymmetric gauge theories and degenerations of Calabi-Yau spaces}},
  journal       = {Nucl. Phys.},
  year          = {1997},
  volume        = {B497},
  pages         = {56-100},
  archiveprefix = {arXiv},
  doi           = {10.1016/S0550-3213(97)00279-4},
  eprint        = {hep-th/9702198},
  primaryclass  = {hep-th},
  reportnumber  = {RU-96-99, IASSNS-HEP-96-112},
  slaccitation  = {%%CITATION = HEP-TH/9702198;%%},
}

@Article{derenthal2014singular,
  author    = {Derenthal, Ulrich},
  title     = {Singular Del Pezzo surfaces whose universal torsors are hypersurfaces},
  journal   = {Proceedings of the London Mathematical Society},
  year      = {2014},
  volume    = {108},
  number    = {3},
  pages     = {638--681},
  publisher = {Wiley Online Library},
}

@Article{Witten:1996qb,
  author        = {Witten, Edward},
  title         = {{Phase transitions in M theory and F theory}},
  journal       = {Nucl. Phys.},
  year          = {1996},
  volume        = {B471},
  pages         = {195-216},
  archiveprefix = {arXiv},
  doi           = {10.1016/0550-3213(96)00212-X},
  eprint        = {hep-th/9603150},
  primaryclass  = {hep-th},
  reportnumber  = {IASSNS-HEP-96-26},
  slaccitation  = {%%CITATION = HEP-TH/9603150;%%},
}

@article{Cecotti:2010bp,
    author = "Cecotti, Sergio and Cordova, Clay and Heckman, Jonathan J. and Vafa, Cumrun",
    title = "{T-Branes and Monodromy}",
    eprint = "1010.5780",
    archivePrefix = "arXiv",
    primaryClass = "hep-th",
    doi = "10.1007/JHEP07(2011)030",
    journal = "JHEP",
    volume = "07",
    pages = "030",
    year = "2011"
}

@Article{Gopakumar:1998jq,
  author        = {Gopakumar, Rajesh and Vafa, Cumrun},
  title         = {{M theory and topological strings. 2.}},
  year          = {1998},
  archiveprefix = {arXiv},
  eprint        = {hep-th/9812127},
  primaryclass  = {hep-th},
  reportnumber  = {HUTP-98-A070},
  slaccitation  = {%%CITATION = HEP-TH/9812127;%%},
}

@Article{Cecotti:2013lda,
  author        = {Cecotti, Sergio and Del Zotto, Michele and Giacomelli, Simone},
  title         = {{More on the N=2 superconformal systems of type $D_p(G)$}},
  journal       = {JHEP},
  year          = {2013},
  volume        = {04},
  pages         = {153},
  archiveprefix = {arXiv},
  doi           = {10.1007/JHEP04(2013)153},
  eprint        = {1303.3149},
  primaryclass  = {hep-th},
}

@Article{Chen:2016bzh,
  author        = {Chen, Bingyi and Xie, Dan and Yau, Shing-Tung and Yau, Stephen S. -T. and Zuo, Huaiqing},
  title         = {{4D $\mathcal{N} = 2$ SCFT and singularity theory. Part II: complete intersection}},
  journal       = {Adv. Theor. Math. Phys.},
  year          = {2017},
  volume        = {21},
  pages         = {121--145},
  archiveprefix = {arXiv},
  doi           = {10.4310/ATMP.2017.v21.n1.a2},
  eprint        = {1604.07843},
  primaryclass  = {hep-th},
}

@Article{Chen:2017wkw,
  author        = {Chen, Bingyi and Xie, Dan and Yau, Stephen S.T. and Yau, Shing-Tung and Zuo, Huaiqing},
  title         = {{4d $\mathcal{N}=2$ SCFT and singularity theory Part III: Rigid singularity}},
  journal       = {Adv. Theor. Math. Phys.},
  year          = {2018},
  volume        = {22},
  pages         = {1885--1905},
  archiveprefix = {arXiv},
  doi           = {10.4310/ATMP.2018.v22.n8.a2},
  eprint        = {1712.00464},
  primaryclass  = {hep-th},
}

@Article{yau2005classification,
  author    = {Yau, Stephen S-T and Yu, Yung},
  title     = {Classification of 3-dimensional isolated rational hypersurface singularities with C*-action},
  journal   = {The Rocky Mountain Journal of Mathematics},
  year      = {2005},
  volume    = {35},
  number    = {5},
  pages     = {1795--1809},
  publisher = {JSTOR},
}

@article{Seiberg:1996ns,
    author = "Seiberg, Nathan and Shenker, Stephen H.",
    title = "{Hypermultiplet moduli space and string compactification to three-dimensions}",
    eprint = "hep-th/9608086",
    archivePrefix = "arXiv",
    reportNumber = "RU-96-68",
    doi = "10.1016/S0370-2693(96)01189-6",
    journal = "Phys. Lett. B",
    volume = "388",
    pages = "521--523",
    year = "1996"
}

@Article{Xie:2015rpa,
  author        = {Xie, Dan and Yau, Shing-Tung},
  title         = {{4d N=2 SCFT and singularity theory Part I: Classification}},
  year          = {2015},
  month         = {10},
  archiveprefix = {arXiv},
  eprint        = {1510.01324},
  primaryclass  = {hep-th},
}

@Article{Shapere:1999xr,
  author        = {Shapere, Alfred D. and Vafa, Cumrun},
  title         = {{BPS structure of Argyres-Douglas superconformal theories}},
  year          = {1999},
  month         = {10},
  archiveprefix = {arXiv},
  eprint        = {hep-th/9910182},
  reportnumber  = {HUTP-99-A057, UKHEP-99-15},
}

@Article{Bourget:2019aer,
  author        = {Bourget, Antoine and Cabrera, Santiago and Grimminger, Julius F. and Hanany, Amihay and Sperling, Marcus and Zajac, Anton and Zhong, Zhenghao},
  title         = {{The Higgs mechanism --- Hasse diagrams for symplectic singularities}},
  journal       = {JHEP},
  year          = {2020},
  volume        = {01},
  pages         = {157},
  archiveprefix = {arXiv},
  doi           = {10.1007/JHEP01(2020)157},
  eprint        = {1908.04245},
  primaryclass  = {hep-th},
  reportnumber  = {Imperial/TP/19/AH/02},
}

@Article{Cabrera:2018ann,
  author        = {Cabrera, Santiago and Hanany, Amihay},
  title         = {{Quiver Subtractions}},
  journal       = {JHEP},
  year          = {2018},
  volume        = {09},
  pages         = {008},
  archiveprefix = {arXiv},
  doi           = {10.1007/JHEP09(2018)008},
  eprint        = {1803.11205},
  primaryclass  = {hep-th},
}

@Article{Gukov:1999ya,
  author        = {Gukov, Sergei and Vafa, Cumrun and Witten, Edward},
  title         = {{CFT's from Calabi-Yau four folds}},
  journal       = {Nucl. Phys. B},
  year          = {2000},
  volume        = {584},
  pages         = {69--108},
  note          = {[Erratum: Nucl.Phys.B 608, 477--478 (2001)]},
  archiveprefix = {arXiv},
  doi           = {10.1016/S0550-3213(00)00373-4},
  eprint        = {hep-th/9906070},
  reportnumber  = {HUTP-99-A034, IASSNS-HEP-99-52, PUPT-1864},
}

@article{Klemm:1996bj,
    author = "Klemm, Albrecht and Lerche, Wolfgang and Mayr, Peter and Vafa, Cumrun and Warner, Nicholas P.",
    title = "{Selfdual strings and N=2 supersymmetric field theory}",
    eprint = "hep-th/9604034",
    archivePrefix = "arXiv",
    reportNumber = "CERN-TH-96-95, HUTP-96-A014, USC-96-008",
    doi = "10.1016/0550-3213(96)00353-7",
    journal = "Nucl. Phys. B",
    volume = "477",
    pages = "746--766",
    year = "1996"
}

@article{Hori:1997zj,
    author = "Hori, Kentaro and Ooguri, Hirosi and Vafa, Cumrun",
    title = "{NonAbelian conifold transitions and N=4 dualities in three-dimensions}",
    eprint = "hep-th/9705220",
    archivePrefix = "arXiv",
    reportNumber = "HUTP-97-A024, LBL-40349, LBNL-40349, UCB-PTH-97-27",
    doi = "10.1016/S0550-3213(97)00529-4",
    journal = "Nucl. Phys. B",
    volume = "504",
    pages = "147--174",
    year = "1997"
}

@article{Argyres:1995jj,
    author = "Argyres, Philip C. and Douglas, Michael R.",
    title = "{New phenomena in SU(3) supersymmetric gauge theory}",
    eprint = "hep-th/9505062",
    archivePrefix = "arXiv",
    reportNumber = "IASSNS-HEP-95-31, RU-95-28",
    doi = "10.1016/0550-3213(95)00281-V",
    journal = "Nucl. Phys. B",
    volume = "448",
    pages = "93--126",
    year = "1995"
}

@article{Ooguri:1996me,
    author = "Ooguri, Hirosi and Vafa, Cumrun",
    title = "{Summing up D instantons}",
    eprint = "hep-th/9608079",
    archivePrefix = "arXiv",
    reportNumber = "HUTP-96-A036, UCB-PTH-96-36, LBL-39220, LBNL-39220",
    doi = "10.1103/PhysRevLett.77.3296",
    journal = "Phys. Rev. Lett.",
    volume = "77",
    pages = "3296--3298",
    year = "1996"
}

@article{Intriligator:1996ex,
    author = "Intriligator, Kenneth A. and Seiberg, N.",
    title = "{Mirror symmetry in three-dimensional gauge theories}",
    eprint = "hep-th/9607207",
    archivePrefix = "arXiv",
    reportNumber = "RU-96-63, IASSNS-HEP-96-80",
    doi = "10.1016/0370-2693(96)01088-X",
    journal = "Phys. Lett. B",
    volume = "387",
    pages = "513--519",
    year = "1996"
}

@article{Cecotti:2010fi,
    author = "Cecotti, Sergio and Neitzke, Andrew and Vafa, Cumrun",
    title = "{R-Twisting and 4d/2d Correspondences}",
    eprint = "1006.3435",
    archivePrefix = "arXiv",
    primaryClass = "hep-th",
    month = "6",
    year = "2010"
}

@article{Xie:2012hs,
    author = "Xie, Dan",
    title = "{General Argyres-Douglas Theory}",
    eprint = "1204.2270",
    archivePrefix = "arXiv",
    primaryClass = "hep-th",
    doi = "10.1007/JHEP01(2013)100",
    journal = "JHEP",
    volume = "01",
    pages = "100",
    year = "2013"
}

@article{Alim:2011ae,
    author = "Alim, Murad and Cecotti, Sergio and Cordova, Clay and Espahbodi, Sam and Rastogi, Ashwin and Vafa, Cumrun",
    title = "{BPS Quivers and Spectra of Complete N=2 Quantum Field Theories}",
    eprint = "1109.4941",
    archivePrefix = "arXiv",
    primaryClass = "hep-th",
    doi = "10.1007/s00220-013-1789-8",
    journal = "Commun. Math. Phys.",
    volume = "323",
    pages = "1185--1227",
    year = "2013"
}

@article{Cecotti:2011rv,
    author = "Cecotti, Sergio and Vafa, Cumrun",
    title = "{Classification of complete N=2 supersymmetric theories in 4 dimensions}",
    eprint = "1103.5832",
    archivePrefix = "arXiv",
    primaryClass = "hep-th",
    month = "3",
    year = "2011"
}

@article{Seiberg:1994rs,
    author = "Seiberg, N. and Witten, Edward",
    title = "{Electric - magnetic duality, monopole condensation, and confinement in N=2 supersymmetric Yang-Mills theory}",
    eprint = "hep-th/9407087",
    archivePrefix = "arXiv",
    reportNumber = "RU-94-52, IASSNS-HEP-94-43",
    doi = "10.1016/0550-3213(94)90124-4",
    journal = "Nucl. Phys. B",
    volume = "426",
    pages = "19--52",
    year = "1994",
    note = "[Erratum: Nucl.Phys.B 430, 485--486 (1994)]"
}

@article{Gaiotto:2009hg,
    author = "Gaiotto, Davide and Moore, Gregory W. and Neitzke, Andrew",
    title = "{Wall-crossing, Hitchin systems, and the WKB approximation}",
    eprint = "0907.3987",
    archivePrefix = "arXiv",
    primaryClass = "hep-th",
    doi = "10.1016/j.aim.2012.09.027",
    journal = "Adv. Math.",
    volume = "234",
    pages = "239--403",
    year = "2013"
}

@article{Shapere:2008zf,
    author = "Shapere, Alfred D. and Tachikawa, Yuji",
    title = "{Central charges of N=2 superconformal field theories in four dimensions}",
    eprint = "0804.1957",
    archivePrefix = "arXiv",
    primaryClass = "hep-th",
    doi = "10.1088/1126-6708/2008/09/109",
    journal = "JHEP",
    volume = "09",
    pages = "109",
    year = "2008"
}

@article{Kapustin:1999ha,
    author = "Kapustin, Anton and Strassler, Matthew J.",
    title = "{On mirror symmetry in three-dimensional Abelian gauge theories}",
    eprint = "hep-th/9902033",
    archivePrefix = "arXiv",
    reportNumber = "IASSNS-HEP-99-15",
    doi = "10.1088/1126-6708/1999/04/021",
    journal = "JHEP",
    volume = "04",
    pages = "021",
    year = "1999"
}

@article{Witten:2003ya,
    author = "Witten, Edward",
    editor = "Shifman, M. and Vainshtein, A. and Wheater, J.",
    title = "{SL(2,Z) action on three-dimensional conformal field theories with Abelian symmetry}",
    eprint = "hep-th/0307041",
    archivePrefix = "arXiv",
    pages = "1173--1200",
    month = "7",
    year = "2003"
}

@article{Benvenuti:2018bav,
    author = "Benvenuti, Sergio",
    title = "{A tale of exceptional $3d$ dualities}",
    eprint = "1809.03925",
    archivePrefix = "arXiv",
    primaryClass = "hep-th",
    doi = "10.1007/JHEP03(2019)125",
    journal = "JHEP",
    volume = "03",
    pages = "125",
    year = "2019"
}

@article{Dedushenko:2019mnd,
    author = "Dedushenko, Mykola and Wang, Yifan",
    title = "{4d/2d $\rightarrow $ 3d/1d: A song of protected operator algebras}",
    eprint = "1912.01006",
    archivePrefix = "arXiv",
    primaryClass = "hep-th",
    reportNumber = "CALT-TH 2019-041, PUPT-2602",
    month = "12",
    year = "2019"
}

@article{Cecotti:2011gu,
    author = "Cecotti, Sergio and Del Zotto, Michele",
    title = "{On Arnold's 14 `exceptional' N=2 superconformal gauge theories}",
    eprint = "1107.5747",
    archivePrefix = "arXiv",
    primaryClass = "hep-th",
    doi = "10.1007/JHEP10(2011)099",
    journal = "JHEP",
    volume = "10",
    pages = "099",
    year = "2011"
}

@book{arnold2012singularities,
  title={Singularities of Differentiable Maps, Volume 2: Monodromy and Asymptotics of Integrals},
  author={Arnold, E. and Gusein-Zade, S.M. and Varchenko, A.N.},
  isbn={9780817683436},
  lccn={2012938547},
  series={Modern Birkh{\"a}user Classics},
   year={2012},
  publisher={Birkh{\"a}user Boston}
}

@article{Buican:2015hsa,
    author = "Buican, Matthew and Nishinaka, Takahiro",
    title = "{Argyres--Douglas theories, S$^1$ reductions, and topological symmetries}",
    eprint = "1505.06205",
    archivePrefix = "arXiv",
    primaryClass = "hep-th",
    reportNumber = "RU-NHETC-2015-02",
    doi = "10.1088/1751-8113/49/4/045401",
    journal = "J. Phys. A",
    volume = "49",
    number = "4",
    pages = "045401",
    year = "2016"
}

@Article{Buican:2015ina,
  author        = {Buican, Matthew and Nishinaka, Takahiro},
  title         = {{On the superconformal index of Argyres--Douglas theories}},
  journal       = {J. Phys. A},
  year          = {2016},
  volume        = {49},
  number        = {1},
  pages         = {015401},
  archiveprefix = {arXiv},
  doi           = {10.1088/1751-8113/49/1/015401},
  eprint        = {1505.05884},
  primaryclass  = {hep-th},
  reportnumber  = {RU-NHETC-2015-01},
}

@article{Beem:2013sza,
    author = "Beem, Christopher and Lemos, Madalena and Liendo, Pedro and Peelaers, Wolfger and Rastelli, Leonardo and van Rees, Balt C.",
    title = "{Infinite Chiral Symmetry in Four Dimensions}",
    eprint = "1312.5344",
    archivePrefix = "arXiv",
    primaryClass = "hep-th",
    reportNumber = "YITP-SB-13-45, CERN-PH-TH-2013-311, HU-EP-13-78",
    doi = "10.1007/s00220-014-2272-x",
    journal = "Commun. Math. Phys.",
    volume = "336",
    number = "3",
    pages = "1359--1433",
    year = "2015"
}

@article{Beem:2017ooy,
    author = "Beem, Christopher and Rastelli, Leonardo",
    title = "{Vertex operator algebras, Higgs branches, and modular differential equations}",
    eprint = "1707.07679",
    archivePrefix = "arXiv",
    primaryClass = "hep-th",
    reportNumber = "YITP-SB-17-27",
    doi = "10.1007/JHEP08(2018)114",
    journal = "JHEP",
    volume = "08",
    pages = "114",
    year = "2018"
}

@Article{Giacomelli:2017ckh,
  author        = {Giacomelli, Simone},
  journal       = {JHEP},
  title         = {{RG flows with supersymmetry enhancement and geometric engineering}},
  year          = {2018},
  pages         = {156},
  volume        = {06},
  archiveprefix = {arXiv},
  doi           = {10.1007/JHEP06(2018)156},
  eprint        = {1710.06469},
  primaryclass  = {hep-th},
}

@article{Tian:2021cif,
    author = "Tian, Jiahua and Wang, Yi-Nan",
    title = "{5D and 6D SCFTs from $\mathbb{C}^3$ orbifolds}",
    eprint = "2110.15129",
    archivePrefix = "arXiv",
    primaryClass = "hep-th",
    doi = "10.21468/SciPostPhys.12.4.127",
    journal = "SciPost Phys.",
    volume = "12",
    number = "4",
    pages = "127",
    year = "2022"
}

@article{Carta:2021dyx,
    author = "Carta, Federico and Giacomelli, Simone and Mekareeya, Noppadol and Mininno, Alessandro",
    title = "{Conformal Manifolds and 3d Mirrors of $(D_n,D_m)$ Theories}",
    eprint = "2110.06940",
    archivePrefix = "arXiv",
    primaryClass = "hep-th",
    reportNumber = "IFT-UAM/CSIC-21-109, ZMP-HH/21-20",
    month = "10",
    year = "2021"
}

@article{Hosseini:2021ged,
    author = "Hosseini, Saghar S. and Moscrop, Robert",
    title = "{Maruyoshi-Song flows and defect groups of $ {\mathrm{D}}_{\mathrm{p}}^{\mathrm{b}} $(G) theories}",
    eprint = "2106.03878",
    archivePrefix = "arXiv",
    primaryClass = "hep-th",
    doi = "10.1007/JHEP10(2021)119",
    journal = "JHEP",
    volume = "10",
    pages = "119",
    year = "2021"
}

@article{Closset:2021lwy,
    author = {Closset, Cyril and Sch\"afer-Nameki, Sakura and Wang, Yi-Nan},
    title = "{Coulomb and Higgs branches from canonical singularities. Part I. Hypersurfaces with smooth Calabi-Yau resolutions}",
    eprint = "2111.13564",
    archivePrefix = "arXiv",
    primaryClass = "hep-th",
    doi = "10.1007/JHEP04(2022)061",
    journal = "JHEP",
    volume = "04",
    pages = "061",
    year = "2022"
}

@article{DeMarco:2022dgh,
    author = "De Marco, Mario and Sangiovanni, Andrea and Valandro, Roberto",
    title = "{5d Higgs branches from M-theory on quasi-homogeneous cDV threefold singularities}",
    eprint = "2205.01125",
    archivePrefix = "arXiv",
    primaryClass = "hep-th",
    doi = "10.1007/JHEP10(2022)124",
    journal = "JHEP",
    volume = "10",
    pages = "124",
    year = "2022"
}

@article{Gopakumar:1998ii,
    author = "Gopakumar, Rajesh and Vafa, Cumrun",
    title = "{M theory and topological strings. 1.}",
    eprint = "hep-th/9809187",
    archivePrefix = "arXiv",
    reportNumber = "HUTP-98-A069",
    month = "9",
    year = "1998"
}

@article{Bennett:2024loi,
    author = "Bennett, Sam and Hanany, Amihay and Kumaran, Guhesh and Li, Chunhao and Liu, Deshuo and Sperling, Marcus",
    title = "{Quiver Subtraction on the Higgs Branch}",
    eprint = "2409.16356",
    archivePrefix = "arXiv",
    primaryClass = "hep-th",
    reportNumber = "Imperial/TP/24/AH/03",
    month = "9",
    year = "2024"
}

@article{Collinucci:2021wty,
    author = "Collinucci, Andr\'es and Sangiovanni, Andrea and Valandro, Roberto",
    title = "{Genus zero Gopakumar-Vafa invariants from open strings}",
    eprint = "2104.14493",
    archivePrefix = "arXiv",
    primaryClass = "hep-th",
    doi = "10.1007/JHEP09(2021)059",
    journal = "JHEP",
    volume = "09",
    pages = "059",
    year = "2021"
}

@article{Saueressig:2007dr,
    author = "Saueressig, Frank and Vandoren, Stefan",
    title = "{Conifold singularities, resumming instantons and non-perturbative mirror symmetry}",
    eprint = "0704.2229",
    archivePrefix = "arXiv",
    primaryClass = "hep-th",
    reportNumber = "ITP-UU-07-21, SPIN-07-14",
    doi = "10.1088/1126-6708/2007/07/018",
    journal = "JHEP",
    volume = "07",
    pages = "018",
    year = "2007"
}

@article{Alexandrov:2007ec,
    author = "Alexandrov, Sergei",
    title = "{Quantum covariant c-map}",
    eprint = "hep-th/0702203",
    archivePrefix = "arXiv",
    reportNumber = "PTA-07-03",
    doi = "10.1088/1126-6708/2007/05/094",
    journal = "JHEP",
    volume = "05",
    pages = "094",
    year = "2007"
}

@article{Carta:2022spy,
    author = "Carta, Federico and Giacomelli, Simone and Mekareeya, Noppadol and Mininno, Alessandro",
    title = "{Dynamical consequences of 1-form symmetries and the exceptional Argyres-Douglas theories}",
    eprint = "2203.16550",
    archivePrefix = "arXiv",
    primaryClass = "hep-th",
    reportNumber = "ZMP-HH/22-7",
    doi = "10.1007/JHEP06(2022)059",
    journal = "JHEP",
    volume = "06",
    pages = "059",
    year = "2022"
}

@article{Xie:2021ewm,
    author = "Xie, Dan",
    title = "{3d mirror for Argyres-Douglas theories}",
    eprint = "2107.05258",
    archivePrefix = "arXiv",
    primaryClass = "hep-th",
    month = "7",
    year = "2021"
}

@article{Collinucci:2022rii,
    author = "Collinucci, Andr\'es and De Marco, Mario and Sangiovanni, Andrea and Valandro, Roberto",
    title = "{Flops of any length, Gopakumar-Vafa invariants and 5d Higgs branches}",
    eprint = "2204.10366",
    archivePrefix = "arXiv",
    primaryClass = "hep-th",
    doi = "10.1007/JHEP08(2022)292",
    journal = "JHEP",
    volume = "08",
    pages = "292",
    year = "2022"
}

@article{Xie:2019vzr,
    author = "Xie, Dan and Yan, Wenbin",
    title = "{4d $\mathcal{N}=2$ SCFTs and lisse W-algebras}",
    eprint = "1910.02281",
    archivePrefix = "arXiv",
    primaryClass = "hep-th",
    doi = "10.1007/JHEP04(2021)271",
    journal = "JHEP",
    volume = "04",
    pages = "271",
    year = "2021"
}

@article{Alim:2011kw,
    author = "Alim, Murad and Cecotti, Sergio and Cordova, Clay and Espahbodi, Sam and Rastogi, Ashwin and Vafa, Cumrun",
    title = "{$\mathcal{N} = 2$ quantum field theories and their BPS quivers}",
    eprint = "1112.3984",
    archivePrefix = "arXiv",
    primaryClass = "hep-th",
    doi = "10.4310/ATMP.2014.v18.n1.a2",
    journal = "Adv. Theor. Math. Phys.",
    volume = "18",
    number = "1",
    pages = "27--127",
    year = "2014"
}

@inproceedings{Aspinwall:2004jr,
    author = "Aspinwall, Paul S.",
    title = "{D-branes on Calabi-Yau manifolds}",
    booktitle = "{Theoretical Advanced Study Institute in Elementary Particle Physics (TASI 2003): Recent Trends in String Theory}",
    eprint = "hep-th/0403166",
    archivePrefix = "arXiv",
    reportNumber = "DUKE-CGTP-04-04",
    doi = "10.1142/9789812775108_0001",
    pages = "1--152",
    month = "3",
    year = "2004"
}

@article{Li:2022njl,
    author = "Li, Bohan and Xie, Dan and Yan, Wenbin",
    title = "{On low rank 4d $ \mathcal{N} $ = 2 SCFTs}",
    eprint = "2212.03089",
    archivePrefix = "arXiv",
    primaryClass = "hep-th",
    doi = "10.1007/JHEP05(2023)132",
    journal = "JHEP",
    volume = "05",
    pages = "132",
    year = "2023"
}

@article{Song:2017oew,
    author = "Song, Jaewon and Xie, Dan and Yan, Wenbin",
    title = "{Vertex operator algebras of Argyres-Douglas theories from M5-branes}",
    eprint = "1706.01607",
    archivePrefix = "arXiv",
    primaryClass = "hep-th",
    reportNumber = "KIAS-P17032",
    doi = "10.1007/JHEP12(2017)123",
    journal = "JHEP",
    volume = "12",
    pages = "123",
    year = "2017"
}

@article{Arakawa:2017aon,
    author = "Arakawa, Tomoyuki",
    title = "{Associated varieties and Higgs branches (a survey)}",
    eprint = "1712.01945",
    archivePrefix = "arXiv",
    primaryClass = "math.RT",
    journal = "Contemp. Math.",
    volume = "711",
    pages = "37--44",
    year = "2018"
}

@article{Xie:2017vaf,
    author = "Xie, Dan and Yau, Shing-Tung",
    title = "{Argyres-Douglas matter and N=2 dualities}",
    eprint = "1701.01123",
    archivePrefix = "arXiv",
    primaryClass = "hep-th",
    month = "1",
    year = "2017"
}

@article{Xie:2017aqx,
    author = "Xie, Dan and Ye, Ke",
    title = "{Argyres-Douglas matter and S-duality: Part II}",
    eprint = "1711.06684",
    archivePrefix = "arXiv",
    primaryClass = "hep-th",
    doi = "10.1007/JHEP03(2018)186",
    journal = "JHEP",
    volume = "03",
    pages = "186",
    year = "2018"
}

@article{Hanany:2011db,
    author = "Hanany, Amihay and Mekareeya, Noppadol",
    title = "{Complete Intersection Moduli Spaces in N=4 Gauge Theories in Three Dimensions}",
    eprint = "1110.6203",
    archivePrefix = "arXiv",
    primaryClass = "hep-th",
    reportNumber = "MPP-2011-93, IMPERIAL-TP-11-AH-07",
    doi = "10.1007/JHEP01(2012)079",
    journal = "JHEP",
    volume = "01",
    pages = "079",
    year = "2012"
}

@article{Benvenuti:2010pq,
    author = "Benvenuti, Sergio and Hanany, Amihay and Mekareeya, Noppadol",
    title = "{The Hilbert Series of the One Instanton Moduli Space}",
    eprint = "1005.3026",
    archivePrefix = "arXiv",
    primaryClass = "hep-th",
    doi = "10.1007/JHEP06(2010)100",
    journal = "JHEP",
    volume = "06",
    pages = "100",
    year = "2010"
}

@article{Hanany:2010qu,
    author = "Hanany, Amihay and Mekareeya, Noppadol",
    title = "{Tri-vertices and SU(2)'s}",
    eprint = "1012.2119",
    archivePrefix = "arXiv",
    primaryClass = "hep-th",
    reportNumber = "IMPERIAL-TP-10-AH-07",
    doi = "10.1007/JHEP02(2011)069",
    journal = "JHEP",
    volume = "02",
    pages = "069",
    year = "2011"
}

@article{Greene:1996dh,
    author = "Greene, Brian R. and Morrison, David R. and Vafa, Cumrun",
    title = "{A Geometric realization of confinement}",
    eprint = "hep-th/9608039",
    archivePrefix = "arXiv",
    reportNumber = "CU-TP-769, DUKE-TH-96-125, HUTP-96-A033",
    doi = "10.1016/S0550-3213(96)00465-8",
    journal = "Nucl. Phys. B",
    volume = "481",
    pages = "513--538",
    year = "1996"
}

@article{Beem:2014rza,
    author = "Beem, Christopher and Peelaers, Wolfger and Rastelli, Leonardo and van Rees, Balt C.",
    title = "{Chiral algebras of class S}",
    eprint = "1408.6522",
    archivePrefix = "arXiv",
    primaryClass = "hep-th",
    reportNumber = "YITP-SB-14-30, CERN-PH-TH-2014-165, YITP-SB-14-30, CERN-PH-TH-2014-165",
    doi = "10.1007/JHEP05(2015)020",
    journal = "JHEP",
    volume = "05",
    pages = "020",
    year = "2015"
}

@article{Arakawa:2010ni,
    author = "Arakawa, Tomoyuki",
    title = "{Associated varieties of modules over Kac-Moody algebras and C(2)-cofiniteness of W-algebras}",
    eprint = "1004.1554",
    archivePrefix = "arXiv",
    primaryClass = "math.QA",
    month = "4",
    year = "2010"
}

@article{Cordova:2015nma,
    author = "Cordova, Clay and Shao, Shu-Heng",
    title = "{Schur Indices, BPS Particles, and Argyres-Douglas Theories}",
    eprint = "1506.00265",
    archivePrefix = "arXiv",
    primaryClass = "hep-th",
    doi = "10.1007/JHEP01(2016)040",
    journal = "JHEP",
    volume = "01",
    pages = "040",
    year = "2016"
}

@article{Kim:2024dxu,
    author = "Kim, Heeyeon and Song, Jaewon",
    title = "{A Family of Vertex Algebras from Argyres-Douglas Theory}",
    eprint = "2412.20015",
    archivePrefix = "arXiv",
    primaryClass = "hep-th",
    month = "12",
    year = "2024"
}

@article{Xie:2012gd,
    author = "Xie, Dan",
    title = "{BPS spectrum, wall crossing and quantum dilogarithm identity}",
    eprint = "1211.7071",
    archivePrefix = "arXiv",
    primaryClass = "hep-th",
    doi = "10.4310/ATMP.2016.v20.n3.a1",
    journal = "Adv. Theor. Math. Phys.",
    volume = "20",
    pages = "405--524",
    year = "2016"
}

@article{Xie:2012jd,
    author = "Xie, Dan",
    title = "{Network, cluster coordinates and N = 2 theory II: Irregular singularity}",
    eprint = "1207.6112",
    archivePrefix = "arXiv",
    primaryClass = "hep-th",
    month = "7",
    year = "2012"
}

@article{DelZotto:2011an,
    author = "Del Zotto, Michele",
    title = "{More Arnold's N = 2 superconformal gauge theories}",
    eprint = "1110.3826",
    archivePrefix = "arXiv",
    primaryClass = "hep-th",
    doi = "10.1007/JHEP11(2011)115",
    journal = "JHEP",
    volume = "11",
    pages = "115",
    year = "2011"
}

@article{Alvarez-Garcia:2023gdd,
    author = "{\'A}lvarez-Garc{\'\i}a, Rafael and Lee, Seung-Joo and Weigand, Timo",
    title = "{Non-minimal elliptic threefolds at infinite distance. Part I. Log Calabi-Yau resolutions}",
    eprint = "2310.07761",
    archivePrefix = "arXiv",
    primaryClass = "hep-th",
    reportNumber = "CTPU-PTC-23-44, ZMP-HH/23-14",
    doi = "10.1007/JHEP08(2024)240",
    journal = "JHEP",
    volume = "08",
    pages = "240",
    year = "2024"
}

@article{Alvarez-Garcia:2023qqj,
    author = "{\'A}lvarez-Garc{\'\i}a, Rafael and Lee, Seung-Joo and Weigand, Timo",
    title = "{Non-minimal elliptic threefolds at infinite distance II: asymptotic physics}",
    eprint = "2312.11611",
    archivePrefix = "arXiv",
    primaryClass = "hep-th",
    reportNumber = "CTPU-PTC-23-54, ZMP-HH/23-22",
    doi = "10.1007/JHEP01(2025)058",
    journal = "JHEP",
    volume = "01",
    pages = "058",
    year = "2025"
}

@article{Beem:2023ofp,
    author = "Beem, Christopher and Martone, Mario and Sacchi, Matteo and Singh, Palash and Stedman, Jake",
    title = "{Simplifying the Type $A$ Argyres-Douglas Landscape}",
    eprint = "2311.12123",
    archivePrefix = "arXiv",
    primaryClass = "hep-th",
    month = "11",
    year = "2023"
}

@article{Xie:2019yds,
    author = "Xie, Dan and Yan, Wenbin",
    title = "{W algebras, cosets and VOAs for 4d $ \mathcal{N} $ = 2 SCFTs from M5 branes}",
    eprint = "1902.02838",
    archivePrefix = "arXiv",
    primaryClass = "hep-th",
    doi = "10.1007/JHEP04(2021)076",
    journal = "JHEP",
    volume = "04",
    pages = "076",
    year = "2021"
}

@article{Bourget:2023dkj,
    author = "Bourget, Antoine and Sperling, Marcus and Zhong, Zhenghao",
    title = "{Decay and Fission of Magnetic Quivers}",
    eprint = "2312.05304",
    archivePrefix = "arXiv",
    primaryClass = "hep-th",
    reportNumber = "UWThPh2023-33",
    doi = "10.1103/PhysRevLett.132.221603",
    journal = "Phys. Rev. Lett.",
    volume = "132",
    number = "22",
    pages = "221603",
    year = "2024"
}

@article{Bourget:2024mgn,
    author = "Bourget, Antoine and Sperling, Marcus and Zhong, Zhenghao",
    title = "{Higgs branch RG flows via decay and fission}",
    eprint = "2401.08757",
    archivePrefix = "arXiv",
    primaryClass = "hep-th",
    reportNumber = "UWThPh2024-4",
    doi = "10.1103/PhysRevD.109.126013",
    journal = "Phys. Rev. D",
    volume = "109",
    number = "12",
    pages = "126013",
    year = "2024"
}

@article{Lawrie:2024wan,
    author = "Lawrie, Craig and Mansi, Lorenzo and Sperling, Marcus and Zhong, Zhenghao",
    title = "{Pathway to decay and fission of orthosymplectic quiver theories}",
    eprint = "2412.15202",
    archivePrefix = "arXiv",
    primaryClass = "hep-th",
    reportNumber = "DESY-24-185, UWThPh 2024-26",
    doi = "10.1103/flb9-6nm3",
    journal = "Phys. Rev. D",
    volume = "112",
    number = "2",
    pages = "026025",
    year = "2025"
}

@article{Arakawa:2018egx,
    author = "Arakawa, Tomoyuki",
    title = "{Chiral algebras of class $\mathcal{S}$ and Moore-Tachikawa symplectic varieties}",
    eprint = "1811.01577",
    archivePrefix = "arXiv",
    primaryClass = "math.RT",
    month = "11",
    year = "2018"
}

@article{Yanagida:2020kim,
    author = "Yanagida, Shintarou",
    title = "{Derived gluing construction of chiral algebras}",
    eprint = "2004.10055",
    archivePrefix = "arXiv",
    primaryClass = "math.QA",
    doi = "10.1007/s11005-021-01394-1",
    journal = "Lett. Math. Phys.",
    volume = "111",
    number = "2",
    pages = "51",
    year = "2021"
}

@article{Bonetti:2018fqz,
    author = "Bonetti, Federico and Meneghelli, Carlo and Rastelli, Leonardo",
    title = "{VOAs labelled by complex reflection groups and 4d SCFTs}",
    eprint = "1810.03612",
    archivePrefix = "arXiv",
    primaryClass = "hep-th",
    doi = "10.1007/JHEP05(2019)155",
    journal = "JHEP",
    volume = "05",
    pages = "155",
    year = "2019"
}

@article{Arakawa:2023cki,
    author = {Arakawa, Tomoyuki and Kuwabara, Toshiro and M{\"o}ller, Sven},
    title = "{Hilbert Schemes of Points in the Plane and Quasi-Lisse Vertex Algebras with $\mathcal{N}=4$ Symmetry}",
    eprint = "2309.17308",
    archivePrefix = "arXiv",
    primaryClass = "math.RT",
    month = "9",
    year = "2023"
}

@article{Carta:2021whq,
    author = "Carta, Federico and Giacomelli, Simone and Mekareeya, Noppadol and Mininno, Alessandro",
    title = "{Conformal manifolds and 3d mirrors of Argyres-Douglas theories}",
    eprint = "2105.08064",
    archivePrefix = "arXiv",
    primaryClass = "hep-th",
    reportNumber = "IFT-UAM/CSIC-21-55",
    doi = "10.1007/JHEP08(2021)015",
    journal = "JHEP",
    volume = "08",
    pages = "015",
    year = "2021"
}

@article{Shan:2023xtw,
    author = "Shan, Peng and Xie, Dan and Yan, Wenbin",
    title = "{Mirror symmetry for circle compactified 4d $\mathcal{N}=2$ SCFTs}",
    eprint = "2306.15214",
    archivePrefix = "arXiv",
    primaryClass = "hep-th",
    month = "6",
    year = "2023"
}

@article{Braden:2014iea,
    author = "Braden, Tom and Licata, Anthony and Proudfoot, Nicholas and Webster, Ben",
    title = "{Quantizations of conical symplectic resolutions II: category $\mathcal O$ and symplectic duality}",
    eprint = "1407.0964",
    archivePrefix = "arXiv",
    primaryClass = "math.RT",
    journal = "Asterisque",
    volume = "384",
    pages = "75--179",
    year = "2016"
}

@article{Kamnitzer:2022nzd,
    author = "Kamnitzer, Joel",
    title = "{Symplectic resolutions, symplectic duality, and Coulomb branches}",
    eprint = "2202.03913",
    archivePrefix = "arXiv",
    primaryClass = "math.RT",
    doi = "10.1112/blms.12711",
    journal = "Bull. London Math. Soc.",
    volume = "54",
    number = "5",
    pages = "1515--1551",
    year = "2024"
}

@article{Cabrera:2017njm,
    author = "Cabrera, Santiago and Hanany, Amihay",
    title = "{Branes and the Kraft-Procesi transition: classical case}",
    eprint = "1711.02378",
    archivePrefix = "arXiv",
    primaryClass = "hep-th",
    doi = "10.1007/JHEP04(2018)127",
    journal = "JHEP",
    volume = "04",
    pages = "127",
    year = "2018"
}

@book{arnold1998singularity,
  title={Singularity theory I},
  author={Arnold, Vladimir Igorevi{\v{c}} and Goryunov, Victor V and Lyashko, OV and Vasil'ev, Valerij Aleksandrovi{\v{c}}},
  volume={1},
  year={1998},
  publisher={Springer Science \& Business Media}
}

@article{gabrielov1973intersection,
  title={Intersection matrices for certain singularities},
  author={Gabrielov, Andrei M},
  journal={Functional Analysis and its Applications},
  volume={7},
  number={3},
  pages={182--193},
  year={1973},
  publisher={Springer}
}

@article{gabrielov1979polar,
  title={Polar curves and intersection matrices of singularities},
  author={Gabrielov, Andrei M},
  journal={Inventiones mathematicae},
  volume={54},
  number={1},
  pages={15--22},
  year={1979},
  publisher={Springer}
}

@article{Cordova:2016uwk,
    author = "Cordova, Clay and Gaiotto, Davide and Shao, Shu-Heng",
    title = "{Infrared Computations of Defect Schur Indices}",
    eprint = "1606.08429",
    archivePrefix = "arXiv",
    primaryClass = "hep-th",
    doi = "10.1007/JHEP11(2016)106",
    journal = "JHEP",
    volume = "11",
    pages = "106",
    year = "2016"
}

@article{Nishinaka:2025ytu,
    author = "Nishinaka, Takahiro and Yoshida, Yutaka",
    title = "{3d Chern--Simons matter theories from generalized Argyres--Douglas theories}",
    eprint = "2512.15201",
    archivePrefix = "arXiv",
    primaryClass = "hep-th",
    month = "12",
    year = "2025"
}

@article{orlik1977monodromy,
  title={The monodromy of weighted homogeneous singularities},
  author={Orlik, Peter and Randell, Richard},
  journal={Inventiones mathematicae},
  volume={39},
  number={3},
  pages={199--211},
  year={1977},
  publisher={Springer-Verlag Berlin/Heidelberg}
}

@article{kac1989classification,
  title={Classification of modular invariant representations of affine algebras. $^1$},
  author={Kac, VG and Wakimoto, M},
  journal={Infinite Dimensional Lie algebras and groups},
  volume={7},
  pages={138},
  year={1989},
  publisher={World Scientific}
}

@article{Gaiotto:2024ioj,
    author = "Gaiotto, Davide and Kim, Heeyeon",
    title = "{3D TFTs from 4d $ \mathcal{N} $ = 2 BPS particles}",
    eprint = "2409.20393",
    archivePrefix = "arXiv",
    primaryClass = "hep-th",
    doi = "10.1007/JHEP03(2025)173",
    journal = "JHEP",
    volume = "03",
    pages = "173",
    year = "2025"
}

@article{Dedushenko:2018bpp,
    author = "Dedushenko, Mykola and Gukov, Sergei and Nakajima, Hiraku and Pei, Du and Ye, Ke",
    title = "{3d TQFTs from Argyres{\textendash}Douglas theories}",
    eprint = "1809.04638",
    archivePrefix = "arXiv",
    primaryClass = "hep-th",
    reportNumber = "CALT-TH-2018-033",
    doi = "10.1088/1751-8121/abb481",
    journal = "J. Phys. A",
    volume = "53",
    number = "43",
    pages = "43LT01",
    year = "2020"
}

@article{Ferrari:2023fez,
    author = "Ferrari, Andrea E. V. and Garner, Niklas and Kim, Heeyeon",
    title = "{Boundary vertex algebras for 3d $\mathcal{N}=4$ rank-0 SCFTs}",
    eprint = "2311.05087",
    archivePrefix = "arXiv",
    primaryClass = "hep-th",
    doi = "10.21468/SciPostPhys.17.2.057",
    journal = "SciPost Phys.",
    volume = "17",
    number = "2",
    pages = "057",
    year = "2024"
}

@article{Arakawa:2010dtu,
    author = "Arakawa, Tomoyuki",
    title = "{A remark on the C 2-cofiniteness condition on vertex algebras}",
    eprint = "1004.1492",
    archivePrefix = "arXiv",
    primaryClass = "math.QA",
    doi = "10.1007/s00209-010-0812-4",
    journal = "Math. Z.",
    volume = "270",
    number = "1",
    pages = "559--575",
    year = "2012"
}

@article{Arakawa:2016hkg,
    author = "Arakawa, Tomoyuki and Kawasetsu, Kazuya",
    title = "{Quasi-lisse vertex algebras and modular linear differential equations}",
    eprint = "1610.05865",
    archivePrefix = "arXiv",
    primaryClass = "math.QA",
    month = "10",
    year = "2016"
}

@article{Kim:2025rog,
    author = "Kim, Minsung and Kim, Sungjoon",
    title = "{3D TFTs and boundary VOAs from BPS spectra of $(G,G')$ Argyres-Douglas theories}",
    eprint = "2511.23194",
    archivePrefix = "arXiv",
    primaryClass = "hep-th",
    reportNumber = "KIAS-Q25021",
    month = "11",
    year = "2025"
}

@article{Go:2025ixu,
    author = "Go, Byeonggi and Jia, Qiang and Kim, Heeyeon and Kim, Sungjoon",
    title = "{From BPS spectra of Argyres-Douglas theories to families of 3d TFTs}",
    eprint = "2502.15133",
    archivePrefix = "arXiv",
    primaryClass = "hep-th",
    doi = "10.1007/JHEP08(2025)012",
    journal = "JHEP",
    volume = "08",
    pages = "012",
    year = "2025"
}

@article{ArabiArdehali:2024ysy,
    author = "Arabi Ardehali, Arash and Dedushenko, Mykola and Gang, Dongmin and Litvinov, Mikhail",
    title = "{Bridging 4D QFTs and 2D VOAs via 3D high-temperature EFTs}",
    eprint = "2409.18130",
    archivePrefix = "arXiv",
    primaryClass = "hep-th",
    reportNumber = "YITP-SB-2024-13",
    month = "9",
    year = "2024"
}

@book{Tachikawa:2013kta,
    author = "Tachikawa, Yuji",
    title = "{N=2 supersymmetric dynamics for pedestrians}",
    eprint = "1312.2684",
    archivePrefix = "arXiv",
    primaryClass = "hep-th",
    reportNumber = "UT-13-42, IPMU-13-0234, UT-13-42, IPMU-13-0234",
    doi = "10.1007/978-3-319-08822-8",
    volume = "890",
    month = "12",
    year = "2013"
}

@article{Akhond:2021xio,
    author = "Akhond, Mohammad and Arias-Tamargo, Guillermo and Mininno, Alessandro and Sun, Hao-Yu and Sun, Zhengdi and Wang, Yifan and Xu, Fengjun",
    title = "{The hitchhiker's guide to 4d $\mathcal{N}=2$ superconformal field theories}",
    eprint = "2112.14764",
    archivePrefix = "arXiv",
    primaryClass = "hep-th",
    reportNumber = "IFT-UAM/CSIC-21-151, ZMP-HH/21-28",
    doi = "10.21468/SciPostPhysLectNotes.64",
    journal = "SciPost Phys. Lect. Notes",
    volume = "64",
    pages = "1",
    year = "2022"
}

@article{Mu:2023uws,
    author = "Mu, Jisheng and Wang, Yi-Nan and Zhang, Hao N.",
    title = "{5d SCFTs from isolated complete intersection singularities}",
    eprint = "2311.05441",
    archivePrefix = "arXiv",
    primaryClass = "hep-th",
    doi = "10.1007/JHEP02(2024)155",
    journal = "JHEP",
    volume = "02",
    pages = "155",
    year = "2024"
}

@article{Gadde:2011uv,
    author = "Gadde, Abhijit and Rastelli, Leonardo and Razamat, Shlomo S. and Yan, Wenbin",
    title = "{Gauge Theories and Macdonald Polynomials}",
    eprint = "1110.3740",
    archivePrefix = "arXiv",
    primaryClass = "hep-th",
    reportNumber = "YITP-SB-11-30",
    doi = "10.1007/s00220-012-1607-8",
    journal = "Commun. Math. Phys.",
    volume = "319",
    pages = "147--193",
    year = "2013"
}

@article{Lemos:2014lua,
    author = "Lemos, Madalena and Peelaers, Wolfger",
    title = "{Chiral Algebras for Trinion Theories}",
    eprint = "1411.3252",
    archivePrefix = "arXiv",
    primaryClass = "hep-th",
    reportNumber = "YITP-SB-14-41",
    doi = "10.1007/JHEP02(2015)113",
    journal = "JHEP",
    volume = "02",
    pages = "113",
    year = "2015"
}

@article{Song:2015wta,
    author = "Song, Jaewon",
    title = "{Superconformal indices of generalized Argyres-Douglas theories from 2d TQFT}",
    eprint = "1509.06730",
    archivePrefix = "arXiv",
    primaryClass = "hep-th",
    doi = "10.1007/JHEP02(2016)045",
    journal = "JHEP",
    volume = "02",
    pages = "045",
    year = "2016"
}

@article{Gadde:2011ik,
    author = "Gadde, Abhijit and Rastelli, Leonardo and Razamat, Shlomo S. and Yan, Wenbin",
    title = "{The 4d Superconformal Index from q-deformed 2d Yang-Mills}",
    eprint = "1104.3850",
    archivePrefix = "arXiv",
    primaryClass = "hep-th",
    reportNumber = "YITP-SB-11-13",
    doi = "10.1103/PhysRevLett.106.241602",
    journal = "Phys. Rev. Lett.",
    volume = "106",
    pages = "241602",
    year = "2011"
}

@article{Buican:2017uka,
    author = "Buican, Matthew and Nishinaka, Takahiro",
    title = "{On Irregular Singularity Wave Functions and Superconformal Indices}",
    eprint = "1705.07173",
    archivePrefix = "arXiv",
    primaryClass = "hep-th",
    reportNumber = "QMUL-PH-17-XX",
    doi = "10.1007/JHEP09(2017)066",
    journal = "JHEP",
    volume = "09",
    pages = "066",
    year = "2017"
}

@article{Mekareeya:2012tn,
    author = "Mekareeya, Noppadol and Song, Jaewon and Tachikawa, Yuji",
    title = "{2d TQFT structure of the superconformal indices with outer-automorphism twists}",
    eprint = "1212.0545",
    archivePrefix = "arXiv",
    primaryClass = "hep-th",
    reportNumber = "MPP-2012-153, UCSD-PTH-12-18, IPMU-12-0219, UT-12-40",
    doi = "10.1007/JHEP03(2013)171",
    journal = "JHEP",
    volume = "03",
    pages = "171",
    year = "2013"
}

@article{Lemos:2012ph,
    author = "Lemos, Madalena and Peelaers, Wolfger and Rastelli, Leonardo",
    title = "{The superconformal index of class $S$ theories of type $D$}",
    eprint = "1212.1271",
    archivePrefix = "arXiv",
    primaryClass = "hep-th",
    reportNumber = "YITP-SB-12-45",
    doi = "10.1007/JHEP05(2014)120",
    journal = "JHEP",
    volume = "05",
    pages = "120",
    year = "2014"
}

@article{Dolan:2002zh,
    author = "Dolan, F. A. and Osborn, H.",
    title = "{On short and semi-short representations for four-dimensional superconformal symmetry}",
    eprint = "hep-th/0209056",
    archivePrefix = "arXiv",
    reportNumber = "DAMTP-02-114",
    doi = "10.1016/S0003-4916(03)00074-5",
    journal = "Annals Phys.",
    volume = "307",
    pages = "41--89",
    year = "2003"
}

@incollection{wemyss2023lockdown,
  title={A lockdown survey on cDV singularities},
  author={Wemyss, Michael},
  booktitle={McKay Correspondence, Mutation and Related Topics},
  volume={88},
  pages={47--94},
  year={2023},
  publisher={Mathematical Society of Japan}
}

@article{ArabiArdehali:2024vli,
    author = "Arabi Ardehali, Arash and Gang, Dongmin and Rajappa, Neville Joshua and Sacchi, Matteo",
    title = "{3d SUSY enhancement and non-semisimple TQFTs from four dimensions}",
    eprint = "2411.00766",
    archivePrefix = "arXiv",
    primaryClass = "hep-th",
    reportNumber = "YITP-SB-2024-26",
    doi = "10.1007/JHEP09(2025)179",
    journal = "JHEP",
    volume = "09",
    pages = "179",
    year = "2025"
}

@article{varolgunes2023seifert,
  title={Seifert form of chain-type invertible singularities},
  author={Varolgunes, Umut},
  journal={Kyoto Journal of Mathematics},
  volume={63},
  number={1},
  pages={155--193},
  year={2023},
  publisher={Duke University Press}
}

\end{document}